\documentclass[11pt]{article}
\usepackage[margin=1in,bottom=1in]{geometry}
\usepackage[utf8]{inputenc}
\usepackage[T1]{fontenc}
\usepackage[english]{babel}

\usepackage{tabularx}
\usepackage{mdframed}
\usepackage{libertine}
\usepackage{amsmath}
\usepackage{amsthm}
\DeclareMathOperator{\symdiff}{\triangle}
\usepackage{amssymb}
\usepackage[only,lightning]{stmaryrd}
\newcommand{\disagrees}{\mathrel{\mkern-5mu\lightning\mkern-7mu}}
\usepackage{mathrsfs} 
\usepackage{mathtools}
\usepackage{thm-restate}
\usepackage[colorlinks = true,linkcolor = blue,urlcolor = blue, citecolor = blue]{hyperref}
\usepackage[capitalize, nameinlink]{cleveref}
\usepackage{url}
\usepackage[disable]{todonotes}
\usepackage{mathrsfs} 
\usepackage{soul}
\usepackage{relsize}
\usepackage{enumitem}
\usepackage[all,defaultlines=3]{nowidow}
\usepackage{multicol}
\usepackage{enumitem}
\usepackage{xspace}
\usepackage{nicefrac}
\usepackage{comment}
\usepackage[linesnumbered,ruled]{algorithm2e}
\usepackage{textpos}
\setlist[itemize]{topsep=4pt,itemsep=3pt,parsep=0pt} 
\setlist[enumerate]{topsep=4pt,itemsep=3pt,parsep=0pt} 

\usetikzlibrary{decorations.markings}

\crefname{algocf}{Algorithm}{Algorithms}
\crefname{claim}{Claim}{Claims}
\crefname{figure}{Figure}{Figures}
\renewcommand{\preceq}{\preccurlyeq}

\renewcommand{\subset}{\subseteq}
\newtheorem{theorem}{Theorem}[section]

\newtheorem*{theorem*}{Theorem}

\newtheorem{corollary}[theorem]{Corollary}
\newtheorem{conjecture}{Conjecture}

\newtheorem{observation}[theorem]{Observation}
\newtheorem{lemma}[theorem]{Lemma}

\newtheorem{claim}[theorem]{Claim}

\theoremstyle{definition}
\newtheorem{definition}[theorem]{Definition}
\theoremstyle{plain}

\theoremstyle{definition}

\newtheorem*{example*}{Example}

\numberwithin{equation}{section}

\renewcommand{\cal}{\mathcal}

\newenvironment{claimproof}[1][Proof of the claim.]{%
  \begin{proof}[#1]%
}{%
  \end{proof}%
}

\newcommand{\adj}{\mathsf{adj}}

\newcommand{\wei}{\mathbf{w}}

\newcommand{\Pp}{\mathcal{P}}
\newcommand{\Qq}{\mathcal{Q}}
\newcommand{\Rr}{\mathcal{R}}
\newcommand{\Ww}{\mathcal{W}}
\newcommand{\Nn}{\mathcal{N}}

\newcommand{\PP}{\mathcal{P}}

\newcommand{\tup}[1]{{\bar{#1}}}

\def\phi{\varphi}
\newcommand{\N}{\mathbb{N}}
\newcommand{\R}{\mathbb{R}}

\newcommand{\Ball}{\mathrm{Ball}}
\newcommand{\dist}{\mathrm{dist}}

\newcommand{\Aa}{\mathcal{A}}

\newcommand{\Tt}{\mathcal{T}}

\newcommand{\Ff}{\mathcal{F}}
\newcommand{\Ss}{\mathcal{S}}

\newcommand{\sdiff}{\Delta}
\newcommand{\spoly}{\mathrm{subpoly}}
\newcommand{\hamm}{\mathrm{ham}}
\newcommand{\hammsub}{\hamm^{\scriptscriptstyle\subseteq}}

\newcommand{\from}{\colon}
\newcommand{\Cc}{\mathscr C}
\newcommand{\CC}{\mathscr C}
\newcommand{\Dd}{\mathscr D}
\def\epsilon{\varepsilon}
\def\eps{\varepsilon}
\newcommand{\Oh}{\mathcal{O}}
\newcommand{\Oht}{\widehat{\Oh}}

\renewcommand{\emptyset}{\varnothing}

\renewcommand{\leq}{\leqslant}
\renewcommand{\geq}{\geqslant}
\renewcommand{\le}{\leq}
\renewcommand{\ge}{\geq}

\renewcommand{\setminus}{-}

\usepackage{tikz}

\SetKwBlock{Repeat}{repeat}{}
\let\oldnl\nl
\newcommand{\nonl}{\renewcommand{\nl}{\let\nl\oldnl}}

\newcommand{\affi}[1]{\textcolor{black!50}{#1}}

\tikzstyle{vertex}=[circle,draw=black,fill=gray,minimum size=0.15cm,inner sep=0pt]

\tikzstyle{svertex}=[circle,draw=black,fill=black,minimum size=0.11cm,inner sep=0pt]

\newcommand{\pleq}{\sqsubseteq}
\newcommand{\pgeq}{\sqsupseteq}

\begin{document}
\pagenumbering{roman}
\pagestyle{empty}

\newcommand{\funding}{
JD: Funded by the Deutsche Forschungsgemeinschaft (DFG, German Research Foundation) -- project number 563448077.
  NM and ST received funding from the European Research Council (ERC) with grant agreement No.\ 101126229 -- BUKA.
  NM also received funding from the European Union through an ERA Fellowship with grant agreement No.\ 101334340 -- LoCoMoDe.
  A large part of this work was done while MP was affiliated with the University of Warsaw and was supported by the project BOBR that has received funding from the European Research Council (ERC) under the European Union's Horizon 2020 research and innovation programme, grant agreement No.\ 948057. During the employment at TU Darmstadt, MP was supported by the Alexander von Humboldt Foundation in the
  framework of the Alexander von Humboldt Professorship, endowed by the Federal Ministry of Research, Technology, and Space, as well as by the European Union through European Research Council project WYDRA, grant agreement no. 101230549.\\[0.2cm] Views and opinions expressed are those of the authors only and do not necessarily reflect those of the European Union or the European Research Council Executive Agency. Neither the European Union nor the granting authority can be held responsible for them.
}

\renewcommand\footnotemark{}
\title{Monadically Dependent Graph Classes Have\\ Almost Bounded Merge-Width\thanks{\funding}}
\date{}

\author{
	Jan Dreier\\{\small \affi{HPI Potsdam}}\\{\small \affi{Germany}}
	\and
	Nikolas Mählmann\\{\small \affi{University of Warsaw}}\\{\small \affi{Poland}}
	\and
	Michał Pilipczuk\\{\small \affi{TU Darmstadt}}\\{\small \affi{Germany}}
	\and
	Szymon Toru\'nczyk\\{\small \affi{University of Warsaw}}\\{\small \affi{Poland}}
}

\maketitle

\begin{abstract}
	A graph class $\Cc$ is monadically dependent if one cannot interpret all graphs in vertex-colored graphs from $\Cc$ using a fixed formula of first-order logic. We prove that every monadically dependent graph class has \emph{almost bounded merge-width}, thereby confirming a conjecture of the first and the last author~[STOC~'25]. 
	Our proof is algorithmic and the witnessing decompositions in the form of \emph{merge sequences} can be computed in nearly-cubic time.
	This is a step towards the proof of fixed-parameter tractability of model checking first-order logic on monadically dependent graph classes, which is achieved in a companion paper.	
\end{abstract}

 \begin{textblock}{20}(-2.65, 3.9)
 \includegraphics[width=100px]{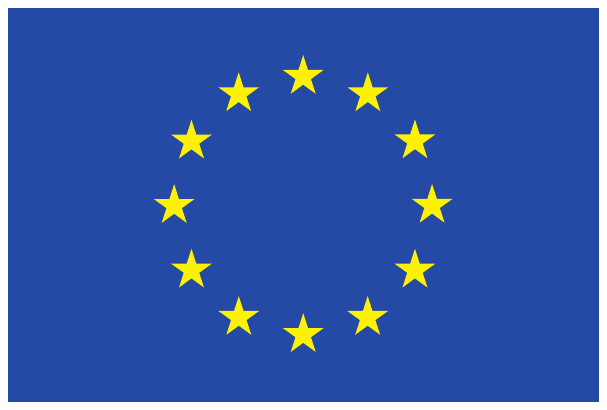}%
 \end{textblock}

\thispagestyle{empty}

\newpage
\setcounter{tocdepth}{1}
\tableofcontents

\clearpage
\pagestyle{plain}
\pagenumbering{arabic}

\section{Introduction}\label{sec:intro}
It is conjectured that various well-studied graph problems---such as the independent set problem, the dominating set problem, or more generally, problems expressible in first-order logic---become tractable,  in the sense of parameterized complexity, on graph classes which are \emph{monadically dependent}.
A graph class $\CC$ is \emph{monadically independent} if it encodes all graphs as follows:
there is some number $k$ and first-order formula $\phi(x,y)$ in the signature of $k$-colored graphs, such that every graph $H$ can be obtained by the following recipe. First pick some graph $G\in\CC$ and its $k$-coloring $G^+$; then construct the graph whose edges are the vertex pairs satisfying $\phi(x,y)\lor \phi(y,x)$ in $G^+$; finally remove some vertices, obtaining the target graph~$H$. Conversely, $\CC$ is \emph{monadically dependent} if it is not monadically independent.
Examples include the class of planar graphs, graph classes with bounded maximum degree, every class excluding a fixed graph as a minor, the class of unit interval graphs, and many other classes of graphs studied in the literature.

Thus, monadic dependence is defined as a natural dividing line for the \emph{expressive power} of graph classes with respect to first-order logic, classifying them into those where arbitrarily complicated relations can be encoded (monadically independent classes) and those where any fixed encoding mechanism is in some way restricted (monadically dependent classes). Monadic dependence has been the subject of intensive investigations over the last 10 years, mostly due to the following conjecture that predicts that this expressive-power dividing line coincides with the dividing line for the algorithmic tractability of first-order logic.

\begin{conjecture}[\cite{Warwick16,GajarskyHOLR20}]\label{conj:main}
	Let $\Cc$ be a monadically dependent class of graphs. Then the model checking problem for first-order logic on $\Cc$ is fixed-parameter tractable: There exists an algorithm that given a graph $G\in \Cc$ and a first-order sentence $\varphi$, verifies whether $\varphi$ holds in $G$ in time $\Oh_{\Cc,\varphi}(|G|^c)$, for some constant $c$.
\end{conjecture}

As observed by Adler and Adler~\cite{AdlerA14} following a result of Podewski and Ziegler~\cite{PodewskiZ78}, on monotone\footnote{A graph class is \emph{monotone} if it is closed under taking subgraphs, and \emph{hereditary} if it is closed under taking induced subgraphs.} graph classes monadic dependence coincides with the concept of \emph{nowhere denseness} of Ne\v{s}et\v{r}il and Ossona de Mendez~\cite{NesetrilM11}. Hence, in those classes \cref{conj:main} is verified by the landmark result of Grohe, Kreutzer, and Siebertz~\cite{GroheKS17}. 
Combined, the results of Dreier, M\"ahlmann, and Siebertz~\cite{dreier2023first} and
Dreier, Eleftheriadis, M\"ahlmann, McCarty, Pilipczuk, and Toru\'nczyk~\cite{DreierEMMPT24} proved \cref{conj:main} in \emph{monadically stable} classes: a property lying strictly between nowhere denseness and monadic dependence that postulates non-interpretability of all \emph{linear orders}, instead of all graphs, using a fixed first-order formula in vertex-colored graphs from the class. The conjecture was also confirmed in classes of ordered graphs by Bonnet, Giocanti, Ossona de Mendez, Simon, Thomass\'e, and Toru\'nczyk~\cite{BonnetGMSTT24}; there, monadic dependence coincides with the boundedness of the graph parameter~\emph{twin-width}. On the negative side, Dreier, M\"ahlmann, and Toru\'nczyk proved that in monadically independent graph classes that are hereditary, the model checking problem for first-order logic is as hard as on general graphs, that is, $\mathsf{AW}[\star]$-hard~\cite{DreierMT24}. Thus, a confirmation of \cref{conj:main} would prove that on hereditary graph classes, monadic dependence precisely delimits fixed-parameter tractability of model checking first-order logic (assuming $\mathsf{FPT}\neq \mathsf{AW}[\star]$).

We refer the reader to the introductory sections of~\cite{DreierEMMPT24,DreierMT24} for a broader introduction to the background of \cref{conj:main} and the already sizeable body of work around it.

Recently, Dreier and Toru\'nczyk~\cite{DreierT25} proposed a line of attack on \cref{conj:main}. They introduced a family of graph parameters called \emph{merge-width}, indexed by a radius parameter $r\in \N$, and conjectured the following two assertions that together would amount to a proof of \cref{conj:main}:
\begin{enumerate}[label={(\arabic*)}]
	\item\label{s:abmw} Monadically dependent graph classes $\CC$ have \emph{almost bounded merge-width}: for every $r\in \N$, the radius-$r$ merge-width of any graph $G\in \Cc$ is bounded by a subpolynomial\footnote{A function $f\colon \N\to \N$ is \emph{subpolynomial} if for every $\eps>0$, $\lim_{n\to \infty} \frac{f(n)}{n^\eps}=0$.} function of the size of $G$. (The function may depend on $\Cc$ and $r$.) Moreover, suitable decompositions (called \emph{merge sequences}) witnessing the almost boundedness of merge-width can be efficiently constructed.

	\newcommand{\footnoteExample}{\footnote{Examples are all classes that are nowhere dense but have unbounded expansion~\cite{DreierT25}, such as the class of graphs whose girth is bigger than their maximum degree~\cite[Lem.\ 8]{pilipczuk2017sparsity-lecture1}.}}
	Note that there are monadically dependent classes that have unbounded merge-width\footnoteExample, which explains why only almost boundedness is conjectured.

	\item\label{s:mc}   Model checking first-order logic can be performed in fixed-parameter time on any hereditary graph class of almost bounded merge-width.
\end{enumerate}
To corroborate the viability of this approach, Dreier and Toru\'nczyk showed the following:
\begin{itemize}
	\item Classes of bounded twin-width as well as classes of bounded expansion have bounded merge-width. (That is, the radius-$r$ merge-width of any member of the class is bounded by a constant depending on $r$ and the class.) Also, nowhere dense classes have almost bounded merge-width.
	All these classes are monadically dependent and are known to admit fixed-parameter algorithms for model checking first-order logic (with the twin-width case requiring access to a suitable decomposition). On the other hand, hereditary classes that are monadically independent do not have almost bounded~merge-width.
	\item The model checking problem for first-order logic can be solved in fixed-parameter time in graph classes of bounded merge-width, provided suitable merge sequences witnessing the boundedness of merge-width can be efficiently constructed. (Note that Step \ref{s:mc} above concerns the more general classes of \emph{almost} bounded merge-width.)
\end{itemize}

Recently, the authors together with Rose McCarty~\cite{DreierMMPT26} proved that monadically dependent classes have almost bounded radius-$1$ merge-width; this implements Step~\ref{s:abmw} of the approach for $r=1$. 
In this paper, we implement Step~\ref{s:abmw} in full by proving the following result.

\begin{restatable}{theorem}{main}\label{thm:main}
	For every $r\in \N$ and monadically dependent graph class $\Cc$, there is a subpolynomial function $f\colon \N\to \N$ such that for every $n$-vertex graph $G\in \Cc$, the radius-$r$ merge-width of $G$ is bounded by $f(n)$.
	
	Moreover, given $G\in \Cc$, a merge sequence for $G$ of radius-$r$ width bounded by $f(n)$ can be computed in $\Oht_{\Cc,r}(n^3)$ time\footnote{The $\Oht(\cdot)$ notation hides subpolynomial~factors.}.
\end{restatable}

A companion paper by three of the authors~\cite{DreierMT26} implements Step~\ref{s:mc} of the approach and thereby completes the proof of \cref{conj:main}.

Both the proof for the radius-$1$ case from~\cite{DreierMMPT26} and our proof of \Cref{thm:main} are based on neighborhood complexity and tools from VC combinatorics.
However, the resulting proofs differ significantly, and we were unable to extend the proof from~\cite{DreierMMPT26} to higher radii.

\paragraph*{Discussion.}
\Cref{thm:main} confirms the conjecture of the first and last author \cite[Conjecture 1.18]{DreierT25}, stating that for hereditary graph classes,
monadic dependence coincides with almost bounded merge-width. Most importantly, this opens the way towards proving \Cref{conj:main}, that the model checking problem is fixed-parameter tractable on monadically dependent graph classes, which is achieved in a companion paper \cite{DreierMT26} by a subset of the authors.
It also confirms a conjecture of the last author \cite[Conjecture 10.7, full~version]{flip-width-focs}, stating the equivalence of monadic dependence and \emph{almost bounded flip-width}, for hereditary graph classes.
Flip-width is a family of graph parameters based on games, similar to the Cops and Robber game characterizing treewidth. It is known that for hereditary graph classes,  almost bounded merge-width implies almost bounded flip-width \cite{DreierT25}, which implies monadic dependence \cite{DreierMT24}. 
Thus, our result proves the equivalence of the three notions, and yields a characterization of monadic dependence in terms of games.

This adds to a list of logical, combinatorial, and algorithmic
characterizations of monadically dependent graph classes, which we state below, referring to the relevant papers for the definitions and proofs.
\begin{theorem}\label{thm:list}
	The following conditions are equivalent for a hereditary graph class $\CC$:
	\begin{enumerate}
		\item\label{it:mdep} $\CC$ is monadically dependent,
		\item $\CC$ is dependent \cite{BraunfeldLaskowski2026},
		\item $\CC$ has almost bounded merge-width \cite{DreierT25},
		\item $\CC$ has almost bounded flip-width \cite{flip-width-focs},
	    \item $\CC$ is flip-breakable \cite{DreierMT24},
	    \item $\CC$ is flip-separable \cite{bonnet2025separability-icalp},
		\item\label{it:fpt} first-order model checking is fixed-parameter tractable on $\CC$ \cite{DreierMT24,DreierMT26}.

		(The implication \eqref{it:fpt}$\rightarrow$\eqref{it:mdep} assumes ${\mathsf{AW}[\star] \neq \mathsf{FPT}}$.)
	\end{enumerate}	
\end{theorem}

\noindent We refer to \cite{braunfeld2021characterizations,BraunfeldLaskowski2026} for a list of model-theoretic characterizations of monadic dependence, in the greater generality of relational structures. We note that the results in the current paper (including the statement of \Cref{thm:list}) easily extend to structures equipped with several \emph{binary} relations, using a standard encoding in graphs \cite[Section 19.3, full version]{DreierMT24}. Lifting them to relational structures is left for future work.

\paragraph{Acknowledgements.}
We thank Rose McCarty for insightful discussions and contributions to the early development of the ideas in this paper.

\section{Overview}\label{sec:overview}

\begin{figure}[!h]
	\centering
	\resizebox{\textwidth}{!}{%
	\begin{tikzpicture}[
			box/.style={draw, rounded corners, align=center, text width=2.1cm, minimum height=1.3cm, font=\small},
			flowarrow/.style={->, >=stealth, thick},
			lbl/.style={font=\footnotesize, align=center}
		]
		\node[box] (nei) at (-4.1,0) {Almost linear neighborhood complexity\\ \cite{DreierMMPT26}};
		\node[box] (pre) at (0,0) {Predamping Lemma\\ \cref{lem:predamping}};
		\node[box] (clo) at (4.1,0) {Damping Theorem\\ \cref{thm:static-s-flip}};
		\node[box] (par) at (8.2,0) {Almost bounded partition merge-width\\ \cref{thm:main-ordering}};
		\node[box] (mer) at (12.3,0) {Almost bounded merge-width\\ \cref{thm:main}};
		\draw[flowarrow] (nei) -- node[lbl, above] {rules\\(R1)--(R3)} node[lbl, below] {Sec.~\ref{sec:ingredients}, \ref{sec:predamping}} (pre);
		\draw[flowarrow] (pre) -- node[lbl, above] {flip} node[lbl, below] {Sec.~\ref{sec:flip}, \ref{sec:damping}} (clo);
		\draw[flowarrow] (clo) -- node[lbl, above] {iterate\\$\log n$ times} node[lbl, below] {Sec.~\ref{sec:merge-width}} (par);
		\draw[flowarrow] (par) -- node[lbl, above] {\cite{DrabikDGNPT26}} node[lbl, below] {Thm.~\ref{thm:partition-mw}} (mer);
	\end{tikzpicture}
	}
	\caption{Structure of the proof of \cref{thm:main}. Each arrow is an implication, labeled by where it is proved.}
	\label{fig:ov-flow}
\end{figure}
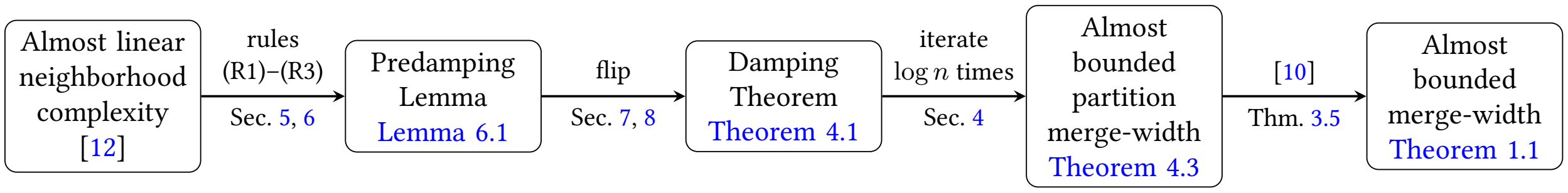

In this section we give a technical overview of the proof of \cref{thm:main}, focusing on presenting the main conceptual ideas. \Cref{fig:ov-flow} shows the structure of the proof. We go through it from right to left.

\subsection{Merge-width}

We start by recalling the definition of merge-width and the relevant intuitions behind it. Actually, throughout this paper we will not rely on the original definition of Dreier and Toru\'nczyk~\cite{DreierT25}, but on a variant called \emph{partition merge-width} proposed recently by Drabik, Dumas, Geniet, Nowakowski, Pilipczuk, and Toru\'nczyk~\cite{DrabikDGNPT26}. On the one hand, partition merge-width puts fewer constraints on the decompositions that need to be constructed, which makes our task technically easier. On the other hand, decompositions witnessing bounds on partition merge-width can be efficiently transformed into decompositions witnessing bounds on (original) merge-width, at the cost of a minor loss in the values of the parameters. See \cref{thm:partition-mw} for a precise formulation. To introduce partition merge-width, we need a few definitions~first.

Let $G$ be a graph. The notion of a decomposition underlying partition merge-width is called a \emph{partition sequence}: it is a sequence
\[\Pp_0\pleq \Pp_1 \pleq \ldots\pleq \Pp_m\]
of partitions of the vertex set of $G$ such that 
\begin{itemize}
	\item $\Pp_0$ is the partition into singletons;
	\item $\Pp_m$ has one part $V(G)$; and
	\item every next partition $\Pp_{i+1}$ is a coarsening of $\Pp_i$ (which is denoted by $\Pp_i\pleq \Pp_{i+1}$).
\end{itemize}
The intuition is that the partitions $\Pp_0,\ldots,\Pp_m$ should describe a process of iterative merging of various parts of the graph $G$, starting from singletons and ending with the whole vertex set being in a single part, so that at every time step of the merging process, the current partition $\Pp_i$ provides good control over the ``information flow'' in $G$. This control is measured by the \emph{width} of the partition sequence; a notion whose definition is quite subtle.

Let $\Pp$ be a partition of the vertex set of $G$. (Think: $\Pp=\Pp_i$ for some $i\in \{0,\ldots,m\}$.) A \emph{$\Pp$-flip} of $G$ is any graph $H$ that can be obtained from $G$ as follows: For any pair of parts $A,B\in \Pp$, either leave all the edges between $A$ and $B$ intact, or complement the adjacency relation between $A$ and $B$; that is, replace all the edges $uv$ with $u\in A$ and $v\in B$ with non-edges, and vice versa. In this definition we allow $A=B$, which corresponds to either leaving the adjacency relation within $A$ intact or complementing it.

With this definition, we introduce the \emph{$\Pp$-metric}~\cite{GajarskyMMOPPSS23} on the vertex set of $G$: for $u,v\in V(G)$, we set
\[\dist_\Pp(u,v)\coloneqq \max \{\dist_H(u,v)\colon H\textrm{ is a }\Pp\textrm{-flip of }G\},\]
where $\dist_H(u,v)$ is the standard shortest-path distance metric in $H$. The intuition behind this definition is that $\dist_\Pp(u,v)$ measures how well $\Pp$ controls short connections between $u$ and $v$ by verifying whether these connections can be destroyed in a $\Pp$-flip.
As observed in~\cite{GajarskyMMOPPSS23}, $\dist_\Pp(\cdot,\cdot)$ is indeed a metric on the vertex set of $G$.

Now, for a radius parameter $r\in \N$, the \emph{radius-$r$ width} of a partition sequence $\Pp_0,\Pp_1,\ldots,\Pp_m$ is the smallest integer $k\in \N$ satisfying the following:
\begin{quote}
	For every $i<m$ and vertex $u$ of $G$, there are at most $k$ parts $P\in \Pp_i$ with $\dist_{\Pp_{i+1}}(u,P)\leq r$.
\end{quote}
Thus, the intuition is that every vertex of the graph is well-separated in the $\Pp_{i+1}$-metric from all but at most $k$ parts of $\Pp_i$. The $+1$ in the indices is important: While we measure distances with respect to the coarser partition $\Pp_{i+1}$ (so the distances in the $\Pp_{i+1}$-metric can be larger than in the $\Pp_i$-metric), we measure the number of close parts in the finer partition $\Pp_i$. This allows a smooth control of the properties of partitions $\Pp_i$ along the whole partition sequence. 

We remark that in the original definition of merge-width~\cite{DreierT25}, distances are measured in ``error graphs'' $R_1,\ldots,R_m$ that accumulate more and more error edges along the partition sequence, and each error graph $R_i$ must be ``compatible'' with a single $\Pp_i$-flip $H_i$. Intuitively, working with partition merge-width allows us to consider all such $\Pp_i$-flips at the same time thanks to the flexibility of the $\Pp_i$-metric, and it also relaxes the inconvenient monotonicity requirement imposed on the error graphs. 

For readers familiar with twin-width, it is also instructive to understand the difference between the mechanics of twin-width and of merge-width. While in both cases, the underlying notion of a decomposition is a partition sequence, in twin-width, if two parts $A,B$ of the current partition $\Pp_i$ are neither complete nor anticomplete to each other, then we consider this an ``error edge'' between whole parts $A$ and $B$ that affects all the vertices of $A$ and $B$. In merge-width, error edges (i.e., edges of the graphs $R_i$ as above) occur and are measured at the level of individual vertices, and thus vertices of a single part $A\in \Pp_i$ can be incident to error edges leading to many other parts of $\Pp_i$ without making the width large. As shown in~\cite{DreierT25}, such a reinterpretation of the mechanics of error edges leads to a notion --- merge-width --- that contrary to twin-width, is also capable of describing key notions of sparsity theory (classes of bounded expansion and nowhere dense classes~\cite{sparsity}). This is because it captures the mechanics of \emph{generalized coloring numbers}, a family of parameters characterizing those classes. Finally, working with the $\Pp_i$-metric in the definition of partition merge-width hides the concept of error edges from the immediate focus, but it should rather be regarded as a technical method to take all possible flips into account at the same time.

In the language of partition merge-width, our main result reads as follows. (In this overview, we write $\spoly_{\tup q}(n)$ to denote an unspecified subpolynomial function of variable $n$ that may depend on parameters~$\tup q$.)

\begin{theorem}[\cref{thm:main-ordering}, simplified]\label{thm:ov-main-ordering}
	Let $r\in \N$ and $\Cc$ be a monadically dependent graph class.
	Then there is an $\Oht_{\Cc,r}(n^3)$-time algorithm that given an $n$-vertex graph $G\in \Cc$, outputs a partition sequence of $G$ of radius-$r$ width bounded by $\spoly_{\Cc,r}(n)$.
\end{theorem}

\cref{thm:main} then follows from combining the statement above with the results of Drabik et al.~\cite{DrabikDGNPT26}.

\subsection{Neighborhood complexity}

Partitions in the partition sequence that we are going to construct in the proof of \cref{thm:ov-main-ordering} will have a very specific form: they will be partitions into \emph{neighborhood classes}, or \emph{$1$-types}, over vertex subsets. More precisely, for a graph $G$ and a subset of vertices $A\subseteq V(G)$, we define the partition $\Tt_1(G,A)$ of $V(G)$ into \emph{$1$-types} over $A$ as follows: 
\begin{itemize}
	\item every vertex  of $A$ is placed into a singleton $1$-type; and
	\item vertices outside of $A$ are partitioned into $1$-types according to their neighborhood in $A$: $u,v\in V(G)\setminus A$ have the same $1$-type over $A$ if and only if $N(u)\cap A=N(v)\cap A$.
\end{itemize} 
Note that in general graphs, there can be as many as $|A|+2^{|A|}$ $1$-types over $A$: $|A|$ singleton types of the vertices $A$, and $2^{|A|}$ corresponding to the $2^{|A|}$ different subsets of $A$. By the Sauer--Shelah Lemma, this bound can be reduced to polynomial in $|A|$ assuming that the set system of neighborhoods in $G$ has bounded VC dimension; this is the case in monadically dependent graph classes (see \cref{obs:mondep-vc}). However, it turns out that monadic dependence actually implies a much stronger bound: monadically dependent classes have \emph{almost linear neighborhood complexity}, meaning that the number of $1$-types over $A$ is almost linear in the size of $A$. This was proved very recently by the authors and Rose McCarty~\cite{DreierMMPT26}.

\begin{theorem}[{\cite[Theorem~2]{DreierMMPT26}}]\label{thm:ov-nei-comp}
	Let $\Cc$ be a monadically dependent graph class. Then for every graph $G\in \Cc$ and non-empty vertex subset $A\subseteq V(G)$, we have
	\[|\{N(u)\cap A\colon u\in V(G)\}|\leq |A|\cdot \spoly_{\Cc}(|A|).\]
\end{theorem}

Note that there are only $|A|$ $1$-types corresponding to the vertices of $A$, hence \cref{thm:ov-nei-comp} implies that $|\Tt_1(G,A)|\leq |A|\cdot \spoly_{\Cc}(|A|)$ (see \cref{thm:mondep-1-types}). 

\cref{thm:ov-nei-comp} is a fundamental result about monadically dependent graph classes on which our whole approach hinges. 
It generalizes analogous results for nowhere dense classes~\cite{EickmeyerGKKPRS17}, monadically stable classes~\cite{DreierEMMPT24}, and classes of bounded merge-width~\cite{BonamyG26}. We refer the reader to the introductory section of~\cite{DreierMMPT26} for more background on neighborhood complexity.

\subsection{Reduction to the Damping Theorem}\label{sec:ov-assembly}

\tikzset{
	ov vertex/.style={circle, fill, inner sep=1.2pt},
	ov edge/.style={gray},
	ov part/.style={draw, rounded corners=2pt, fill=gray!22},
	ov refined/.style={draw, dashed, rounded corners=2pt},
	ov set/.style={draw, rounded corners=3pt},
	ov marked/.style={circle, draw=red!75!black, thick, inner sep=1.9pt},
	ov marked alt/.style={rectangle, draw=blue!70!black, thick, minimum size=6pt, inner sep=0pt},
	ov label/.style={font=\scriptsize, align=center}
}
\DeclareRobustCommand{\ovmarked}{\tikz[baseline=-0.6ex]{\node[ov vertex] at (0,0) {}; \node[ov marked] at (0,0) {};}}
\DeclareRobustCommand{\ovmarkedalt}{\tikz[baseline=-0.6ex]{\node[ov vertex] at (0,0) {}; \node[ov marked alt] at (0,0) {};}}

In the proof of \cref{thm:ov-main-ordering}, we will actually construct a sequence of vertex subsets
\[V(G)=B_0\supseteq B_1\supseteq \ldots\supseteq B_{m-1}\supseteq B_m=\emptyset\]
and put
\[\Pp_i\coloneqq \Tt_1(G,B_i)\qquad\textrm{for all }i\in \{0,1,\ldots,m\}.\]
Thus, it is clear that $\Pp_0$ is a partition into singletons, $\Pp_m$ has one part $V(G)$, and every next partition $\Pp_{i+1}$ is a coarsening of $\Pp_i$; so $\Pp_0,\Pp_1,\ldots,\Pp_m$ is a partition sequence. Our goal is to construct the sequence $B_0,B_1,\ldots,B_m$ so that this partition sequence has subpolynomial radius-$r$ width.

For the rest of this overview it is convenient to phrase the width in terms of balls. For a partition $\Pp$ of $V(G)$, a vertex $u$, and $r\in \N$, let $\Ball^r_\Pp[u]\coloneqq \{v\colon \dist_\Pp(u,v)\leq r\}$; that is, $\Ball^r_\Pp[u]$ is the set of vertices that are at distance at most $r$ from $u$ in every $\Pp$-flip of $G$. Similarly, for a graph $H$, we let $\Ball^r_H[u]$ be the set of vertices at distance at most $r$ from $u$ in $H$. The radius-$r$ width of the partition sequence above is then the least $k$ such that for every $i\in [m]\coloneqq \{1,\ldots,m\}$ and every vertex $u$, the ball $\Ball^r_{\Tt_1(G,B_i)}[u]$ intersects at most $k$ parts of $\Tt_1(G,B_{i-1})$. Hence \cref{thm:ov-main-ordering} follows immediately from the following statement.

\begin{theorem}[\cref{thm:main-ordering}, simplified]\label{thm:ov-chain}
	Let $r\in \N$ and $\Cc$ be a monadically dependent graph class. Then for every $n$-vertex graph $G\in \Cc$ one can compute in time $\Oht_{\Cc,r}(n^3)$ a sequence of vertex subsets $V(G)=B_0\supseteq B_1\supseteq \ldots\supseteq B_m=\emptyset$ such that for every $i\in [m]$ and vertex $u$, the ball $\Ball^r_{\Tt_1(G,B_i)}[u]$ intersects at most $\spoly_{\Cc,r}(n)$ parts of $\Tt_1(G,B_{i-1})$. In particular, $\Tt_1(G,B_0)\pleq \Tt_1(G,B_1)\pleq \ldots\pleq \Tt_1(G,B_m)$ is a partition sequence of radius-$r$ width bounded by $\spoly_{\Cc,r}(n)$.
\end{theorem}

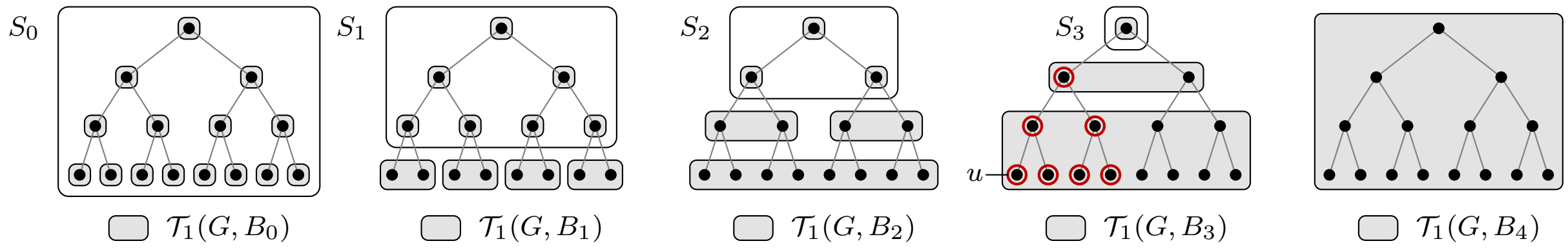
\begin{figure}[t]
	\centering
	\begin{tikzpicture}
		\foreach \i in {0,...,4} {
				\begin{scope}[xshift=\i*3.2cm]
					\ifcase\i
					\or
						\foreach \j in {0,...,3} {\draw[ov part] (0.64*\j-0.12,-1.65) rectangle (0.64*\j+0.44,-1.35);}
					\or
						\foreach \j in {0,1} {\draw[ov part] (1.28*\j+0.01,-1.15) rectangle (1.28*\j+0.95,-0.85);}
						\draw[ov part] (-0.15,-1.65) rectangle (2.39,-1.35);
					\or
						\draw[ov part] (0.33,-0.65) rectangle (1.91,-0.35);
						\draw[ov part] (-0.15,-1.65) rectangle (2.39,-0.85);
					\or
						\draw[ov part] (-0.15,-1.65) rectangle (2.39,0.15);
					\fi
					\ifcase\i
						\draw[ov set] (-0.22,0.22) rectangle (2.46,-1.72);
						\node[ov label, anchor=east] at (-0.27,0) {$S_0$};
					\or
						\draw[ov set] (-0.06,0.22) rectangle (2.3,-1.22);
						\node[ov label, anchor=east] at (-0.11,0) {$S_1$};
					\or
						\draw[ov set] (0.26,0.22) rectangle (1.98,-0.72);
						\node[ov label, anchor=east] at (0.21,0) {$S_2$};
					\or
						\draw[ov set] (0.9,0.22) rectangle (1.34,-0.22);
						\node[ov label, anchor=east] at (0.85,0) {$S_3$};
					\or
					\fi
					\foreach \k in {0,...,3} {
							\pgfmathtruncatemacro{\c}{2^\k-1}
							\pgfmathtruncatemacro{\inb}{\k+\i<=3 ? 1 : 0}
							\ifnum\inb=1
								\foreach \j in {0,...,\c} {
										\pgfmathsetmacro{\x}{0.32*(2^(3-\k)*\j+(2^(3-\k)-1)/2)}
										\draw[ov part] (\x-0.11,-0.5*\k-0.11) rectangle (\x+0.11,-0.5*\k+0.11);
									}
							\fi
						}
					\foreach \k in {0,...,2} {
							\pgfmathtruncatemacro{\c}{2^\k-1}
							\pgfmathtruncatemacro{\kk}{\k+1}
							\foreach \j in {0,...,\c} {
									\pgfmathsetmacro{\x}{0.32*(2^(3-\k)*\j+(2^(3-\k)-1)/2)}
									\pgfmathsetmacro{\xa}{0.32*(2^(2-\k)*(2*\j)+(2^(2-\k)-1)/2)}
									\pgfmathsetmacro{\xb}{0.32*(2^(2-\k)*(2*\j+1)+(2^(2-\k)-1)/2)}
									\draw[ov edge] (\x,-0.5*\k) -- (\xa,-0.5*\kk);
									\draw[ov edge] (\x,-0.5*\k) -- (\xb,-0.5*\kk);
								}
						}
					\foreach \k in {0,...,3} {
							\pgfmathtruncatemacro{\c}{2^\k-1}
							\foreach \j in {0,...,\c} {
									\pgfmathsetmacro{\x}{0.32*(2^(3-\k)*\j+(2^(3-\k)-1)/2)}
									\node[ov vertex] at (\x,-0.5*\k) {};
								}
						}
					\draw[ov part] (0.3,-2.17) rectangle (0.7,-1.92);
					\node[ov label, anchor=west] at (0.75,-2.05) {$\Tt_1(G,B_{\i})$};
					\ifnum\i=3
						\foreach \x/\y in {0.48/-0.5, 0.16/-1, 0.8/-1, 0/-1.5, 0.32/-1.5, 0.64/-1.5, 0.96/-1.5} {\node[ov marked] at (\x,\y) {};}
						\node[ov label, anchor=east, inner sep=0.5pt] at (-0.32,-1.5) {$u$};
						\draw (-0.32,-1.5) -- (-0.09,-1.5);
					\fi
				\end{scope}
			}
	\end{tikzpicture}
	\caption{Construction of the sets $S_i$ and $B_i$ on a complete binary tree $G$, where $S_0 = V(G)$ and each set $S_i$ is obtained from $S_{i-1}$ by removing the lowest level. In this example, the sets $S_i$ are nested, so $B_i=\bigcup_{j\geq i} S_j=S_i$. (In general, the sets $S_i$ need not be nested.) The panels show the resulting partition sequence $\Tt_1(G,B_0)\pleq \Tt_1(G,B_1)\pleq \ldots\pleq \Tt_1(G,B_4)$. In the fourth panel, the \ovmarked-marked set is $\Ball^r_{\Tt_1(G,B_3)}[u]$ for every $r\geq 4$; it intersects three parts of $\Tt_1(G,B_2)$, so it contributes three to the radius-$r$ width of this partition sequence.}
	\label{fig:ov-assembly}
\end{figure}

We construct this sequence in $m$ steps, and each step contributes $\spoly_{\Cc,r}(n)$ to the width. Hence, the final width grows linearly in $m$. The key is to make every step halve the remaining work, leading to $m\leq \Oh(\log n)$, which is absorbed by $\spoly_{\Cc,r}(n)$. More precisely, we maintain a set $S_i$ of vertices that still need to be processed. Initially, $S_0=V(G)$. In a single step, we replace $S_i$ by a new set $S_{i+1}$ of vertices that still need to be processed, with $|S_{i+1}|\leq |S_i|/2$, so that after at most $1+\log_2 n$ steps no vertices are left. The sets $S_i$ need not be nested, that is, a vertex may need to be processed again later. Accordingly, we let $B_i\coloneqq S_i\cup S_{i+1}\cup \ldots$ be the set of all vertices that were processed at step $i$ or later. \Cref{fig:ov-assembly} illustrates this on a complete binary tree.
A single step of the construction is captured by the following theorem, which takes $A=S_i$ and yields $S=S_{i+1}$. 

\begin{theorem}[Damping Theorem; \cref{thm:static-s-flip}, simplified]\label{thm:ov-damping}
	Let $r\in \N$ and $\Cc$ be a monadically dependent graph class. Then for every $n$-vertex graph $G\in \Cc$ and $A\subseteq V(G)$, one can compute in time $\Oht_{\Cc,r}(n^3)$ a set $S\subseteq V(G)$ with $|S|\leq |A|/2$ and a $\Tt_1(G,S)$-flip $H$ of $G$ such that for every vertex $u$, the ball $\Ball^r_H[u]$ intersects at most $\spoly_{\Cc,r}(n)$ parts of $\Tt_1(G,A\cup S)$.
\end{theorem}

\begin{figure}[t]
	\centering
	\begin{tikzpicture}
		\foreach \p/\dx in {g/0, h/6} {
				\begin{scope}[xshift=\dx cm]
					\draw[ov refined] (-0.8,2.33) -- (3.31,2.33) -- (3.31,-0.33) -- (-0.8,-0.33) -- (-0.8,0.33) -- (2.65,0.33) -- (2.65,1.67) -- (-0.8,1.67) -- cycle;
					\draw[ov set] (-0.35,-0.25) rectangle (2.55,0.25);
					\node[ov label, anchor=east] at (-0.4,0) {$A$};
					\draw[ov part] (-0.3,1.8) rectangle (2.5,2.2);
					\draw[ov refined] (-0.55,0.68) rectangle (2.45,1.32);
					\draw[ov refined] (-1.95,0.67) rectangle (-0.92,1.33);
					\draw[ov set] (-1.5,0.75) rectangle (-1,1.25);
					\node[ov label, anchor=east] at (-1.55,1) {$S$};
					\draw[ov part] (-1.4,0.85) rectangle (-1.1,1.15);
					\node[ov vertex] (\p s) at (-1.25,1) {};
					\foreach \j/\x in {1/0,2/1.1,3/2.2} {
							\draw[ov part] (\x-0.15,-0.15) rectangle (\x+0.15,0.15);
							\draw[ov part] (\x-0.15,0.85) rectangle (\x+0.15,1.15);
							\node[ov vertex] (\p l\j) at (\x,0) {};
							\node[ov vertex] (\p k\j) at (\x,1) {};
							\node[ov vertex] (\p p\j) at (\x,2) {};
						}
					\node[ov label, anchor=south east] at (-0.15,0.99) {$u$};
				\end{scope}
			}
		\foreach \i in {1,2,3} {\foreach \j in {1,2,3} {
						\ifnum\i=\j\relax
							\draw[ov edge] (hk\i) -- (hl\j); \draw[ov edge] (hk\i) -- (hp\j);
						\else
							\draw[ov edge] (gk\i) -- (gl\j); \draw[ov edge] (gk\i) -- (gp\j);
						\fi
					}}
		\draw[ov edge] (gs) -- (gk1);
		\draw[ov edge] (gs) to[bend left=30] (gk2);
		\draw[ov edge] (gs) to[bend left=35] (gk3);
		\foreach \x in {gk1,gs,gl2,gl3,gp2,gp3} {\node[ov marked] at (\x) {};}
		\foreach \x in {hk1,hl1,hp1} {\node[ov marked alt] at (\x) {};}
		\node[ov label] at (0.6,-0.7) {$G$};
		\node[ov label] at (6.6,-0.7) {$H$};
		\draw[ov part] (-0.35,-1.27) rectangle (0.05,-1.02);
		\node[ov label, anchor=west] at (0.1,-1.15) {$\Tt_1(G,A\cup S)$};
		\draw[ov refined] (2.56,-1.27) rectangle (2.96,-1.02);
		\node[ov label, anchor=west] at (3.01,-1.15) {$\Tt_1(G,S)$};
		\node[ov vertex] at (4.84,-1.15) {};
		\node[ov marked] at (4.84,-1.15) {};
		\node[ov label, anchor=west] at (4.99,-1.15) {$\Ball^1_G[u]$};
		\node[ov vertex] at (6.85,-1.15) {};
		\node[ov marked alt] at (6.85,-1.15) {};
		\node[ov label, anchor=west] at (7.0,-1.15) {$\Ball^r_H[u]$};
	\end{tikzpicture}
	\caption{An example of the Damping Theorem.
	The graph consists of three rows of vertices of equal length~$k$, 
	with edges joining vertices in different columns and adjacent rows; additionally a vertex $s$ adjacent to the middle row is added.
The  input set $A$ is the bottom row and the output is $S=\{s\}$. The shaded sets are the parts of $\Tt_1(G,A\cup S)$, and the dashed sets are the parts of $\Tt_1(G,S)$. The flip $H$ complements the adjacency between $S$ and the middle row, and between the middle row and the union of the top and bottom rows. While the original ball $\Ball^1_G[u]$ (marked~\ovmarked) intersects $\Omega(k)$ parts of $\Tt_1(G,A\cup S)$, the flipped ball $\Ball^r_H[u]$ (marked~\ovmarkedalt) intersects only three parts of $\Tt_1(G,A\cup S)$, for every $r\geq 1$.}
	\label{fig:ov-damping}
\end{figure}
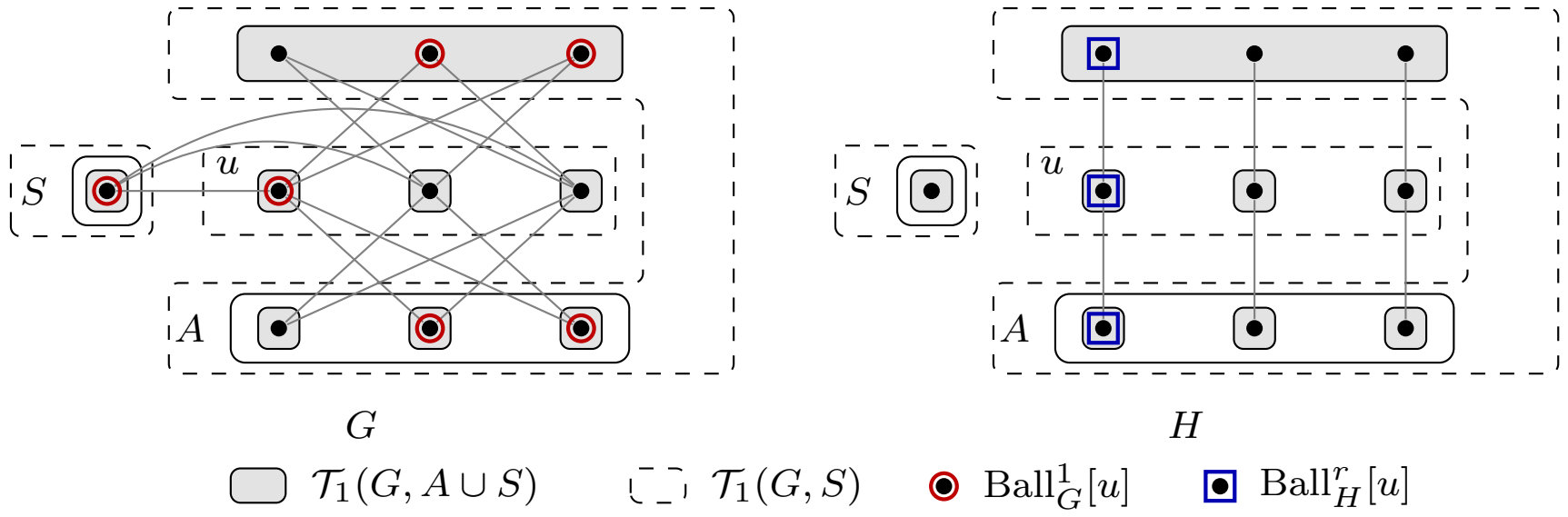

Note that two different partitions appear in the statement. The flip $H$ is taken with respect to the coarser partition $\Tt_1(G,S)$, but the parts intersected by the balls are counted in a finer one, ${\Tt_1(G,A\cup S)}$. This mirrors the $+1$ offset in the definition of radius-$r$ width, where distances are measured in the coarser $\Pp_{i+1}$-metric and close parts are counted in the finer partition $\Pp_i$.

We call the output $S$ a \emph{damper} of the input $A$. \Cref{fig:ov-damping} shows an example of $A$, $S$, the flip $H$, and the partitions $\Tt_1(G,S)$ and $\Tt_1(G,A\cup S)$.
The proof of the Damping Theorem is the key technical substance of this paper, and we sketch it in the next subsection.

\begin{proof}[Proof sketch of \cref{thm:ov-chain} using \cref{thm:ov-damping} (Damping Theorem)]
We let each $S_i$ be a damper of $S_{i-1}$, witnessed by a $\Tt_1(G,S_i)$-flip $H_i$. Since dampers have at most half the size, $m\leq 1+\log_2 n$. 

Fix $i\in [m]$ and a vertex $u$, and let $U\coloneqq \Ball^r_{\Tt_1(G,B_i)}[u]$. (For $i=3$ and $r\geq 4$, this is the \ovmarked-marked set in \cref{fig:ov-assembly}.) For every $j\geq i$ we have $S_j\subseteq B_i$, so $H_j$ is also a $\Tt_1(G,B_i)$-flip, and therefore $U\subseteq \Ball^r_{H_j}[u]$. By the Damping Theorem, $U$ intersects at most $\spoly_{\Cc,r}(n)$ parts of $\Tt_1(G,S_{j-1}\cup S_j)$, and hence also of the coarser partition $\Tt_1(G,S_{j-1})$. However, to prove the theorem, it remains to show that $U$ intersects at most $\spoly_{\Cc,r}(n)$ parts of $\Tt_1(G,B_{i-1})$, which is the common refinement of the partitions $\Tt_1(G,S_{j-1})$ for $j\geq i$.
This is not immediate: a priori the number of parts of $\Tt_1(G,B_{i-1})$ intersected by $U$ is bounded only by the product of the numbers of parts of the partitions $\Tt_1(G,S_{j-1})$ intersected by $U$. This is resolved by an additivity property of $1$-types (\cref{lem:1-types-additive}), which is again a consequence of the almost linear number of types: if $U$ intersects $k_j$ parts of $\Tt_1(G,S_{j-1})$, for each $j\geq i$, then $U$ intersects at most $\spoly_{\Cc}(n)\cdot \sum_j k_j$ parts of $\Tt_1(G,\bigcup_{j\geq i} S_{j-1})=\Tt_1(G,B_{i-1})$.
Since each $k_j$ is bounded by $\spoly_{\Cc,r}(n)$ and there are at most $m\leq 1+\log_2 n$ indices $j$, this is $\spoly_{\Cc,r}(n)$, as desired.
For the full proof see \cref{sec:merge-width}.
\end{proof}

\subsection{Proving the Damping Theorem}\label{sec:ov-damping}

We now discuss the proof of the Damping Theorem (\cref{sec:ingredients,sec:predamping,sec:flip,sec:damping}). Its key building block is the following routine.
\begin{center}
	\fbox{\begin{minipage}{0.85\textwidth}
		\textbf{Neighborhood Damping routine}\\[2pt]
		\emph{Input:} a set $A\subseteq V(G)$.\\
		\emph{Output:} a set $S$ with $|S|\leq |A|/2$ and a $\Tt_1(G,S)$-flip $H$ of $G$ in which the neighborhood of every vertex intersects few parts of a partition $\Rr$ satisfying certain properties.
	\end{minipage}}
\end{center}
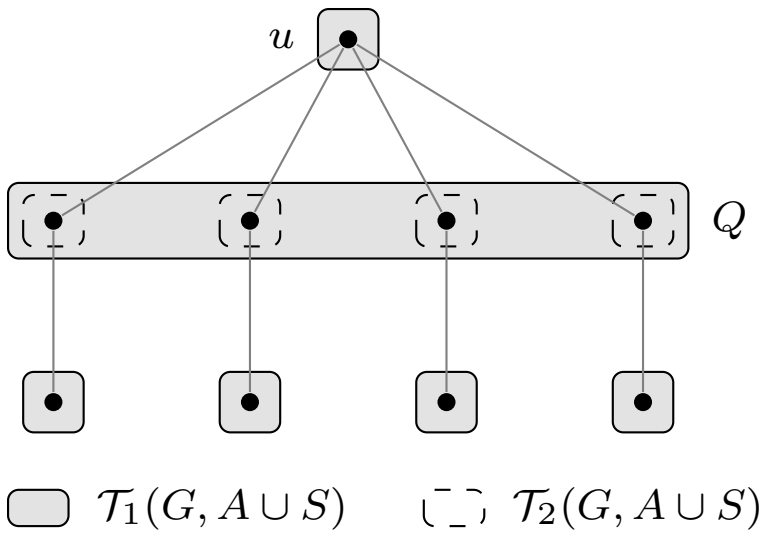
\begin{figure}[t]
	\centering
	\begin{tikzpicture}
		\draw[ov part] (1.75,2.2) rectangle (2.15,2.6);
		\node[ov vertex] (u) at (1.95,2.4) {};
		\node[ov label, anchor=east] at (1.73,2.4) {$u$};
		\draw[ov part] (-0.3,0.95) rectangle (4.2,1.45);
		\node[ov label, anchor=west] at (4.22,1.2) {$Q$};
		\foreach \k/\x in {1/0, 2/1.3, 3/2.6, 4/3.9} {
				\draw[ov refined] (\x-0.2,1.03) rectangle (\x+0.2,1.37);
				\node[ov vertex] (x\k) at (\x,1.2) {};
				\draw[ov part] (\x-0.2,-0.2) rectangle (\x+0.2,0.2);
				\node[ov vertex] (r\k) at (\x,0) {};
				\draw[ov edge] (u) -- (x\k); \draw[ov edge] (x\k) -- (r\k);
			}
		\draw[ov part] (-0.3,-0.82) rectangle (0.1,-0.58);
		\node[ov label, anchor=west] at (0.15,-0.7) {$\Tt_1(G,A\cup S)$};
		\draw[ov refined] (2.45,-0.82) rectangle (2.85,-0.58);
		\node[ov label, anchor=west] at (2.9,-0.7) {$\Tt_2(G,A\cup S)$};
	\end{tikzpicture}
	\caption{A $\Tt_1(G,S)$-flip $H$ of $G$. The neighborhood of $u$ intersects only a single part of $\Tt_1(G,A\cup S)$, while $\Ball^2_H[u]$ unfortunately intersects many. The finer partition $\Tt_2(G,A\cup S)$ into $2$-types (dashed, shown within $Q$, defined below) exposes this conflict already at distance $1$.}
	\label{fig:ov-balls}
\end{figure}

For instance, $\Rr=\Tt_1(G,A\cup S)$ is allowed, which gives the Damping Theorem for $r=1$. For larger $r$, this is not enough, as few parts per neighborhood do not imply few parts per ball: in \cref{fig:ov-balls}, the neighborhood of $u$ intersects a single part of $\Tt_1(G,A\cup S)$, but its neighbors lead to different parts, so $\Ball^2_H[u]$ intersects many. This conflict only shows at distance $2$, out of reach of the Neighborhood Damping routine, which only controls neighborhoods. Fortunately, the routine also works for finer partitions $\Rr$ (\cref{lem:predamping,cor:mismatch-symmetric}). So we look for a partition $\Rr$ that exposes such conflicts already at distance $1$: few parts of $\Rr$ per neighborhood should imply few parts of $\Tt_1(G,A\cup S)$ per ball of radius $r$. A natural way to achieve this is for $\Rr$ to separate vertices that lead to different parts, as the dashed partition in \cref{fig:ov-balls} does for the neighbors of $u$.

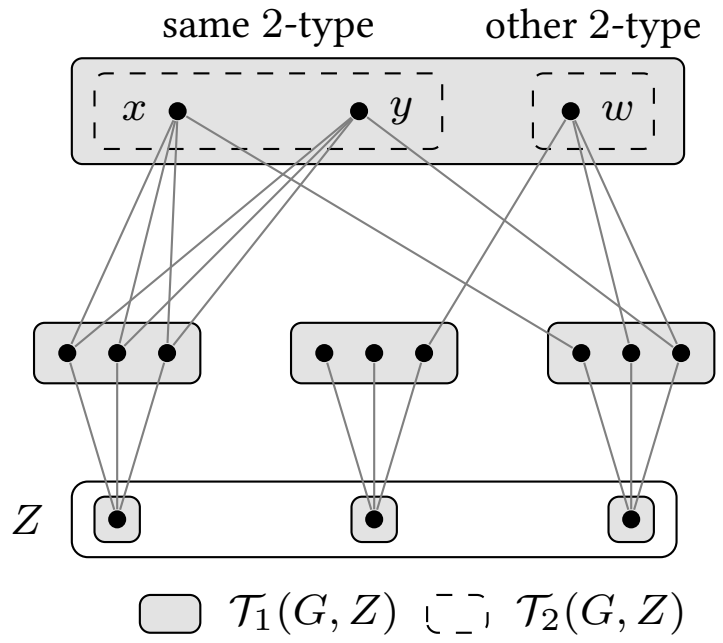
\begin{figure}[t]
	\centering
	\begin{tikzpicture}
		\draw[ov set] (-0.3,-0.25) rectangle (3.7,0.25);
		\node[ov label, anchor=east] at (-0.35,0) {$Z$};
		\foreach \d/\x in {1/0,2/1.7,3/3.4} {
				\draw[ov part] (\x-0.15,-0.15) rectangle (\x+0.15,0.15);
				\node[ov vertex] (b\d) at (\x,0) {};
			}
		\foreach \x/\q in {0/1, 1.7/2, 3.4/3} {
				\draw[ov part] (\x-0.55,0.9) rectangle (\x+0.55,1.3);
				\foreach \d in {1,2,3} {\node[ov vertex] (q\q\d) at (\x-0.66+0.33*\d,1.1) {};}
			}
		\foreach \q in {1,2,3} {\foreach \d in {1,2,3} {\draw[ov edge] (q\q\d) -- (b\q);}}
		\draw[ov part] (-0.3,2.35) rectangle (3.75,3.05);
		\draw[ov refined] (-0.15,2.45) rectangle (2.15,2.95);
		\draw[ov refined] (2.75,2.45) rectangle (3.55,2.95);
		\node[ov vertex, label={[ov label]left:$x$}] (x) at (0.4,2.7) {};
		\node[ov vertex, label={[ov label]right:$y$}] (y) at (1.6,2.7) {};
		\node[ov vertex, label={[ov label]right:$w$}] (w) at (3,2.7) {};
		\node[ov label] at (1,3.25) {same $2$-type};
		\node[ov label] at (3.15,3.25) {other $2$-type};
		\foreach \x in {x,y} {\foreach \d in {1,2,3} {\draw[ov edge] (\x) -- (q1\d);}}
		\draw[ov edge] (x) -- (q31);
		\draw[ov edge] (y) -- (q33);
		\draw[ov edge] (w) -- (q23); \draw[ov edge] (w) -- (q32); \draw[ov edge] (w) -- (q33);
		\draw[ov part] (0.15,-0.75) rectangle (0.55,-0.5);
		\node[ov label, anchor=west] at (0.6,-0.62) {$\Tt_1(G,Z)$};
		\draw[ov refined] (2.05,-0.75) rectangle (2.45,-0.5);
		\node[ov label, anchor=west] at (2.5,-0.62) {$\Tt_2(G,Z)$};
	\end{tikzpicture}
	\caption{The $1$-types over a set $Z$, and the $2$-types over $Z$ within the top $1$-type. In the middle row, $x$ and $y$ are complete to the left part, anticomplete to the middle part, and mixed to the right part, and therefore have the same $2$-type, while $w$ is anticomplete to the left one and mixed to the other two.}
	\label{fig:ov-types}
\end{figure}

\paragraph*{Higher types.}
To construct such a partition, we split each $1$-type according to how its vertices attach to the $1$-types, and iterate. Formally, fix a set $Z$ of vertices and $i\geq 1$. A vertex $v$ attaches to an $i$-type $Q$ over $Z$ in one of three ways:
\begin{itemize}
	\item $v$ is \emph{complete} to $Q$: it is adjacent to all vertices of $Q$;
	\item $v$ is \emph{anticomplete} to $Q$: it is adjacent to no vertex of $Q$; or
	\item $v$ is \emph{mixed} to $Q$: neither of the above.
\end{itemize}
Two vertices have the same \emph{$(i+1)$-type} over $Z$ if they have the same $i$-type over $Z$ and attach in the same way to every $i$-type over $Z$. Thus the partition $\Tt_{i+1}(G,Z)$ of $V(G)$ into $(i+1)$-types over $Z$ refines $\Tt_i(G,Z)$. See \cref{fig:ov-types} for an example.

The $r$-types are exactly what we were looking for in the Neighborhood Damping routine (\cref{lem:mismatch-ball}): For every set $Z\subseteq V(G)$, every $\Tt_1(G,Z)$-flip $H$ of $G$, and every $m\in \N$, we have
\begin{equation}\label{eq:ov-propagation}
	\begin{array}{c}
		\text{for every vertex } v\text{, } N_H(v) \text{ intersects}\\
		\text{at most } m \text{ parts of } \Tt_r(G,Z)
	\end{array}
	\quad\Longrightarrow\quad
	\begin{array}{c}
		\text{for every vertex } u\text{, } \Ball^r_H[u] \text{ intersects}\\
		\text{at most } (m+1)^r \text{ parts of } \Tt_1(G,Z).
	\end{array}
\end{equation}
Hence, running the Neighborhood Damping routine with $\Rr=\Tt_r(G,A\cup S)$ and applying \eqref{eq:ov-propagation} with $Z=A\cup S$ to the resulting $\Tt_1(G,S)$-flip $H$ (which is in particular a $\Tt_1(G,A\cup S)$-flip, as $S\subseteq A\cup S$) proves the Damping Theorem.

\paragraph*{Predampers.}
The Neighborhood Damping routine is implemented in two steps: first we construct the set $S$, using the following lemma, and then we construct the flip $H$. The lemma measures how differently two vertices $u,v$ look from the perspective of a partition $\Rr$ of $V(G)$: let $\sdiff[u,v]\coloneqq (N(u)\symdiff N(v))\cup \{u,v\}$ for distinct $u,v$, and $\sdiff[u,u]\coloneqq \emptyset$, and let $\hamm_\Rr(u,v)$ be the number of parts of $\Rr$ that intersect $\sdiff[u,v]$.

\begin{lemma}[Predamping Lemma; \cref{lem:predamping}, simplified]\label{lem:ov-predamping}
	Let $r\in \N$ and $\Cc$ be a monadically dependent graph class. Then for every $n$-vertex graph $G\in \Cc$ and every sufficiently large set $A\subseteq V(G)$, one can compute in time $\Oht_{\Cc,r}(n^3)$ a set $S\subseteq V(G)$ such that, denoting $\Rr\coloneqq \Tt_r(G,A\cup S)$:
	\begin{enumerate}[label={(A\arabic*)}]
		\item\label{p:ov-size} $|S|\leq |A|/2$;
		\item\label{p:ov-reps} every vertex $u$ has a vertex $s_u\in S$ with $\hamm_\Rr(u,s_u)\leq \spoly_{\Cc,r}(n)$; and
		\item\label{p:ov-diff} any two vertices $u,v$ in the same $1$-type over $S$ satisfy $\hamm_\Rr(u,v)\leq \spoly_{\Cc,r}(n)$.
	\end{enumerate}
\end{lemma}

\noindent A set $S$ as above shall be called a \emph{predamper} of $A$.

\begin{proof}[Proof sketch of \cref{thm:ov-damping} (Damping Theorem) using \cref{lem:ov-predamping} (Predamping Lemma)]\ \\
Let $S$ be a predamper of $A$. We construct a $\Tt_1(G,S)$-flip $H$ in which the neighborhood of every vertex intersects only $\spoly_{\Cc,r}(n)$ parts of $\Rr=\Tt_r(G,A\cup S)$. This is the output of the Neighborhood Damping routine, so together with \eqref{eq:ov-propagation}, it proves the Damping Theorem. For every $1$-type $Q$ over $S$, pick $x\in Q$ and let $s_Q\coloneqq s_x$. By \ref{p:ov-reps}, \ref{p:ov-diff}, and the triangle inequality, $\hamm_\Rr(v,s_Q)\leq \spoly_{\Cc,r}(n)$ for all $v\in Q$. As $s_Q\in S$, the set $N(s_Q)$ is a union of parts of $\Tt_1(G,S)$, so we may flip the adjacency between $Q$ and every $1$-type over $S$ contained in $N(s_Q)$. This turns the neighborhood of every $v\in Q$ into $N(v)\symdiff N(s_Q)$, which intersects only $\spoly_{\Cc,r}(n)$ parts of $\Rr$. The choices for $Q$ and $Q'$ may disagree on whether to flip the pair $(Q,Q')$; a majority vote over the pairs $(R,R')$ of parts of $\Rr$ with $R\subseteq Q$ and $R'\subseteq Q'$ resolves this at the cost of a factor $3$ (\cref{lem:symmetrization}). Let $H$ be the resulting flip. For the full proof, see \cref{sec:flip,sec:damping}.
\end{proof}

\subsection{Proving the Predamping Lemma}\label{sec:ov-predamping}
It remains to sketch the proof of the Predamping Lemma, which is the combinatorial core of this paper: given a set $A$, we construct a predamper $S$ of $A$. We start with $S\coloneqq \emptyset$ and add vertices to $S$ according to three rules (R1)--(R3), as long as one of them applies. The rules refer to the current partition $\Rr=\Tt_r(G,A\cup S)$, which gets finer as $S$ grows. We fix an arbitrary ordering of $V(G)$, call the minimum of each part of $\Rr$ its \emph{representative}, and let $X$ be the set of representatives. When a part is split, its representative remains the representative of one of the new parts, so $X$ only grows as $S$ grows. We state the rules in a simplified form. The precise statements are given in the proof of \cref{lem:predamping}. Once no rule applies, \ref{p:ov-reps} and \ref{p:ov-diff} hold. For \ref{p:ov-size}, we show for each rule that it is applied at most $|A\cup S|/9$ times, where $S$ is the final set. This gives $|S|\leq |A\cup S|/3$, and hence $|S|\leq |A|/2$. 

\paragraph*{Crossing.}
Let $\gamma_i$ be the least number such that $|\Tt_i(G,Z)|\leq \gamma_i\cdot |Z|$ for every non-empty $Z\subseteq V(G)$. For $1$-types, \cref{thm:mondep-1-types} gives a bound $\gamma_1\leq \spoly_{\Cc}(n)$. This bound can actually be lifted to $i$-types (\cref{thm:mondep-r-types}) because they are first-order definable (\cref{cl:types-definable}): the $(i+1)$-type of a vertex over $Z$ is determined by three fixed first-order formulas evaluated against a set containing one representative of each $i$-type over $Z$, and in monadically dependent classes the number of types of a fixed finite set of formulas over a vertex set is again almost linear in its size (\cref{lem:mondep-formula-types}). By induction on $i$, we get $\gamma_i\leq \spoly_{\Cc,i}(n)$ (\cref{thm:mondep-r-types}). As we are ultimately interested in $\Rr=\Tt_r(G,A\cup S)$, denote $\gamma\coloneqq \gamma_r$; then $\gamma\leq \spoly_{\Cc,r}(n)$.

The subpolynomial bound on $\gamma$ implies an almost linear bound on $|\Rr|$ in terms of $|A\cup S|$, which is the main property of $\Rr$ that the Neighborhood Damping routine needs. A vertex $u$ \emph{crosses} a part $R\in \Rr$ if $u\notin R$ and $u$ is neither complete nor anticomplete to $R$. The first rule is the following.
\begin{description}[font=\normalfont]
	\item[(R1, simplified)] If a vertex $w$ crosses more than $9\gamma$ parts of $\Rr$, add $w$ to $S$.
\end{description}
Observe that each application of this rule splits all the crossed parts, because $\Rr$ refines $\Tt_1(G,A\cup S)$, so it increases the number of parts of $\Rr$ by more than $9\gamma$. Since $\Rr$ only gets finer as $S$ grows, if (R1) is applied $t$ times, then the final partition has more than $t\cdot 9\gamma$ parts. As it also has at most $\gamma\cdot |A\cup S|$ parts, we get $t\cdot 9\gamma<\gamma\cdot |A\cup S|$, which implies $t<|A\cup S|/9$.

Rule (R1) lets us ignore crossed parts. Precisely, let $\hammsub_\Rr(u,v)$ be the number of parts of $\Rr$ entirely contained in $N(u)\symdiff N(v)$. Every other part that intersects $\sdiff[u,v]$ contains $u$ or $v$, or is crossed by $u$ or by~$v$. So once (R1) no longer applies,
\[\hamm_\Rr(u,v)\leq \hammsub_\Rr(u,v)+18\gamma+2.\]
Hence it suffices to establish the following revised assertions:
\begin{enumerate}[label={(A\arabic*$'$)}, start=2]
	\item\label{p:ov-reps-t} Every vertex $u$ has a vertex $s_u\in S$ with $\hammsub_\Rr(u,s_u)\leq \spoly_{\Cc,r}(n)$.
	\item\label{p:ov-diff-t} Any two vertices $u,v$ in the same $1$-type over $S$ satisfy $\hammsub_\Rr(u,v)\leq \spoly_{\Cc,r}(n)$.
\end{enumerate}

\paragraph*{Packing.}
Towards ensuring \ref{p:ov-reps-t}, we introduce the following rule:
\begin{description}[font=\normalfont]
	\item[(R2, simplified)] If a vertex $w$ satisfies $\hammsub_\Rr(w,s)>\ell$ for all $s\in S$, add $w$ to $S$.
\end{description}
Once (R2) no longer applies, \ref{p:ov-reps-t} holds with the bound $\ell$ (which we will soon set to the desired $\ell = \spoly_{\Cc,r}(n)$).
To bound the number of applications, let $Y$ be the set of vertices added by (R2). Let $S$ and $X$ be the final sets. As $X$ contains one representative of each part of $\Rr=\Tt_r(G,A\cup S)$, we have $|X|\leq \gamma\cdot |A\cup S|$. Consider distinct $u,v\in Y$, where $v$ was added after $u$. When $v$ was added, we had $\hammsub_\Rr(v,u)>\ell$, and as refining $\Rr$ can only increase $\hammsub_\Rr$, this also holds for the final partition. So $N(u)\symdiff N(v)$ contains more than $\ell$ parts of the final partition, and hence also more than $\ell$ of their representatives: $|\sdiff[u,v]\cap X|>\ell$. Bounding $|Y|$ is therefore a packing problem. The \emph{packing ratio} of $G$ is the least $\beta$ such that for every $\ell\in \N$ and $X,Y\subseteq V(G)$ with $|Y|\geq 2$ and $|\sdiff[u,v]\cap X|\geq \ell$ for all distinct $u,v\in Y$, we have $|Y|\leq \beta\cdot |X|/\ell$. Combined with the almost linear bound $|\Tt_1(G,Z)|\leq \gamma\cdot|Z|$ on the number of $1$-types, Haussler's Packing Lemma~\cite{Haussler95} yields $\beta\leq \spoly_{\Cc,r}(n)$ (\cref{lem:haussler-application}). Hence $|Y|\leq \beta\cdot |X|/\ell\leq \beta\gamma\cdot |A\cup S|/\ell$. So if we choose $\ell\coloneqq 9\beta\gamma\leq \spoly_{\Cc,r}(n)$, we conclude that (R2) is applied at most $|A\cup S|/9$ times.

\paragraph*{Hitting symmetric differences.}
Two vertices $u,v\notin S$ have the same $1$-type over $S$ if and only if $S$ is disjoint from $N(u)\symdiff N(v)$. So \ref{p:ov-diff-t} asks that $S$ hits $N(u)\symdiff N(v)$ whenever $\hammsub_\Rr(u,v)\geq k$, for a threshold $k=\spoly_{\Cc,r}(n)$ chosen below. As $X$ contains one representative of each part of $\Rr$, such a set contains at least $k$ vertices of $X$; so it suffices that $S$ hits every set $N(u)\symdiff N(v)$ containing at least $k$ vertices of~$X$. That is, $S$ should be an $\eps$-net for these sets with respect to the uniform measure on $X$, with $\eps=k/|X|$. If $X$ were fixed, a random sample would do: including every vertex of $X$ with probability $1/(9\gamma)$ yields about $|X|/(9\gamma)\leq |A\cup S|/9$ vertices and, with high probability, hits every set with $k$ vertices of $X$, provided $k\geq \Omega(\gamma\log n)$. However, this argument is circular, for adding the sample to $S$ refines~$\Rr$, and the new representatives were never sampled. What we need is an online version of $\eps$-nets.
\begin{description}[font=\normalfont]
	\item[(Online $\eps$-net)] Let $\Ff$ be a set system on a set $V$, and let $k\in \N$. An adversary repeatedly adds elements to a set $P\subseteq V$. After each addition, we may add elements of $V$ to a set $S$, but never remove any, so that $S$ again intersects every $F\in \Ff$ with at least $k$ elements of $P$. The goal is to keep $|S|\leq c\cdot |P|/k$ for a small factor $c$.
\end{description}
In our setting, $\Ff$ consists of the sets $N(u)\symdiff N(v)$, and $P=X$. Every vertex added to $S$, whether by (R1), by (R2), or by the net itself, refines $\Rr$ and may thereby add new representatives to $X$. Using again the almost linear bound on the number of $1$-types and Welzl's theorem on orderings with low crossing number~\cite{Welzl89}, we propose a construction of an online $\eps$-net with $c\leq \spoly_{\Cc,r}(n)$. The third rule follows this net.
\begin{description}[font=\normalfont]
	\item[(R3, simplified)] If the online $\eps$-net, given the current set $X$, adds a vertex $w$, add $w$ to $S$.
\end{description}
Once (R3) no longer applies, $S$ hits every set $N(u)\symdiff N(v)$ with at least $k$ vertices of $X$, which gives~\ref{p:ov-diff-t}. By the guarantee of the online $\eps$-net, and as $|X|\leq \gamma\cdot |A\cup S|$ in the end, (R3) is applied at most $c\cdot |X|/k\leq c\gamma\cdot |A\cup S|/k$ times. Choosing $k\coloneqq 9c\gamma\leq \spoly_{\Cc,r}(n)$, this is at most $|A\cup S|/9$. 

We remark that the above description using an online $\eps$-net serves only as an intuition. The proof of \cref{lem:predamping} does not isolate it as a separate statement, but implements (R3) directly via a vertex ordering of low alternation obtained using Welzl's method (\cref{cor:precise-welzl}). We also note that while we are not aware that this concrete construction was known in the literature, orderings with low alternation were used before to construct (generalizations of) $\eps$-nets, see e.g.~\cite[Theorem~1.3]{AlonJKSY22}.

\section{Preliminaries}\label{sec:prelims}

\paragraph*{Basic notation.} We write $\N$ for the set of positive integers. For $k\in \N$ we denote $[k]\coloneqq \{1,\ldots,k\}$. The \emph{symmetric difference} of sets $X$ and $Y$ is the set $X\symdiff Y\coloneqq (X\setminus Y)\cup (Y\setminus X)$. 
For two partitions $\Pp$ and~$\Qq$ of the same set, we say that $\Pp$ is \emph{finer} than, or is a \emph{refinement} of $\Qq$, and that $\Qq$ is \emph{coarser} than, or is a \emph{coarsening} of $\Pp$, if every part of $\Pp$ is contained in a single part of $\Qq$. We denote it by $\Pp \pleq \Qq$.

\paragraph*{Asymptotics.} We use the standard $\Oh(\cdot)$ notation that hides constant multiplicative factors. For a tuple of objects $\tup q$, we may decorate this notation as $\Oh_{\tup q}(\cdot)$ to signify that the hidden factors may depend on $\tup q$. Typically, $\tup q$ will consist of a monadically dependent graph class $\Cc$ and a distance parameter $r\in \N$, both fixed in the~context.

Call a function $f\colon \N\to \N$ \emph{subpolynomial} if for every real $\varepsilon>0$, we have
\[\lim_{n\to \infty} \frac{f(n)}{n^\eps}=0.\]
We write $\spoly(\cdot)$ for an anonymous subpolynomial function, and in particular $f(n)\leq \spoly(n)$ means that $f(n)$ is bounded by a subpolynomial function of $n$. If $f,g\from\N\to\N$  satisfy 
$f(n),g(n)\le \spoly(n)$, then also $f(n)+g(n)\le \spoly(n)$ and $f(n)\cdot g(n)\le \spoly(n)$. We sometimes use the notation $\spoly_{\tup q}(n)$ to denote a subpolynomial function that may depend on a tuple of objects $\tup q$. Also, we use notation $\Oht(\cdot)$ to hide subpolynomial multiplicative factors, and we may write $\Oht_{\tup q}(\cdot)$ to signify that those factors may depend on $\tup q$.

\paragraph*{Graphs.}
We use standard graph notation. All graphs considered in this paper are finite, undirected, and simple (without loops or parallel edges). For the purpose of computation, we assume that all graphs are represented by their adjacency matrices.

For a vertex $u$ of a graph $G$, by $N_G(u)$ we denote the \emph{open neighborhood}, which is the set of all neighbors of $u$ in $G$, and by $N_G[u]$ we denote the \emph{closed neighborhood}, defined as $N_G(u)\cup \{u\}$. We also write $N_G[A]\coloneqq \bigcup_{u\in A} N_G[u]$. Further, the \emph{open} and \emph{closed symmetric difference} of distinct vertices $u$ and $v$ are defined~as
\[\sdiff_G(u,v)\coloneqq (N_G(u)\symdiff N_G(v))\setminus \{u,v\}\qquad\textrm{and}\qquad \sdiff_G[u,v]\coloneqq (N_G(u)\symdiff N_G(v))\cup \{u,v\}.\]
We define $\sdiff_G(v,v)=\sdiff_G[v,v]=\emptyset$.
In this notation, we may drop the graph in the subscript if it is clear from the context.

The (shortest path) distance between vertices $u$ and $v$ in a graph $G$ is denoted by $\dist_G(u,v)$.
The \emph{ball} with center $u$ of radius $r\in \N\cup \{0\}$ in $G$ is the set of all the vertices that are at distance at most $r$ from $u$ in $G$. We denote it by $\Ball^r_G[u]$.

A vertex $u$ in a graph $G$ is \emph{complete} to a vertex subset $A\subseteq V(G)$ if $u\notin A$ and $u$ is adjacent to every member of $A$. Similarly, $u$ is \emph{anticomplete} to $A$ if $u\notin A$ and $u$ is non-adjacent to every member of $A$. We say that $u$ is \emph{homogeneous} to $A$ if $u$ is either complete or anticomplete to $A$. Note that if $u\in A$, then $u$ cannot be homogeneous to $A$. More generally, for disjoint $A,B\subseteq V(G)$, we say that $A$ and $B$ are \emph{complete} (resp. \emph{anticomplete}) to each other if every vertex of $A$ is complete (resp. \emph{anticomplete}) to $B$; and $A,B$ are \emph{homogeneous} to each other if they are  complete or anticomplete to each other.

A \emph{vertex ordering} of a graph $G$ is a total order on the vertices of $G$.

\paragraph*{VC dimension.} A \emph{set system} over a set $U$ is a family of subsets of $U$. The \emph{VC dimension} of a set system $\Ff$ over $U$ is the maximum cardinality of a set $X\subseteq U$ that is \emph{shattered} by $\Ff$ in the following sense: For every $Y\subseteq X$ there exists $F\in \Ff$ such that $F\cap X=Y$.

\paragraph*{Transducibility and monadic dependence.} We use first-order logic on graphs, with adjacency represented by the relation $\adj(x,y)$. For $k\in\N$, a \emph{$k$-colored graph} is a graph equipped with $k$ unary predicates $U_1,\ldots,U_k$. A \emph{$k$-coloring}, or \emph{expansion}, $G^+$ of a graph $G$ assigns interpretations of these predicates on $V(G)$; the predicates need not be disjoint.

For a $k$-colored graph $H$ and a first-order formula $\varphi(x,y)$ in its signature, define $\varphi(H)$ to be the graph with vertex set $V(H)$ in which distinct vertices $u,v$ are adjacent if and only if $\varphi(u,v)\vee\varphi(v,u)$ holds in~$H$.

For two graph classes $\Cc,\Dd$, we say that $\Cc$ \emph{transduces} $\Dd$, or that $\Dd$ is \emph{transducible} from $\Cc$, if there are $k\in\N$ and a first-order formula $\varphi(x,y)$ in the signature of $k$-colored graphs such that every graph $H\in\Dd$ can be obtained from some graph $G\in\Cc$ by first choosing a $k$-coloring $G^+$ of $G$ and then taking an induced subgraph of $\varphi(G^+)$. The central concept studied in this paper is the following.

\begin{definition}
	A graph class $\Cc$ is \emph{monadically dependent} if $\Cc$ does not transduce the class of all graphs.
\end{definition}

It is easy to see that transducibility is a transitive relation on graph classes~\cite[Lemma~2]{Pilipczuk25}. Therefore, any graph class transducible from a monadically dependent class is again monadically dependent. Also, note the following.

\begin{observation}\label{obs:mondep-vc}
For every monadically dependent graph class $\Cc$ there is a constant $d$ such that for every $G\in \Cc$, the set system $\{N_G[u]\colon u\in V(G)\}$ over $V(G)$ has VC dimension at most $d$.
\end{observation}
\begin{proof}
	Suppose that the set system $\{N_G[u]\colon u\in V(G)\}$ for some graph $G\in\Cc$ shatters a set $X$. Pick any graph $H$ with vertex set $X$. For every edge $uv\in E(H)$, choose a vertex $w_{uv}\in V(G)$ such that $N_G[w_{uv}]\cap X=\{u,v\}$, and color all chosen vertices with a distinguished predicate $U$, obtaining a $1$-coloring $G^+$ of $G$.
	Let $\varphi(x,y)$ be the first-order formula with one unary predicate $U$  asserting that there is a vertex $z\in U$ with $x,y\in N[z]$. Then $H$ is the subgraph of $\varphi(G^+)$ induced by $X$. If the VC dimensions of these set systems over graphs in $\Cc$ were unbounded, the fixed formula $\varphi$ would therefore witness that $\Cc$ transduces the class of all graphs, contradicting monadic dependence.
\end{proof}

\paragraph{$1$-types.}
Let $G$ be a graph and $S\subset V(G)$ a set of its vertices.
We say that two vertices $a,b\in V(G)$ have the same \emph{$1$-type over $S$} if either $a=b\in S$, or $a,b\notin S$ and $N_G(a)\cap S=N_G(b)\cap S$
(the more commonly used term is \emph{atomic type}, but later we will use an extension where $1$ is replaced with a parameter).
We call the corresponding partition of $V(G)$ the partition into \emph{$1$-types over $S$}, and denote it by $\Tt_1(G,S)$. Note that if $R\subset S$ then $\Tt_1(G,R)$ is coarser than $\Tt_1(G,S)$.

\paragraph{Neighborhood complexity.}
As mentioned in \cref{sec:overview}, we will heavily rely on the recent result of Dreier et al.~\cite{DreierMMPT26} stating that in monadically dependent graph classes, the number of $1$-types over a vertex subset $A$ is almost linear in the cardinality of $A$.

\begin{theorem}[\cite{DreierMMPT26}]\label{thm:mondep-1-types}
	For every monadically dependent graph class $\Cc$, there is a function $\gamma\colon \N\to \N$ with $\gamma(n)\le \spoly_{\Cc}(n)$ such that for any graph $G\in \Cc$ and non-empty set $A\subseteq V(G)$, we have 
	\[|\Tt_1(G,A)|\leq \gamma(|A|)\cdot |A|.\]
\end{theorem}

\cref{thm:mondep-1-types} is closely related to the following concept: the \emph{shatter function} or the \emph{neighborhood complexity}\footnote{In the literature, the shatter function/neighborhood complexity typically counts the number of distinct neighborhoods of the form $N_G(u)\cap A$ for $u\in V(G)$. Here, we find it more convenient to measure $|\Tt_1(G,A)|$ instead. The two quantities differ by an additive term of at most $|A|$, which is immaterial for our purposes.} of a graph $G$ is the function $\pi_G\colon \N\to \N$ that maps every $k\in \N$ to the maximum of $|\Tt_1(G,A)|$ among vertex subsets $A$ of cardinality~$k$. Thus, \cref{thm:mondep-1-types} asserts that in monadically dependent classes, the shatter function $\pi_G(k)$ is always bounded by $k\cdot \gamma(k)$ for a subpolynomial function~$\gamma$ that depends only on the class $\Cc$.

\paragraph{Flips and definable flips.}

Let $G$ be a graph with vertex set $V$ and $\Pp$ be a partition of its vertex set.
A \emph{$\Pp$-flip} of $G$ is a graph $H$ with the same vertex set as $G$ such that 
for every pair of parts $X,Y\in\Pp$ (possibly $X=Y$) one of the following cases holds:
\begin{itemize}
	\item $xy\in E(G)\Leftrightarrow xy\in E(H),$ \quad for all distinct $x\in X$ and $y\in Y$, or
\item  $xy\in E(G)\Leftrightarrow xy\notin E(H),$ \quad for all distinct $x\in X$ and $y\in Y$.
\end{itemize}
Thus, a $\Pp$-flip $H$ of $G$ can be fully described by $G$ and a symmetric binary relation $\sigma\subset \Pp\times \Pp$ which distinguishes those pairs 
$X,Y\in\Pp$ for which the second case above occurs; then $H$ is obtained by complementing the adjacency relation between $X$ and $Y$, for every pair of parts $X,Y\in \Pp$ with $(X,Y)\in \sigma$.

For $S\subset V(G)$,
an \emph{$S$-definable flip} -- or more briefly, an \emph{$S$-flip} of $G$ -- is a $\Tt_1(G,S)$-flip of $G$, where $\Tt_1(G,S)$ is the partition of $V(G)$ into $1$-types over $S$.  The following observation is immediate.

\begin{observation}\label{obs:flips-monotone}
	Let $G$ be a graph and $\Pp,\Qq$ be partitions of the vertex set of $G$ such that $\Pp$ is a refinement of $\Qq$. Then every $\Qq$-flip of $G$ is also a $\Pp$-flip of $G$. Consequently, if $S\subseteq T\subseteq V(G)$ are two vertex subsets, then every $S$-flip of $G$ is also a $T$-flip of $G$.
\end{observation}

\paragraph{Flip metric.}
Next, we recall the concept of \emph{flip-metric}, introduced by Gajarsk\'y et al. in~\cite{GajarskyMMOPPSS23}. Let $G$ be a graph and $\Pp$ be a partition of the vertex set of $G$. For vertices $u,v\in V(G)$, we define
\[\dist_\Pp(u,v)\coloneqq \max \{\dist_H(u,v)\colon H\textrm{ is a }\Pp\textrm{-flip of }G\}.\]
The idea behind this definition is to measure the distance between $u$ and $v$ relative to a partition $\Pp$, but instead of committing to doing so in a single $\Pp$-flip, to consider all the possible $\Pp$-flips at once. 

As observed in~\cite{GajarskyMMOPPSS23}, $\dist_\Pp(\cdot,\cdot)$ is a quasi-metric\footnote{A \emph{quasi-metric} is a metric that can take value $+\infty$.} on the vertex set of $G$. Note that radius-$r$ balls in this quasi-metric take the following form:
\[\Ball^r_\Pp[u]\coloneqq \bigcap \{\Ball^r_H[u]\colon H\textrm{ is a }\Pp\textrm{-flip of }G\}.\]

\subsection*{Merge-width}

We first recall the definition of merge-width as introduced by Dreier and Toru\'nczyk~\cite{DreierT25}.
Let $G$ be a graph with vertex set $V$. By $\binom{V}{2}$ we denote the set of unordered pairs of distinct vertices of $G$. A \emph{merge sequence} for $G$ is a sequence
\[(\Pp_1,R_1),(\Pp_2,R_2),\ldots,(\Pp_m,R_m)\]
such that
\begin{itemize}
	\item $\Pp_1\pleq \Pp_2\pleq \ldots \pleq \Pp_m$ is a sequence of ever coarser partitions of the vertex set of $G$, where $\Pp_1$ is the partition into singletons and $\Pp_m$ has one part $V$;
	\item $R_1\subseteq R_2\subseteq \ldots\subseteq R_m\subseteq \binom{V}{2}$ is an inclusion chain of sets of pairs of vertices of $G$;
	\item for every $i\in [m]$ and pair of parts $P,P'\in \Pp_i$ (possibly $P=P'$), either all pairs $uu'\in \binom{V}{2}$ with $u\in P$, $u'\in P'$, $uu'\notin R_i$, are edges in $G$, or they are all non-edges in $G$.
\end{itemize}
For $r\in \N$, the \emph{radius-$r$ width} of the merge sequence $(\Pp_1,R_1),(\Pp_2,R_2),\ldots,(\Pp_m,R_m)$ is the smallest integer $k$ satisfying the following property:
\begin{quote}
	For every $i\in [m-1]$ and  $u\in V$, $\Ball^r_{(V,R_{i+1})}[u]$ intersects at most $k$ parts of $\Pp_i$.
\end{quote}
Note the $+1$ in this definition: It is important that while radius-$r$ balls are considered with respect to the larger edge set $R_{i+1}$, we measure the number of intersected parts of the finer partition $\Pp_i$.

The \emph{radius-$r$ merge-width} of $G$ is the smallest $k$ such that $G$ admits a merge sequence of radius-$r$ width at most $k$. A graph class $\Cc$ has \emph{almost bounded merge-width} if for every $r\in \N$, every $n$-vertex graph in $\Cc$ has radius-$r$ merge-width bounded by $f(n)$ for some function $f(n)\le \spoly_{\Cc,r}(n)$. Having \emph{bounded merge-width} is more restrictive: we require that the radius-$r$ merge-width of any $G\in \Cc$ is bounded by a constant depending only on $\Cc$ and $r$.

\paragraph*{Partition merge-width.}
In~\cite{DrabikDGNPT26}, Drabik, Dumas, Geniet, Nowakowski, Pilipczuk, and Toru\'nczyk studied several variants of the definition of merge-width. They showed that some relaxations of the conditions lead to functionally equivalent families of parameters. We will use their concept of \emph{partition merge-width}.
Intuitively, partition merge-width avoids the need of explicitly providing the monotone sequence of sets of pairs $R_i$, and works with balls in the flip-metric instead.

A \emph{partition sequence} for a graph $G$ is a sequence 
\[\Pp_1\pleq \Pp_2 \pleq \ldots\pleq \Pp_m\]
of ever coarser partitions of the vertex set of $G$ such that $\Pp_1$ is the partition into singletons and $\Pp_m$ has one part $V(G)$.
For $r\in \N$, the \emph{radius-$r$ width} of $\Pp_1,\ldots,\Pp_m$ is the smallest integer $k$ satisfying the following:
\begin{quote}
	For every $i\in [m-1]$ and $u\in V(G)$, $\Ball_{\Pp_{i+1}}^r[u]$ intersects at most $k$ parts of $\Pp_i$.
\end{quote}
Again, the \emph{radius-$r$ partition merge-width} of $G$ is the least $k$ such that $G$ admits a partition sequence of radius-$r$ width at most $k$.

The radius-$r$ partition merge-width of a graph $G$ is never larger than its radius-$r$ merge-width: The last condition in the definition of a merge sequence is equivalent to saying that there is a $\Pp_i$-flip of $G$ that is a subgraph of $(V(G),R_i)$, so the bound on the reachable parts of $\Pp_{i+1}$ transfers; see \cite[Lemmas 3.3 and 3.6]{DrabikDGNPT26}. The implication in the other direction is far from obvious, but as proved by Drabik et al., can be obtained at the cost of a minor loss in the radius.

\begin{theorem}[{\cite[Theorem~5.1]{DrabikDGNPT26}}]\label{thm:partition-mw}
	Let $r\in \N$, $G$ be a graph, and $\pi_G\colon \N\to \N$ be the shatter function of $G$. Let $k$ be the radius-$(6r+1)$ partition merge-width of $G$. Then $G$ has radius-$r$ merge-width at most $k\cdot \pi_G(k)$. 
	
	Moreover, given a partition sequence for $G$ of radius-$(6r+1)$ width $k$, a merge sequence for $G$ of radius-$r$ width at most $k\cdot \pi_G(k)$ can be computed in $\Oh(n^3)$ time, where $n$ is the vertex count of $G$.
\end{theorem}

We note that for the algorithmic statement, Drabik et al.~\cite{DrabikDGNPT26} only claim that the running time is polynomial. The $\Oh(n^3)$ bound stated above follows from a careful inspection of the proof combined with a simple amortization argument;  we outline this in \cref{app:partition-mw}.

\section{Damping Theorem, and derivation of the main result}\label{sec:merge-width}

In this section we reduce the proof of \cref{thm:main} to an auxiliary statement called the Damping Theorem. The subsequent sections will be devoted to the proof of the Damping Theorem.

We now state the key technical ingredient of the proof of \cref{thm:main}. 

\begin{restatable}[Damping Theorem]{theorem}{sct}\label{thm:static-s-flip}
	Let $r\in \N$ and $\Cc$ be a monadically dependent graph class. Then there exists a function $h\colon \N\to \N$ with $h(n)\le \spoly_{\Cc,r}(n)$ such that for any $q\in \N$, $n$-vertex graph $G\in \Cc$, and vertex subset $A\subseteq V(G)$, there exist a set $S\subseteq V(G)$ and an $S$-flip $H$ of $G$ satisfying the following properties:
	\begin{itemize}
		\item $|S|\leq |A|/q$; and
		\item for every vertex $u\in V(G)$, $\Ball_H^r[u]$ intersects at most $q^r\cdot h(n)$ parts of $\Tt_1(G,A\cup S)$.
	\end{itemize}
	Also, there is an algorithm that, given $q$, $r$, $G$, and~$A$, computes such a set $S$ and an $S$-flip $H$ in time~$\Oht_{\Cc,r}(n^3)$. 
\end{restatable}

The set $S$ and the flip $H$ act as a ``damper'' for $A$, by reducing its impact on outside vertices $v$. Crucially, $S$ is only a fixed fraction of the size of $A$.
We note that for the purpose of proving \cref{thm:main}, it is sufficient to focus on the case $q=2$ in \cref{thm:static-s-flip}; that is, require that $S$ is at least twice smaller than $A$. 
However, we go the extra inch to prove a more ``tunable'' version of \cref{thm:static-s-flip} that allows a trade-off between the size of $S$ and the quality of the flip~$H$. The purpose of it is to derive a variant of \cref{thm:main} that will be used in~\cite{DreierMT26}. This larger generality poses no additional conceptual obstacles in the proof. We also note that the algorithm in \Cref{thm:static-s-flip} does not depend on the class $\CC$.

In the remaining sections of this paper we work towards the proof of \cref{thm:static-s-flip}. Before we proceed with it, we present how \cref{thm:main} can be derived from \cref{thm:static-s-flip}. We first need a lemma providing a certain additivity property of $1$-types.

\begin{lemma}\label{lem:1-types-additive}
	Let $G$ be a graph and $\gamma\in\N$ be 
	such that $|\Tt_1(G,A)|\le \gamma|A|$ for every non-empty $A\subset V(G)$.
	Let $S_1,\ldots,S_t,X\subseteq V(G)$ be vertex subsets. Further, for every $i\in [t]$, let $k_i$ be the number of $1$-types over $S_i$ that are intersected by $X$. Then the number of $1$-types over $S\coloneqq \bigcup_{i\in [t]} S_i$ intersected by $X$ is bounded by $\gamma\cdot \sum_{i\in [t]} k_i$.
\end{lemma}
\begin{proof}
	Consider any $i\in [t]$. 
	We claim that there exists $R_i\subseteq S_i$ with $|R_i|\leq k_i-1$ such that whenever two elements of $X$ have different $1$-types over $S_i$, they also have different $1$-types over $R_i$.
	Let $X_i$ be a subset of $X$ containing one representative of each $1$-type over $S_i$ intersected by $X$; thus $|X_i|=k_i$. It suffices to find $R_i\subseteq S_i$ with $|R_i|\leq k_i-1$ such that
	the elements of $X_i$ have pairwise different $1$-types over $R_i$.
	Start with $R_i\coloneqq \emptyset$ and as long as there exist two distinct vertices $x,x'\in X_i$ with the same $1$-type over $R_i$, add to $R_i$ any vertex $y\in S_i$ that distinguishes $x$ and $x'$; that is, $y\in \{x,x'\}$ or $y$ is non-homogeneous to $\{x,x'\}$. (Such a vertex $y$ always exists because $x,x'$ have different $1$-types over~$S_i$.) Every vertex added to $R_i$ strictly increases the number of $1$-types over $R_i$ intersected by $X_i$, hence after at most $|X_i|-1=k_i-1$ iterations we obtain the desired set $R_i$.
	
	Let $R\coloneqq \bigcup_{i\in [t]} R_i$; thus $R\subseteq S$ and $|R|\leq \sum_{i\in [t]} k_i$. We claim that if vertices $x,x'\in X$ have different $1$-types over $S$, then they also have different $1$-types over $R$. Indeed, since $x,x'\in X$ have different $1$-types over $S$, there exists $y\in S$ that distinguishes $x$ and $x'$ in the sense from the previous paragraph. Then $y\in S_i$ for some $i\in [t]$. Consequently, $x$ and $x'$ have different $1$-types over $S_i$, so they also have different $1$-types over $R_i$. As $R_i\subseteq R$, it follows that $x$ and $x'$ have different $1$-types over $R$.
	
	From the observation above it follows that the number of $1$-types over $S$ intersected by $X$ is equal to the number of $1$-types over $R$ intersected by $X$.  
	We conclude by noting that the total number of $1$-types over $R$ is bounded by $\gamma|R|\leq \gamma\cdot \sum_{i\in [t]} k_i$.
\end{proof}

We now prove a variant of \cref{thm:main} that provides a very concrete shape of a partition sequence of the given graph. We note that this shape is actually quite close to the concept of \emph{definable merge-width}, also proposed by Drabik et al.~\cite{DrabikDGNPT26}. But in order not to overcomplicate things, we choose to stick to the terminology of partition merge-width.

\begin{theorem}\label{thm:main-ordering}
	Let $r\in \N$ and $\Cc$ be a monadically dependent graph class. Then there is a function $k\colon \N\to \N$ with $k(n)\le \spoly_{\Cc,r}(n)$ such that for any integer $q\geq 2$ and any $n$-vertex graph $G\in \Cc$, there exists a sequence of vertex~subsets
	\[\emptyset=B_m\subseteq B_{m-1}\subseteq \ldots\subseteq B_1\subseteq B_0=V(G)\]
	with $m\leq 1+\log_q n$ satisfying the following condition. For every $i\in [m]$ and $u\in V(G)$, 
	\begin{quote}
		$\Ball^r_{\Pp_i}[u]$ \quad intersects at most \quad $k'(n) \coloneqq q^r\cdot (\log_q n+1)\cdot k(n)$ \quad parts of \quad $\Pp_{i-1}$,
	\end{quote}
	where $\PP_i \coloneqq \Tt_1(G,B_i)$. In particular, $\PP_0 \pleq \PP_1 \pleq \ldots \pleq \PP_m$ is a partition sequence for $G$ of radius-$r$ width at most $k'(n)$.

	Moreover, given $q$, $r$, and $G$, such a sequence $B_0,B_1,\ldots,B_m$ can be computed in $\Oht_{\Cc,r}(n^3)$ time.
\end{theorem}
\begin{proof}[Proof of \cref{thm:main-ordering} assuming \cref{thm:static-s-flip}]
	Let $\gamma,h\colon \N\to \N$ be the functions provided by \cref{thm:mondep-1-types,thm:static-s-flip} for the class $\Cc$ and radius $r$. Thus, $\gamma(n)\le \spoly_\Cc(n)$ and $h(n)\le \spoly_{\Cc,r}(n)$. Denote $\gamma\coloneqq \gamma(n)$ and $h\coloneqq h(n)$, where $n$ is the vertex count of the considered graph $G$.
	
	We iteratively define sets $S_0,S_1,S_2,\ldots$ and graphs $H_1,H_2,\ldots$ as follows:
	\begin{itemize}
		\item Set $S_0\coloneqq V(G)$.
		\item For $i\geq 1$, apply \cref{thm:static-s-flip} with parameter $q$ to the set $A=S_{i-1}$. Thus, in $\Oht_{\Cc,r}(n^3)$ time we obtain a set $S_i\subseteq V(G)$ satisfying $|S_i|\leq |S_{i-1}|/q$ and an $S_i$-flip $H_i$ such that for every $u\in V(G)$, $\Ball_{H_i}^r[u]$ intersects at most $q^r\cdot h$ parts of $\Tt_1(G,S_{i-1}\cup S_i)$. If $S_i=\emptyset$, we finish the iteration and set $m\coloneqq i$. Otherwise, we proceed to the next step with the set $S_i$ defined.
	\end{itemize}
	Note that we have $|S_i|\leq n/q^i$ for every $i\in \N$, hence the iteration must stop after at most $1+\log_q n$ steps, yielding $m\leq 1+\log_q n$. 
	In particular we have $m\leq 1+\log_2 n\leq \Oht(1)$ due to $q\geq 2$, so sets $S_0,S_1,S_2,\ldots$ and graphs $H_1,H_2,\ldots$ can be computed in total time $\Oht_{\Cc,r}(n^3)$.
	
	For $i\in \{0,1,\ldots,m\}$, define
	\[B_i\coloneqq \bigcup_{i \leq j \leq m} S_j.\]
	Thus $V(G)=B_0\supseteq B_1\supseteq \ldots \supseteq B_m=\emptyset$, and sets $B_0,\ldots,B_m$ can be computed from the sets $S_0,\ldots,S_m$ in $\Oh(n^2)$ time. Denote $\Pp_i\coloneqq \Tt_1(G,B_i)$.
	
	Consider any $i\in [m]$ and $u\in V(G)$. Let $X\coloneqq \Ball^r_{\Pp_i}[u]$. Note that $S_j$ is a subset of $B_i$ for every $j\geq i$, hence by \cref{obs:flips-monotone}, $H_j$ is also a $B_i$-flip of $G$. Consequently, we have
	\[X=\Ball^r_{\Pp_i}[u]\subseteq \Ball^r_{H_j}[u],\qquad \textrm{for every }j\geq i.\]
	Since $\Ball^r_{H_j}[u]$ intersects at most $q^r\cdot h$ parts of $\Tt_1(G,S_{j-1}\cup S_j)$, it also intersects at most $q^r\cdot h$ parts of $\Tt_1(G,S_{j-1})$, for this partition can only be coarser. We conclude that
	\[X\textrm{ intersects at most }q^r\cdot h\textrm{ parts of }\Tt_1(G,S_{j-1}),\qquad \textrm{for every }j\geq i.\]
	We now apply \cref{lem:1-types-additive} to the sets $S_{i-1},S_i,S_{i+1},\ldots,S_{m-1},X$. Noting that $B_{i-1}=\bigcup_{i-1\le j\le m-1}S_j$, we thus conclude that
	\[X\textrm{ intersects at most }\gamma\cdot m\cdot q^r\cdot h\textrm{ parts of }\Tt_1(G,B_{i-1})=\Pp_{i-1}.\]
	Therefore, we may set $k(n)\coloneqq \gamma(n)\cdot h(n)$ and note that $k(n)\le \spoly_{\Cc,r}(n)$.
\end{proof}

\cref{thm:main} now follows from a straightforward combination of \cref{thm:main-ordering,thm:partition-mw}.

\main*
\begin{proof}
	Applying \cref{thm:main-ordering} for $q=2$ and radius $r'\coloneqq 6r+1$ to the considered graph $G\in \Cc$ yields, in $\Oht_{\Cc,r}(n^3)$ time, a partition sequence for $G$ of radius-$r'$ width bounded by $k'(n)\coloneqq 2^{6r+1}\cdot (\log_2 n+1)\cdot k(n)\le \spoly_{\Cc,r}(n)$. We may now apply \cref{thm:partition-mw} to this sequence and thus obtain, in $\Oh(n^3)$ time, a merge sequence for $G$ of radius-$r$ width bounded by $k'(n)\cdot \pi_G(k'(n))$. By \cref{thm:mondep-1-types}, $\pi_G$ is bounded by a polynomial (even almost linear) function, hence $k'(n)\cdot \pi_G(k'(n))\le \spoly_{\Cc,r}(n)$.
\end{proof}

Finally, we derive a variant of \cref{thm:main} that will be used in~\cite{DreierMT26}.

\begin{theorem}
	Let $r\in \N$ and $\Cc$ be a monadically dependent graph class. Then for any rational $\eps>0$ and any $n$-vertex graph $G\in \Cc$, there is a sequence of vertex subsets
	\[\emptyset=B_m\subseteq B_{m-1}\subseteq \ldots\subseteq B_1\subseteq B_0=V(G)\]
	with $m\leq \Oh(r/\eps)$, which satisfies the following condition. For every $i\in [m]$ and $u\in V(G)$,
	\begin{quote}
		$\Ball^r_{\Tt_1(G,B_i)}[u]$ \quad intersects at most \quad $\Oh_{\Cc,r,\eps}(n^\eps)$ \quad parts of \quad $\Tt_1(G,B_{i-1})$.
	\end{quote}
	Moreover, there is an algorithm (independent of $\CC$ and $r$) which, given $\eps$, $r$, and $G$, computes such a sequence $B_0,B_1,\ldots,B_m$, and runs	in time $\Oht_{\Cc,r}(n^{3})$ for graphs $G\in\CC$.
\end{theorem}
\begin{proof}
	We may assume $n\geq 2$. We apply \cref{thm:main-ordering} with $q=\left\lceil n^{\nicefrac{\eps}{2r}}\right\rceil$ and radius parameter $r$. This yields the desired sequence of vertex subsets $B_0,B_1,\ldots,B_m$ with $m\leq 1+\log_q n\leq 1+2r/\eps\le \Oh(r/\eps)$. The bound on the number of $1$-types over $B_{i-1}$ intersected by a radius-$r$ ball in the $\Tt_1(G,B_i)$-metric becomes
	\[q^r\cdot (\log_q n+1)\cdot k(n)\leq (2n^{\nicefrac{\eps}{2r}})^r\cdot (1+2r/\eps)\cdot k(n)\leq 2^r(1+2r/\eps)\cdot n^{\nicefrac{\eps}{2}}\cdot k(n).\]
	Since $k(n)\le \spoly_{\Cc,r}(n)$, this quantity is bounded by $\Oh_{\Cc,r,\eps}(n^\eps)$. Finally, the running time is bounded by $\Oht_{\Cc,r}(n^3)$ by \cref{thm:main-ordering}.
\end{proof}

\section{Ingredients}\label{sec:ingredients}

In this section we gather several concepts and tools that we will need for the proof of \cref{thm:static-s-flip}. We also introduce some terminology that will be useful for stating some of the intermediate results on a suitable level of generality.

\subsection{$\mathbf{r}$-types}
We introduce the key notion behind the proof of \Cref{thm:static-s-flip}.
It extends the concept of $1$-types to larger distances.
Recall that
given a graph $G$ and $A\subseteq V(G)$, two vertices $u,v$ of $G$ have the same $1$-type over $A$ if either $u=v\in A$, or $u,v\notin A$ and $N_G(u)\cap A=N_G(v)\cap A$.

For an integer $r\geq 2$, we define the partition $\Tt_r(G,A)$ into \emph{$r$-types} over $A$ inductively as follows. Two vertices $u,u'\in V(G)$ have the same $r$-type if and only if the following conditions are satisfied:
	\begin{itemize}
		\item $u$ and $u'$ have  the same $(r-1)$-type over $A$;
		\item for each $Q\in \Tt_{r-1}(G,A)$, $u$ is complete to $Q\setminus \{u\}$ if and only if $u'$ is complete to~$Q\setminus \{u'\}$;~and
		\item for each $Q\in \Tt_{r-1}(G,A)$, $u$ is anticomplete to $Q\setminus \{u\}$ if and only if $u'$ is anticomplete~to~$Q\setminus \{u'\}$.
	\end{itemize}
Note that the exclusion of $u$ or $u'$ from $Q$ in the definition is relevant only if $u,u'\in Q$; otherwise, the conditions just speak about (anti)completeness to $Q$.
Also observe that if $A\subset B\subset V(G)$ then $\Tt_r(G,A)$ is coarser than $\Tt_r(G,B)$.

In the next lemma we extend \cref{thm:mondep-1-types} to $r$-types. The proof is a reduction to \cref{thm:mondep-1-types} that exploits first-order definability of the partition into $r$-types.

\begin{lemma}\label{thm:mondep-r-types}
	For every monadically dependent graph class $\Cc$ and $r\in \N$, there is a function $\gamma_r\colon \N\to \N$ with $\gamma_r(n)\le \spoly_{\Cc,r}(n)$ such that for any graph $G\in \Cc$ and non-empty set $A\subseteq V(G)$, we have 
	\[|\Tt_r(G,A)|\leq \gamma_r(|A|)\cdot |A|.\]
\end{lemma}
\begin{proof}
	We first observe that the partitions into $r$-types are definable in the following sense.
	
	\begin{claim}\label{cl:types-definable}
		For every $r\in \N$ there is a first-order formula $\tau_r(x,y)$ over $\{A\}$-colored graphs such that for any graph $G$, vertex subset $A$, and vertices $u,u'\in V(G)$, we have
		\[\tau_r(u,u')\textrm{ holds in }(G,A)\qquad\textrm{if and only if}\qquad u\textrm{ and }u'\textrm{ have the same }r\textrm{-type over }A.\]
	\end{claim}
	\begin{claimproof}
		This follows from a straightforward encoding of the inductive definition of $r$-types in first-order logic. Precisely, for $r=1$ we may use the formula
		\[\tau_1(x,x')\coloneqq (x=x')\vee [x\notin A\wedge x'\notin A\wedge \forall y\, (y\in A\rightarrow (\adj(x,y)\leftrightarrow \adj(x',y)))].\]
		For $r\geq 2$, formula $\tau_r(x,x')$ can be constructed from $\tau_{r-1}(x,x')$ as follows:
		\begin{align*}
			\tau_r(x,x') \coloneqq\, & \tau_{r-1}(x,x')\wedge \\\ & \forall y\Big[
			 ((\forall z\ z\neq x\wedge \tau_{r-1}(y,z)\rightarrow \phantom{\neg}\adj(x,z))\leftrightarrow (\forall z\ z\neq x'\wedge \tau_{r-1}(y,z)\rightarrow\phantom{\neg} \adj(x',z))) \wedge \\
			& \phantom{\forall y\Big[}((\forall z\ z\neq x\wedge \tau_{r-1}(y,z)\rightarrow \neg\adj(x,z))\leftrightarrow (\forall z\ z\neq x'\wedge \tau_{r-1}(y,z)\rightarrow \neg\adj(x',z))) \Big].\quad \qedhere
		\end{align*}
	\end{claimproof}
	
	We use the following bound on types of formulas in monadically dependent graph classes, which is a generalization and  easy consequence of \Cref{thm:mondep-1-types}. For a finite set $\Gamma$ of first-order formulas $\varphi(x,y)$ over colored graphs, a colored graph $G$, and a set $B\subseteq V(G)$, let $\Tt_\Gamma(G,B)$ denote the partition of $V(G)$ into \emph{$\Gamma$-types over $B$}: vertices $u,v$ belong to the same part if and only if, for every $b\in B$ and $\varphi\in\Gamma$, we have that $\varphi(u,b)\leftrightarrow\varphi(v,b)$ holds in $G$.

	\begin{lemma}\label{lem:mondep-formula-types}
		Fix a finite set $\Gamma$ of first-order formulas $\varphi(x,y)$, possibly using a fixed finite set of vertex colors, and a monadically dependent graph class $\Cc$. There is a function $\gamma_\Gamma\colon\N\to\N$ with $\gamma_\Gamma(n)\leq\spoly_{\Cc,\Gamma}(n)$ such that, for every graph $G\in\Cc$, every expansion $G^+$ of $G$ by the colors occurring in $\Gamma$, and every non-empty set $B\subseteq V(G)$, we have
		\[|\Tt_\Gamma(G^+,B)|\leq |B|\cdot\gamma_\Gamma(|B|).\]
	\end{lemma}
	\noindent The proof of \cref{lem:mondep-formula-types} proceeds by lifting \Cref{thm:mondep-1-types} in a standard way, and is
	deferred to \cref{sec:formula-types}.

	\medskip
	For $r\geq 2$, let $\Gamma_r$ consist of the three formulas asserting, in the colored graph $G^+:=(G,A)$, that
	\begin{itemize}
		\item $x$ and $y$ have the same $(r-1)$-type over $A$;
		\item $x$ is complete to the $(r-1)$-type of $y$ over $A$, with $x$ itself removed; and
		\item $x$ is anticomplete to the $(r-1)$-type of $y$ over $A$, with $x$ itself removed.
	\end{itemize}
	These formulas exist by \cref{cl:types-definable} and depend only on $r$. For each $i\geq 1$, choose a set $A_i$ containing exactly one representative of each $i$-type over $A$.
	Then
	\[\Tt_{\Gamma_r}(G^+,A_{r-1})=\Tt_r(G,A):\]
	the first formula identifies the $(r-1)$-type of a vertex, while the other two record its attachments to every $(r-1)$-type. For $r=1$, \cref{thm:mondep-1-types} gives $|A_1|\leq |A|\cdot\spoly_\Cc(|A|)$. Applying \cref{lem:mondep-formula-types} inductively now yields
	\[
		|A_r|=|\Tt_{\Gamma_r}(G^+,A_{r-1})|
		\leq |A_{r-1}|\cdot\spoly_{\Cc,r}(|A_{r-1}|)
		\leq |A|\cdot\spoly_{\Cc,r}(|A|),
	\]
	as required.
\end{proof}

\paragraph{Partition selectors.}
We introduce some terminology to conveniently capture some of the key properties of the partitions $\cal T_r(G,A)$.
A \emph{partition selector} is a function $\Pp(\cdot,\cdot)$ that to a graph $G$ and a set of vertices $A\subseteq V(G)$ assigns a partition $\Pp(G,A)$ of the vertex set of $G$. We will only be interested in the partition selectors $\cal T_r$ for $r\in\N$ in this paper.

We call a partition selector $\Pp$ \emph{monotone} if  $\Pp(G,A)$ is a coarsening of $\Pp(G,B)$
whenever $A\subseteq B$.
 Also, $\Pp$ is \emph{$t(n)$-time computable} if there is a $t(n)$-time algorithm computing $\Pp(G,A)$ given $G$ and $A$, where $n$ is the number of vertices of $G$.

It is clear from the definition that $\Tt_r$ is monotone and it is always the case that $\Tt_r(G,A)$ is a refinement of $\Tt_{r-1}(G,A)$. More generally, we call a partition selector $\Pp$ \emph{respectful} if for any graph $G$ and $A\subseteq V(G)$, $\Pp(G,A)$ is a refinement of $\Tt_1(G,A)$; thus $\Tt_r$ is respectful.

\begin{definition}
	The \emph{growth ratio} of a partition selector $\Pp$ on a graph $G$ is the least integer $\gamma$ such that
	\[|\Pp(G,A)|\leq \gamma\cdot |A|\qquad \textrm{for every non-empty set } A\subseteq V(G).\]
\end{definition}

In particular, \cref{thm:mondep-r-types} immediately implies the following.

\begin{theorem}\label{thm:mondep-t-complexity}
	For every monadically dependent graph class $\Cc$ and $r\in \N$, there is a function $\gamma_r\colon \N\to \N$~with $\gamma_r(n)\le \spoly_{\Cc,r}(n)$ such that for any $n$-vertex graph $G\in \Cc$, $\Tt_r$ has growth ratio at most~$\gamma_r(n)$ on~$G$.
\end{theorem}
\begin{proof}
	It suffices to note that in the context of \cref{thm:mondep-r-types}, $|A|\leq |V(G)|=n$  and we may assume without loss of generality that $\gamma_r$ is non-decreasing.
\end{proof}

Finally, we note that the partition selectors $\Tt_r$ are computable in quadratic time.

\begin{lemma}\label{lem:compute-Tr}
	There is an $\Oh(rn^2)$-time algorithm that given $r\in \N$, an $n$-vertex graph $G$, and a vertex subset $A\subseteq V(G)$, computes the partition $\Tt_r(G,A)$.
\end{lemma}
\begin{proof}
	For $r=1$, we may compute partition $\Tt_1(G,A)$ in $\Oh(n^2)$ time by first computing for every vertex $v\in V(G)\setminus A$ a Boolean vector of length $|A|$ representing its neighborhood in $A$, and then applying radix sort to those vectors. For larger $r$, it suffices to show how to compute $\Tt_r(G,A)$ from $\Tt_{r-1}(G,A)$ in $\Oh(n^2)$ time, for any $r\geq 2$. For this, we compute, for every vertex $u\in V(G)$ and part $Q\in \Tt_{r-1}(G,A)$, the number of vertices $v\in Q$ that are adjacent to $u$; call this number $c_{u,Q}$. This can be done in total time $\Oh(n^2)$ by considering every edge $uv$ of $G$ and incrementing numbers $c_{u,Q}$ and $c_{v,P}$, where $P,Q\in \Tt_{r-1}(G,A)$ are such that $u\in P$ and $v\in Q$. Once the numbers $c_{u,Q}$ are computed, we may easily determine, for every vertex $u$ and part $Q\in \Tt_{r-1}(G,A)$, whether $u$ is anticomplete, complete, or non-homogeneous to $Q\setminus \{u\}$. This yields, for every vertex $u$, a vector of length $|\Tt_{r-1}(G,A)|$ that describes for every part $Q\in \Tt_{r-1}(G,A)$ one of six possible states: (i) whether $u$ belongs to $Q$ or not, and (ii) whether $u$ is anticomplete, complete, or non-homogeneous to $Q\setminus \{u\}$. Sorting these vectors using radix sort and grouping consecutive duplicates yields the desired partition $\Tt_r(G,A)$  in time $\Oh(n\cdot |\Tt_{r-1}(G,A)|)\leq \Oh(n^2)$.
\end{proof}

\subsection{Packing ratio}

We will use the well-known result due to Haussler~\cite{Haussler95}. The bound in the following formulation follows from the proof\footnote{The constant $C_2$ on page 158 of \cite{matouvsek1999geometric} can be taken to be $c\cdot(4d)^\rho$
 in our notation, and the last inequality on page 159, for $\mathcal P= \mathcal F$,
translates to the inequality \eqref{eq:haussler}.} in~\cite{matouvsek1999geometric}.

\begin{theorem}[Haussler's Packing Lemma]\label{thm:haussler}
	Let $\Ff$ be a set system over a finite set $U$ with $|\Ff|\ge 2$ 
	such that the following properties hold for some reals $c,\rho\geq 1$ and numbers $k,d\in\N$:
	\begin{itemize}
		\item 
	$|\{F\cap X\colon F\in \Ff\}|\leq c\cdot |X|^\rho\qquad\textrm{for all }\emptyset\neq X\subseteq U,$
		\item  $\Ff$ has VC dimension at most $d$, and
		\item $|F\triangle F'|\ge k$ for all distinct $F,F'\in\cal F$.
	\end{itemize}		
	Then
	\begin{align}\label{eq:haussler}
	|\cal F|\le 2c\cdot \Bigl(\frac{4d\cdot |U|}k\Bigr)^\rho\le \Oh_{d,\rho,k}\Bigl(c\cdot |U|^\rho\Bigr).	
	\end{align}
	
\end{theorem}

It will be convenient to encapsulate Haussler's Packing Lemma in the following definition.

\begin{definition}
	The \emph{packing ratio} of a graph $G$ is the least integer $\beta$ with the following property:
	\begin{quote}
		For all sets $A,X\subseteq V(G)$ with $|X|\ge 2$ and every $k\in \N$ such that $|\sdiff[u,v]\cap A|\geq k$ for all distinct $u,v\in X$, we have $|X|\leq \beta\cdot \frac{|A|}{k}$.
	\end{quote}
\end{definition}

The observation is that Haussler's Packing Lemma together with a bound on the number of $1$-types implies a bound on the packing ratio.

\begin{lemma}\label{lem:haussler-application}
	Let $G$ be a graph, $\gamma$ be the growth ratio of $\Tt_1$ on $G$, and $d$ be the maximum of $1$ and the VC dimension of the set system $\{N[u]\colon u\in V(G)\}$. Then  the packing ratio $\beta$ of $G$ is at most
	\[\beta\le 10d\gamma\leq 20\gamma\log_2(2\gamma).\]
\end{lemma}
\begin{proof}
We first show $d\leq 2\log_2(2\gamma)$.
	This is immediate if $d=1$. Otherwise, let $B$ be a shattered set of size $d$. Then
	\[2^d=|\{N[u]\cap B\colon u\in V(G)\}|\leq |\Tt_1(G,B)|\leq \gamma d.\]
	Using the elementary inequality $d\leq 2^{d/2+1}$, we obtain $2^{d/2}\leq 2\gamma$, and hence $d\leq 2\log_2(2\gamma)$.

	Denote $\beta\coloneqq 10d\gamma$.
	Consider vertex subsets $A,X\subseteq V(G)$ with $|X|\ge 2$ and $k\in \N$ such that $|\sdiff[u,v]\cap A|\geq k$ for all distinct $u,v\in X$. Note that since $k$ is positive, for all distinct $u,v\in X\setminus A$ we have $N[u]\cap A\neq N[v]\cap A$, so
	\[|X|\leq |\Tt_1(G,A)|\leq \gamma\cdot |A|.\]
	In particular, the bound $|X|\leq \beta\cdot \frac{|A|}{k}$ holds whenever $k\leq 9$, since $d\geq 1$. So from now on we may assume that $k\geq 10$.
	
	Let
	\[\Ff\coloneqq \{N[u]\cap A\colon u\in X\}.\]
	Note that the assumption $|\sdiff[u,v]\cap A|\geq k\geq 10$ for distinct $u,v\in X$ implies that the sets $N[u]\cap A$ for $u\in X$ are pairwise different (so $|\Ff|=|X|$) and
	\begin{equation}\label{eq:osmiornica}
		|F\symdiff F'|\geq k-2,\qquad\textrm{for all distinct }F,F'\in \Ff.
	\end{equation}
	For every non-empty $B\subseteq A$, we have
	\[|\{F\cap B\colon F\in\Ff\}|\leq |\Tt_1(G,B)|\leq \gamma\cdot |B|,\]
	and the VC dimension of $\Ff$ is at most $d$. Thus we may apply \cref{thm:haussler} to $\Ff$ over $A$ with $c=\gamma$, $\rho=1$, and separation parameter $k-2$, using \eqref{eq:osmiornica}. This gives
	\[
		|X|=|\Ff|\leq 8d\gamma\cdot \frac{|A|}{k-2}
		\leq 10d\gamma\cdot \frac{|A|}{k}
		=\beta\cdot \frac{|A|}{k},
	\]
	where the second inequality follows from $k\geq 10$.
\end{proof}

By combining \cref{thm:mondep-1-types} with \cref{lem:haussler-application} we immediately get the following.

\begin{corollary}\label{cor:mondep-twinness}
	For every monadically dependent graph class $\Cc$ there is a function $\beta\colon \N\to \N$ with $\beta(n)\le \spoly_\Cc(n)$ such that every $n$-vertex graph $G\in \Cc$ has packing ratio at most $\beta(n)$.
\end{corollary}

\subsection{Orders with low alternation}

Further, we will rely on orderings of low \emph{alternation}, also known as orderings with low \emph{crossing number}.
The following definition is already tailored to the graph setting.

\begin{definition}
	Let $G$ be a graph and $\preceq$ be a vertex ordering of $G$. The \emph{alternation} of $\preceq$ is the least integer $\alpha$ with the following property: 
	\begin{quote}
		For every vertex $u$ of $G$, there are at most $\alpha$ pairs of vertices $w,w'$ that are consecutive in $\preceq$ and satisfy $u\in \sdiff[w,w']$.
	\end{quote}
\end{definition}

We will rely on the fundamental result of Welzl~\cite{Welzl89} about the existence of paths with low crossing number in set systems of bounded VC dimension. The precise formulation we use is due to Bonnet, Duron, Sylvester, and Zamaraev~\cite{BonnetD0Z25}, and is already adapted to the graph setting.

\begin{theorem}[follows from {\cite[Theorem~20]{BonnetD0Z25}}]\label{thm:alternation-general}
	Let $G$ be an $n$-vertex graph and $f\colon \R_{\geq 0}\to \R_{\geq 0}$ be a strictly increasing function such that $|\{N[u]\cap A\colon u\in V(G)\}|\leq f(|A|)$ for every non-empty $A\subseteq V(G)$. Further, let $d$ be the VC dimension of $\{N[u]\colon u\in V(G)\}$, treated as a set system over $V(G)$. Then there exists a vertex ordering of $G$ with alternation bounded by
	\[\alpha\coloneqq 2+2\log_2 n+10d\cdot \sum_{i=1}^{n}\frac{1}{f^{-1}(i/2)}.\]
	Moreover, there is an $\Oh(\alpha^2 n^3)$-time algorithm that given $G$, computes a vertex ordering with alternation bounded as above.
\end{theorem} 

We remark that the formulation in~\cite{BonnetD0Z25} does not specify the running time of the construction procedure. In \cref{app:welzl} we show that with minimal effort, the method described in \cite{BonnetD0Z25} can be implemented so that it runs in $\Oh(\alpha^2 n^3)$ time; this proves the algorithmic claim of \cref{thm:alternation-general}. We also remark that instead of alternation, Bonnet, Duron, Sylvester, and Zamaraev~\cite{BonnetD0Z25} consider a closely related parameter \emph{contiguity} and prove a bound of $\alpha/2-1$ on it; the bound on the alternation follows by the same reasoning, as explained in \cref{app:welzl}.

Note that the algorithm of \cref{thm:alternation-general} does not need to be provided access to the function $f$; it can be employed on any graph and it produces a vertex ordering satisfying the alternation bound for any function $f$ that satisfies the prerequisites.
Also, when $f(x)=\gamma\cdot x$ for some $\gamma>0$, we have
\[\sum_{i=1}^{n}\frac{1}{f^{-1}(i/2)}=2\gamma\cdot\sum_{i=1}^{n}\frac{1}{i}\leq 2\gamma(1+\log_2 n).\]
Then, the bound of \cref{thm:alternation-general} can be upper bounded by
\[2+20d\gamma+(20d\gamma+2)\cdot \log_2 n.\]
Together with \cref{thm:mondep-1-types,obs:mondep-vc}, we immediately conclude the following.

\begin{corollary}\label{cor:precise-welzl}
	Let $G$ be an $n$-vertex graph, $\gamma$ be the growth ratio of $\Tt_1$ on $G$, and $d$ be the VC dimension of the set system $\{N[u]\colon u\in V(G)\}$. Then $G$ admits a vertex ordering of alternation bounded by $2+20d\gamma+(20d\gamma+2)\cdot \log_2 n$, and such a vertex ordering can be computed in $\Oh(d^2\gamma^2\cdot n^3\log^2 n)$ time.
\end{corollary}

\begin{corollary}\label{cor:mondep-welzl}
	For every monadically dependent graph class $\Cc$ there is a function $\alpha\colon \N\to \N$ with $\alpha(n)\le \spoly_\Cc(n)$ such that every $n$-vertex graph $G\in \Cc$ admits a vertex ordering of alternation at most $\alpha(n)$. Further, such a vertex ordering can be computed in time $\Oht(n^3)$, given $G$ alone.
\end{corollary}

\section{Predamping Lemma}\label{sec:predamping}

In this section we take the first step towards the Damping Theorem (\cref{thm:static-s-flip}) by proving the Predamping Lemma (\cref{lem:predamping}) below. Intuitively, the goal of this step is to find a set $S$ that is already suitable, but whose relevant properties are expressed in the language of $r$-types and symmetric differences of neighborhoods, rather than in terms of the existence of a flip. To describe those properties, we need some~definitions.

Let $G$ be a graph and $A$ be a set of its vertices. We say that a vertex $u$ \emph{crosses} $A$ if $u\notin A$ and $u$ is not homogeneous to $A$.
For a partition $\Qq$ of $V(G)$ and two vertices $u,v$ of $G$, we define their \emph{Hamming distance} with respect to $\Qq$ as
\[\hamm_\Qq(u,v)\coloneqq \big|\{Q\in \Qq \colon \sdiff[u,v]\cap Q\neq \emptyset \}\big|.\]
Note that for any two vertices $u,v$,
we have that $\sdiff[u,v]=\emptyset$ 
if and only if $u=v$, by our definition. It follows that $\hamm_\Qq(u,v)=0$ if and only if $u=v$. 
Moreover, since $\sdiff[u,w]\subseteq \sdiff[u,v]\cup \sdiff[v,w]$ for any triple of vertices $u,v,w$, we have that $\hamm_\Qq(\cdot,\cdot)$ is symmetric and satisfies the triangle inequality; thus it defines a metric.
Observe also that if $\Rr$ is a refinement of $\Qq$, then we have
\[\hamm_{\Rr}(u,v)\geq \hamm_{\Qq}(u,v)\qquad \textrm{for all }u,v\in V(G).\]

With all these definitions in place, the goal of this section is to prove the Predamping Lemma formulated below. Note that compared to the simplified formulation from \cref{sec:overview}, we choose to express all the bounds in terms of three parameters: 
\begin{itemize}[nosep]
	\item $\gamma$: the growth ratio of the considered partition selector;
	\item $\beta$: the packing ratio; and
	\item $\alpha$: the alternation of a given vertex ordering.
\end{itemize}
The point of this is to make the formulation more robust with respect to various applications of the method; see the discussion in \cref{sec:conclusions}.

\begin{lemma}[Predamping Lemma]\label{lem:predamping}
	Let $\Pp$ be a monotone and respectful partition selector, $G$ be a graph, and $\alpha,\beta,\gamma\in \N$ be such~that
	\begin{itemize}
		\item $\Pp$ has growth ratio at most $\gamma$ on $G$;
		\item $G$ has packing ratio at most $\beta$; and
		\item $G$ admits a vertex ordering $\preceq$ of alternation at most $\alpha$.
	\end{itemize}
	Then for every $q\in \N$ and set $A\subseteq V(G)$ with $|A|\geq 6(q+1)$ there exists a set $S\subseteq V(G)$ such that
	\begin{enumerate}[label={(A\arabic*)}]
		\item\label{a:size} $|S|\leq |A|/q$;
		\item\label{a:reps} for every $u\in V(G)$ there exists $s_u\in S$ such that $\hamm_{\Pp(G,A\cup S)}(u,s_u)\leq 15(q+1)\cdot \beta\gamma$; and
		\item\label{a:diff} for any two vertices $u,v\in V(G)$ belonging to the same $1$-type over $S$, we have \[\hamm_{\Pp(G,A\cup S)}(u,v)\leq 44(q+1)\cdot \alpha\gamma.\]
	\end{enumerate}
	Moreover, assuming that $\Pp$ is computable in time $t(n)$, there is an $\Oh((\beta+\gamma)|A|\cdot (\beta\gamma n^2+t(n)))$-time algorithm computing such a set $S$ given $q$, $G$, $A$, and $\preceq$. The algorithm does not need to be provided the values of $\alpha,\beta,\gamma$.
\end{lemma}
\begin{proof}
	We describe an algorithmic procedure that constructs a suitable set $S$. Before we proceed with the description, let us argue that at a mild cost, we may actually assume that the procedure is provided the values of $\alpha,\beta,\gamma$. For $\alpha$ this is clear, since the alternation of the provided vertex ordering $\preceq$ can be easily computed in $\Oh(n^2)$ time. For $\beta$ and $\gamma$, the procedure we are going to present has the following feature: It can be run for any supposed values of $\beta$ and $\gamma$, and either it succeeds in finding a set $S$ satisfying the postulated properties for these supposed values, or it produces a witness that the growth ratio of $\Pp$ on $G$ is larger than the supposed value of $\gamma$, or a witness that the packing ratio of $G$ is larger than the supposed value of $\beta$. Therefore, we apply the procedure starting with $\beta=\gamma=1$ and each time a witness is found, we increment the corresponding parameter $\beta$ or $\gamma$ and run the procedure again. As a result, within at most $\beta+\gamma$ re-runs we eventually find a set $S$ that satisfies the postulated bounds for the supposed values of $\beta$ and $\gamma$ upper-bounded by their actual values. Consequently, from now on our goal is to design an $\Oh(|A|\cdot (\beta\gamma n^2+t(n)))$-time procedure that is given the values of $\alpha,\beta,\gamma$.

	With the assumption above made, we proceed to the description of the procedure. Throughout the proof we denote
	\[q'\coloneqq 3(q+1)\qquad \textrm{and}\qquad \zeta\coloneqq 2q'\gamma.\]
	We first focus on presenting the procedure itself and verifying the combinatorial properties of its output. We relegate the discussion of the algorithmic aspects to the end of the proof.
	
	The procedure works as follows.
	We start with $S\coloneqq \emptyset$ and iteratively add vertices to $S$ one by one according to rules that we will explain in a moment. During this process, we maintain the partition $\Rr\coloneqq \Pp(G,A\cup S)$, which, due to the monotonicity of $\Pp$, can only get finer with each vertex added to~$S$. Moreover, for every part $R\in \Rr$ we let $x_R$ be the $\preceq$-minimum vertex belonging to $R$, and call $x_R$ the \emph{representative} of $R$. We define the set of representatives
	\[X\coloneqq \{x_R\colon R\in \Rr\}.\]
	Note that when a vertex is added to $S$ during the procedure, every part $R$ of $\Rr$ can possibly break into several new parts. The new part that contains $x_R$ retains $x_R$ as its representative, while every other new part is assigned a new representative. Therefore, $X$ can only grow by addition of new vertices.
	
	We now describe the rules according to which vertices are added to $S$. As long as any of the following rules can be applied, apply the first one which is applicable.
	\begin{enumerate}[label={(R\arabic*)}]
		\item If there is a vertex $w$ that crosses more than $q' \gamma$ parts of $\Rr$, then add $w$ to $S$.\label{r:types}
		\item If there is a vertex $w$ such that $\hamm_\Rr(w,s)>5q'\beta\gamma$ for all $s\in S$, then add $w$ to $S$.\label{r:reps}
		\item If there is a vertex $w$ such that for every $s\in S$, there are at least $\zeta$ vertices of $X$ lying strictly between $w$ and $s$ in $\preceq$, then add $w$ to $S$.\label{r:symdiff}
	\end{enumerate}
	Observe that in any application of any of the rules above, we must have $w\notin S$. Indeed, to argue this for Rules \ref{r:reps} and \ref{r:symdiff} it suffices to consider $s=w$ supposing $w\in S$, while for Rule \ref{r:types}, we note that a vertex of $S$ crosses no parts of $\Rr=\Pp(G,A\cup S)$ due to $\Pp$ being respectful. Thus, $S$ grows by one vertex with every application of a rule, and clearly no rule is applicable when $S=V(G)$. Therefore, the procedure eventually reaches a state in which no rule is applicable, yielding the final set $S$, which we shall denote by~$S^\star$. Our claim is that $S^\star$ satisfies all the assertions from the lemma statement. Denote $\Rr^\star\coloneqq \Pp(G,A\cup S^\star)$ and $X^\star\coloneqq \{x_R\colon R\in \Rr^\star\}$.
	
	Assertion~\ref{a:reps} follows immediately from the inapplicability of Rule~\ref{r:reps}. Assertion~\ref{a:diff} is verified by the following claim and the observation that $(4\alpha+3)\zeta+2=(4\alpha+3)\cdot 6(q+1)\gamma+2\leq 44(q+1)\cdot \alpha\gamma.$
	
	\begin{claim}
		Suppose vertices $u$ and $v$ have the same $1$-type over $S^\star$. Then \[\hamm_{\Rr^\star}(u,v)\leq (4\alpha+3)\zeta+2.\]
	\end{claim}
	\begin{claimproof}
		We may assume that $u,v\notin S^\star$, for otherwise the assumption that $u$ and $v$ have the same $1$-type over $S^\star$ implies that $u=v$ and the claim is trivial.
		
		Suppose for contradiction that $\hamm_{\Rr^\star}(u,v)>(4\alpha+3)\zeta+2$, which means that there are more than $(4\alpha+3)\zeta+2$ parts of $\Rr^\star$ that intersect $\sdiff[u,v]$. At most two parts of $\Rr^\star$ contain $u$ or $v$, and, by the inapplicability of Rule \ref{r:types}, at most $q'\gamma+q'\gamma=\zeta$ parts of $\Rr^\star$ are crossed by $u$ or by~$v$. Therefore, there are more than $(4\alpha+2)\zeta=(2\alpha+1)\cdot 2\zeta$ parts $R\in \Rr^\star$ such that $R\cap \{u,v\}=\emptyset$, both $u$ and $v$ are homogeneous to $R$, and $R$ intersects $\sdiff[u,v]$; so in fact $R\subseteq \sdiff(u,v)$. Let $\Ff\subseteq \Rr^\star$ be the set of those~parts. 
		
		Since the alternation of $\preceq$ is at most $\alpha$, there are at most $2\alpha$ pairs of vertices $w,w'$ that are consecutive in $\preceq$ and satisfy $u\in \sdiff[w,w']$ or $v\in \sdiff[w,w']$. Therefore, the vertex set of $G$ can be partitioned into at most $2\alpha+1$ intervals in $\preceq$ such that for every interval $I$ of the partition, either $I$ consists of a single vertex $u$ or $v$, or both $u$ and $v$ are homogeneous to $I$. Since $|\Ff|>(2\alpha+1)\cdot 2\zeta$, by the pigeonhole principle there exists an interval $I$ of the partition such that
		\begin{equation}\label{eq:dziobak}
			|I\cap \{x_R\colon R\in \Ff\}|>2\zeta.
		\end{equation}
		Note that $|I|>2$ due to $\zeta\geq 1$, so both $u$ and $v$ are homogeneous to $I$. However, since $x_R\in \sdiff[u,v]$ for each $R\in \Ff$, it must be that $u$ is complete to $I$ and $v$ is anticomplete to $I$ or vice versa; so in particular $I\subseteq \sdiff[u,v]$. 
		Since $u,v\notin S^\star$ and $u,v$ have the same $1$-type over $S^\star$, no vertex of~$\sdiff[u,v]$ belongs to~$S^\star$. We conclude~that 
		\begin{equation}\label{eq:pizmak}
			I\cap S^\star=\emptyset.
		\end{equation}
		
		From \eqref{eq:dziobak} and \eqref{eq:pizmak} it follows that there exists a vertex $w\in I$ such that for every $s\in S^\star$, there are at least $\zeta$ vertices of 
		 $\{x_R\colon R\in \Ff\}\subseteq X^\star$ lying strictly between $w$ and $s$ in $\preceq$. Indeed, one can take for instance $w$ to be a median vertex of $I\cap \{x_R\colon R\in \Ff\}$ in $\preceq$. This means that Rule~\ref{r:symdiff} is applicable to $w$, a contradiction.
	\end{claimproof}
	
	We are left with verifying Assertion \ref{a:size}, that is, bounding the size of $S^\star$. Note that $|S^\star|$ is equal to the total number of times Rules \ref{r:types}, \ref{r:reps}, and \ref{r:symdiff} are applied during the procedure. We bound each of these quantities separately. We use the following observation: as $\Pp$ has growth ratio at most $\gamma$ on~$G$, we have
	\begin{equation}\label{eq:slimak}
		|\Rr^\star|\leq \gamma\cdot |A\cup S^\star|.
	\end{equation}
	
	\begin{claim}\label{cl:r:types}
		Rule \ref{r:types} is applied at most $|A\cup S^\star|/q'$ times.
	\end{claim}
	\begin{claimproof}
		Observe that since $\Pp$ is respectful and monotone, every time Rule \ref{r:types} is applied on some vertex $w$, every part of $\Rr$ that is crossed by $w$ breaks into two or more new parts. Consequently, the number of parts of $\Rr$ increases by at least $q'\gamma$. So if Rule \ref{r:types} is applied $\ell$ times in total during the procedure, then we have
		\[|\Rr^\star|\geq q'\gamma\ell.\]
		Combining this with \eqref{eq:slimak} yields $\ell\leq |A\cup S^\star|/q'$.
	\end{claimproof}

	\begin{claim}\label{cl:r:reps}
		Rule \ref{r:reps} is applied at most $|A\cup S^\star|/q'$ times.
	\end{claim}
	\begin{claimproof}
		Let $S_2\subseteq S^\star$ be the set of vertices added to $S$ by Rule \ref{r:reps}. 
		
		We first prove that for every two distinct vertices $u,v\in S_2$, we have
		\begin{equation}\label{eq:skorupiak}
		|\sdiff[u,v]\cap X^\star|>q'\beta\gamma.
		\end{equation}
		Indeed, assuming without loss of generality that $v$ was added to $S$ after $u$, let us consider the time moment just before $v$ was added. At this point we already have $u\in S$, so since Rule~\ref{r:reps} adds $v$ to $S$, we must have 
		\[\hamm_\Rr(u,v)>5q'\beta\gamma,\]
		where $\Rr$ denotes the partition $\Pp(G,A\cup S)$ at the considered time moment. 
		This means that there are more than $5q'\cdot \beta\gamma$ parts of $\Rr$ that intersect $\sdiff[u,v]$. Among these, at most two parts contain $u$ or $v$, and at most $q'\gamma+q'\gamma=2q'\gamma$ parts may be crossed by $u$ or $v$, because otherwise the lower-indexed Rule~\ref{r:types} would be triggered instead of \ref{r:reps}. This leaves us with a set $\Ff$ consisting of more than \[5q'\beta\gamma-2-2q'\gamma\geq q'\beta\gamma\] parts of $\Rr$ that do not contain $u$ or $v$ and intersect $\sdiff(u,v)$ while being homogeneous both to $u$ and to $v$; so they are, in fact, contained in $\sdiff(u,v)$. 
		In particular, $\{x_R\colon R\in \Ff\}\subseteq \sdiff(u,v)\cap X\subseteq \sdiff(u,v)\cap X^\star$.
		We conclude that
		\[|\sdiff[u,v]\cap X^\star|\geq |\Ff|>q'\beta\gamma;\]
		this establishes \eqref{eq:skorupiak}.
		
		Noting that $X^\star$ is non-empty, the assumption that the packing ratio of $G$ is at most $\beta$ implies that
		\[|S_2|\leq \beta\cdot \frac{|X^\star|}{q'\beta\gamma}=\frac{|X^\star|}{q'\gamma}.\]
		By combining this with \eqref{eq:slimak}, we get
		\[|S_2|
		\leq 
		\frac{|X^\star|}{q'\gamma}
		=
		\frac{|\Rr^\star|}{q'\gamma}
		\leq
		\frac{\gamma\cdot |A\cup S^\star|}{q'\gamma}=|A\cup S^\star|/q'.\qedhere\] 
	\end{claimproof}
	
	\begin{claim}\label{cl:r:symdiff}
		Rule \ref{r:symdiff} is applied at most $|A\cup S^\star|/q'$ times.
	\end{claim}
	\begin{claimproof}
		Let $S_3\subseteq S^\star$ be the set of vertices added to $S$ by Rule \ref{r:symdiff}. Observe that for any two distinct vertices $u,v\in S_3$, there are at least $\zeta$ vertices of $X^\star$ lying strictly between $u$ and $v$ in $\preceq$. Indeed, assuming without loss of generality that $v$ was added later than $u$, at the time of adding $v$ there were at least $\zeta$ vertices of $X$ lying strictly between $u$ and $v$, and $X\subseteq X^\star$. Therefore, if we consider the vertices of $S_3$ ordered by $\preceq$, then between any two consecutive such vertices there are at least $\zeta$ vertices of $X^\star$. It follows that
		\[|S_3|\leq 1+|X^\star|/\zeta.\]
		By combining this with \eqref{eq:slimak} we conclude that 
		\[|S_3|
		\leq 
		1+|X^\star|/\zeta
		=
		1+|\Rr^\star|/\zeta
		\leq
		1+\gamma/\zeta\cdot |A\cup S^\star|=1+|A\cup S^\star|/(2q')\leq  |A\cup S^\star|/q',\]
		where the last inequality follows from the assumption $|A|\geq 6(q+1)=2q'$.
	\end{claimproof}
	
	By \cref{cl:r:types,cl:r:reps,cl:r:symdiff}, the total number of applications of any of the rules is at most $3|A\cup S^\star|/q'=|A\cup S^\star|/(q+1)$. Hence
	\[|S^\star|\leq \frac{|A\cup S^\star|}{q+1}\leq \frac{|A|}{q+1}+\frac{|S^\star|}{q+1},\qquad \textrm{implying}\qquad |S^\star|\leq |A|/q.\]
	This verifies Assertion \ref{a:size} and finishes the verification of the requested properties of $S$.
	
	\paragraph{Runtime.}
	We are left with showing that the procedure can be implemented so that it runs in $\Oh(|A|\cdot (\beta\gamma n^2+t(n)))$ time. As argued, the number of times\footnote{Precisely, if this bound were ever to be exceeded, then the procedure must have detected a witness that the supposed value of $\beta$ or $\gamma$ is too small. This results in restarting the procedure with the appropriate value incremented.
	Note that this requires detecting which of the two bounds $\beta$ or $\gamma$ was violated.
	If \ref{r:types} or \ref{r:symdiff} were executed too often, we know that $\gamma$ needs to be increased.
	If \ref{r:reps} was executed too often, then we increase $\beta$ when $X^\star$ has the desired size, and $\gamma$ otherwise.
	} the procedure applies any of Rules \ref{r:types}, \ref{r:reps}, \ref{r:symdiff} is $|S^\star|\leq |A|/q$. 
	Hence, it suffices to argue that finding an applicable rule and applying it can be done in $\Oh(q\beta\gamma n^2+t(n))$ time, because then the total running time will be bounded by
	\[\Oh(|A|/q\cdot (q\beta\gamma n^2+t(n)))\leq \Oh(|A|\cdot (\beta\gamma n^2+t(n))),\]
	as promised. 
	Applying a rule always amounts to adding a vertex $w$ to $S$, recomputing the partition $\Rr=\Pp(G,A\cup S)$, and augmenting the set of representatives $X$, which takes time $t(n)+\Oh(n^2)$. So we only need to show how to find an applicable rule in time $\Oh(q\beta\gamma n^2)$. For this, we consider the rules one by one.
	
	As for Rule \ref{r:types}, for any given vertex $w$, we may easily count the number of parts of $\Rr$ crossed by $w$ in time $\Oh(n)$ by counting the number of neighbors of $w$ in every part of $\Rr$. Hence, finding a vertex $w$ to which Rule \ref{r:types} is applicable can be done in time $\Oh(n^2)$.
	
	As for Rule \ref{r:symdiff}, by scanning the ordering $\preceq$ twice, in opposite directions, in time $\Oh(n)$ we may compute the following two counts for every vertex $w$: the number of elements of $X$ lying between $w$ and the $\preceq$-smallest $s\in S$ satisfying $w\preceq s$ (or $+\infty$ if such $s$ does not exist), and the number of elements of $X$ lying between $w$ and the $\preceq$-largest $s\in S$ satisfying $s\preceq w$ (or $+\infty$ if such $s$ does not exist). Finding $w$ to which Rule \ref{r:symdiff} can be applied boils down to finding $w$ for which both those counts are at least $\zeta$. So verifying the applicability of this rule can be done in $\Oh(n)$ time.
	
	Finally, a straightforward implementation of Rule \ref{r:reps} --- for every pair of vertices $(u,s)\in V(G)\times S$ compute $\hamm_\Rr(u,s)$ in time $\Oh(n)$ --- runs in total time $\Oh(n^3)$. We now show how this complexity can be reduced to $\Oh(q\beta\gamma n^2)$.
	
	First, we will need the following data structure result.
	
	\begin{claim}[{\cite[Theorem~16]{BonnetDPST26}}]\label{cl:lce-trick}
		Given $k\in \N$ and a list $\Aa$ consisting of $n$ binary strings of length $n$ each, one can in $\Oh(n^2)$ time construct a data structure that allows answering the following query in $\Oh(k)$ time: given two strings $a,a'\in \Aa$, decide whether $\hamm(a,a')\leq k$. Here, $\hamm(a,a')$ is the Hamming distance between strings $a$ and $a'$: the number of positions on which $a$ and $a'$ differ.
	\end{claim} 
	
	We remark that in \cite{BonnetDPST26}, this result is formulated in graph-theoretic terms, but the formulation there easily implies the above.
	
	Second, we also need the following data structure observation.
	
	\begin{claim}\label{cl:same-nei}
		We may in time $\Oh(n^2)$ construct a data structure that allows answering the following query in $\Oh(1)$ time: for given $u,v\in V(G)$ and part $R\in \Rr$, decide whether $N(u)\cap R=N(v)\cap R$.
	\end{claim}
	\begin{claimproof}
		For every vertex $u\in V(G)$ and part $R\in \Rr$, compute a binary vector $x_{u,R}$ of length $|R|$ representing the neighborhood of $u$ in $R$. For every part $R\in \Rr$, sort all the vectors $\{x_{u,R}\colon u\in V(G)\}$ using radix sort in time $\Oh(n|R|)$, partition the obtained list into groups of consecutive duplicates (also in time $\Oh(n|R|)$), and for every $u$ record the group on the list of $R$ to which $x_{u,R}$ belongs. Now, deciding whether $N(u)\cap R=N(v)\cap R$ boils down to verifying whether $x_{u,R}$ and $x_{v,R}$ belong to the same group on the list of $R$.
	\end{claimproof}

	\newcommand{\Zz}{\mathcal{Z}}
	
	Armed with \cref{cl:lce-trick,cl:same-nei}, we proceed as follows. Similarly to the proof of \cref{lem:compute-Tr}, in total time $\Oh(n^2)$ we may compute, for every $u\in V(G)$ and part $R\in \Rr$, the number of neighbors of $u$ in $R$. This allows us to deduce whether $u$ is anticomplete, complete, or non-homogeneous to $R$. Note that for every vertex $u$ there are at most $1+q'\gamma$ parts $R\in \Rr$ to which $u$ is non-homogeneous: the one part to which $u$ belongs, and the at most $q'\gamma$ parts that $u$ crosses. (Otherwise, the lower-indexed Rule~\ref{r:types} would be applicable.) For each vertex $u$ we compute the set $\Zz_u\subseteq \Rr$ consisting of those at most $1+q'\gamma$ parts. Also, we compute a Boolean vector $a_u$ of length $|\Rr|$ where the coordinate corresponding to a part $R\in \Rr$ is equal to $1$ if and only if $u\notin R$ and $u$ is complete to $R$. 

	Consider now any two vertices $u,v\in V(G)$; we design a procedure to efficiently decide whether $\hamm_\Rr(u,v)\leq \mu\coloneqq 5q'\beta\gamma$. This can be done as follows:
	\begin{itemize}
		\item If $u=v$, then $\hamm_\Rr(u,v)=0$, so assume $u\neq v$ below.
		\item Inspect every part $R\in \Zz_u\cup \Zz_v$ and determine whether $R$ intersects $\sdiff[u,v]$ by checking whether $u\in R$, or $v\in R$, or $N(u)\cap R\neq N(v)\cap R$. The last check can be done in constant time provided we set up the data structure of \cref{cl:same-nei} beforehand. Since $|\Zz_u\cup \Zz_v|\leq \Oh(q'\gamma)$, this can be done in time $\Oh(q'\gamma)$. We let $c_{u,v}$ be the number of parts $R\in \Zz_u\cup \Zz_v$ that intersect $\sdiff[u,v]$.
		\item Similarly to the above, by inspecting the parts of $\Zz_u\cup \Zz_v$ one by one we can, in time $\Oh(q'\gamma)$,  compute the number of parts $R\in \Zz_u\cup \Zz_v$ such that exactly one of $u$ and $v$ is complete to $R$. Call this number~$d_{u,v}$.
		\item We now observe that
		\begin{equation}\label{eq:aksolotl}
			\hamm_{\Rr}(u,v)=|\{R\in \Rr \colon \sdiff[u,v]\cap R\neq \emptyset\}|=\hamm(a_u,a_v)-d_{u,v}+c_{u,v}
		\end{equation}
		where again, $\hamm(a_u,a_v)$ is the Hamming distance between strings $a_u$ and $a_v$. 
		Indeed, the quantity $\hamm(a_u,a_v)$ counts the number of parts $R\in \Rr$ such that exactly one of $u$ and $v$ is complete to $R$. For parts $R\notin \Zz_u\cup \Zz_v$, both $u$ and $v$ are homogeneous to $R$, so exactly one of $u,v$ being complete to $R$ is equivalent to $R$ having a non-empty intersection with $\sdiff[u,v]$. For parts $R\in \Zz_u\cup \Zz_v$, subtracting $d_{u,v}$ and adding $c_{u,v}$ rectifies the contribution of those parts so that it corresponds to the number of parts in $\Zz_u\cup \Zz_v$ that intersect $\sdiff[u,v]$.
		\item By \eqref{eq:aksolotl}, determining whether $\hamm_\Rr(u,v)\leq \mu$ reduces to determining whether $\hamm(a_u,a_v)\leq k$, where $k\coloneqq \mu+d_{u,v}-c_{u,v}.$
		Noting that $k\leq \mu+2(1+q'\gamma)\leq 2\mu$, this query can be answered in time $\Oh(\mu)$ provided we set up the data structures of \cref{cl:lce-trick} for $\Aa\coloneqq \{a_u\colon u\in V(G)\}$ and all $k\in \{1,\ldots,2\mu\}$ beforehand. (Formally, the strings in $\Aa$ are of length $|\Rr|\leq n$, but we can pad them to length $n$ with zeroes.)
	\end{itemize}
	The above procedure determines whether $\hamm_\Rr(u,v)\leq \mu$ in time $\Oh(\mu)$, so applying it for every $u\in V(G)$ and $v\in S$ allows verifying the applicability of Rule~\ref{r:reps} in time $\Oh(\mu n^2)=\Oh(q\beta\gamma n^2)$. This is conditioned on the data structures of  \cref{cl:lce-trick,cl:same-nei} being initialized beforehand (the former for every $k\in \{1,\ldots,2\mu\}$), which can be again done in time $\Oh(\mu n^2)=\Oh(q\beta\gamma n^2)$.
\end{proof}
\section{Finding the flip}\label{sec:flip}

In this section we make the second step towards the proof of the Damping Theorem (\cref{thm:static-s-flip}). \cref{lem:predamping} already provides a suitable set $S$; now the goal is to find a suitable $S$-flip. However, instead of talking directly about $S$-flips, we find it convenient to express the goal of the current step using the following terminology.

Let $G$ be a graph and $\Qq,\Rr$ be partitions of the vertex set of $G$ such that $\Rr$ is a refinement of $\Qq$. (Later, it will typically be the case that $\Qq=\Tt_1(G,S)$ and $\Rr=\Pp(G,A\cup S)$ for some $A,S\subseteq V(G)$ and a monotone, respectful partition selector $\Pp$.) For a binary relation $\rho\subseteq \Qq\times \Qq$ and a part $Q\in \Qq$, we~denote
\[N_\rho(Q)\coloneqq \bigcup \{Q'\in \Qq\mid (Q,Q')\in \rho\}.\]
Note that in this definition we do \emph{not} require $\rho$ to be symmetric. The intuition is that $N_\rho(Q)$ is the \emph{outneighborhood} of $Q$ with respect to $\rho$. We define the \emph{mismatch} of $\rho$ with respect to $\Rr$ as the smallest integer $m$ satisfying the following:
\begin{quote}
	For every part $Q$ of $\Qq$ and vertex $v\in Q$, the set $N(v)\symdiff N_\rho(Q)$ intersects at most $m$ parts of~$\Rr$.
\end{quote}

\noindent We first observe that \cref{lem:predamping} yields the following statement expressed in the terminology of mismatch.

\begin{lemma}\label{lem:mismatch-asymmetric}
	Let $\Pp$ be a monotone and respectful partition selector, $G$ be a graph, and $\alpha,\beta,\gamma\in \N$ be such~that
	\begin{itemize}
		\item $\Pp$ has growth ratio at most $\gamma$ on $G$;
		\item $G$ has packing ratio at most $\beta$; and
		\item $G$ admits a vertex ordering $\preceq$ of alternation at most $\alpha$.
	\end{itemize}
	Then for every $q\in \N$ and a non-empty set $A\subseteq V(G)$ there exists a set $S\subseteq V(G)$ with $|S|\leq |A|/q$ and a relation $\rho\subseteq \Tt_1(G,S)\times \Tt_1(G,S)$ such that the mismatch of $\rho$ with respect to $\Pp(G,A\cup S)$ is at most $88q\cdot (\alpha+\beta)\gamma$.

	Moreover, assuming that $\Pp$ is computable in time $t(n)$, there is an $\Oh((\beta+\gamma)|A|\cdot (\beta\gamma n^2+t(n)))$-time algorithm computing  such a set $S$ and relation $\rho$, given $q$, $G$, $A$, and $\preceq$.
\end{lemma}
\begin{proof}
	We start by considering the corner case when $|A|<6(q+1)$. Then we set $S\coloneqq \emptyset$ and $\rho\coloneqq \emptyset$. Note that since the growth ratio of $\Pp$ on $G$ is at most $\gamma$, we have $|\Pp(G,A\cup S)|=|\Pp(G,A)|\leq \gamma|A|<6(q+1)\gamma$. As $6(q+1)\gamma<44(q+1)\cdot (\alpha+\beta)\gamma$, the postulated bound on the mismatch of $\rho$ with respect to $\Pp(G,A\cup S)$ holds trivially. Hence, from now on we may assume that $|A|\geq 6(q+1)$.
	
	Let $S$ be the set obtained by applying \cref{lem:predamping} to $A$. For every $1$-type $Q\in \Tt_1(G,S)$, we fix an arbitrary vertex $x_Q\in Q$. By Assertion \ref{a:reps} of \cref{lem:predamping}, we may find a vertex $s_Q\in S$ such that
	\begin{equation}\label{eq:mieczak}
		\hamm_\Rr(x_Q,s_Q)\leq 15(q+1)\cdot \beta\gamma,
	\end{equation}
	where $\Rr\coloneqq \Pp(G,A\cup S)$. Note that computing $\Rr$ takes time $t(n)$, while computing the vertices $s_Q$ can be done in total time $\Oh(q\beta\gamma n^2)\leq \Oh(|A|\cdot \beta\gamma n^2)$ using the method for verifying the applicability of Rule~\ref{r:reps} presented at the end of the proof of \cref{lem:predamping}.

	We define a relation $\rho\subseteq \Tt_1(G,S)\times \Tt_1(G,S)$ as follows: For any $Q,Q'\in \Tt_1(G,S)$, set
	\[(Q,Q')\in \rho\qquad\textrm{if and only if}\qquad s_Q\textrm{ is complete to }Q'.\]
	Since verifying whether $s_Q$ is complete to $Q'$ can be done in time $\Oh(|Q'|)$, the relation $\rho$ can be computed in total time $\Oh(|\Tt_1(G,S)|\cdot \sum_{Q'\in \Tt_1(G,S)} |Q'|)\leq \Oh(n^2)$.
	Also note that since $s_Q\in S$ and $Q'$ is a $1$-type over~$S$, we have that $Q'=\{s_Q\}$ or $s_Q$ is homogeneous to~$Q'$. Therefore, $N(s_Q)$ consists of the union of those $1$-types $Q'\in \Tt_1(G,S)$ to which $s_Q$ is complete. By combining this observation with the definition of $\rho$ we conclude that
	\begin{equation}\label{eq:pierwotniak}
		N_\rho(Q)=N(s_Q).
	\end{equation}

	We claim that the relation $\rho$ defined in this way satisfies the postulated bound on the mismatch with respect to $\Rr$.
	For this, consider any vertex $v$ and let $Q\in \Tt_1(G,S)$ be the $1$-type over $S$ to which $v$ belongs. By assertion \ref{a:diff} of \cref{lem:predamping}, we have
	\[\hamm_\Rr(v,x_Q)\leq 44(q+1)\cdot \alpha\gamma.\]
	By \eqref{eq:mieczak} and the triangle inequality, we infer that
	\[\hamm_\Rr(v,s_Q)\leq \hamm_\Rr(v,x_Q)+\hamm_\Rr(x_Q,s_Q)\leq 44(q+1)\cdot \alpha\gamma+15(q+1)\cdot \beta\gamma\leq 44(q+1)\cdot (\alpha+\beta)\gamma.\]
	This means that the set $\sdiff[v,s_Q]\supseteq N(v)\symdiff N(s_Q)$ intersects at most $44(q+1)\cdot (\alpha+\beta)\gamma\leq 88q\cdot (\alpha+\beta)\gamma$ parts of $\Rr$.
	As $N(s_Q)=N_\rho(Q)$ by \eqref{eq:pierwotniak}, this proves the desired bound on the mismatch of $\rho$.
\end{proof}

Recall that to define an $S$-flip of the considered graph $G$, we have to provide a \emph{symmetric} binary relation $\sigma\subseteq \Tt_1(G,S)\times \Tt_1(G,S)$. The relation $\rho$ provided by \cref{lem:mismatch-asymmetric} is not necessarily symmetric, and hence it does not immediately define an $S$-flip. Fortunately, it turns out that in the context of bounding the mismatch, any relation can be adjusted to a symmetric one at the cost of a minor loss in the mismatch. We prove this in the following lemma.

\begin{lemma}[Symmetrization Lemma]\label{lem:symmetrization}
	Let $G$ be a graph and $\Qq,\Rr$ be partitions of the vertex set of $G$ such that $\Rr$ is a refinement of $\Qq$.
	Let $\rho\subseteq \Qq\times \Qq$ be a relation such that the mismatch of $\rho$ with respect to $\Rr$ is at most $m$, for some $m\in \N$.
	Then there is also a symmetric relation $\sigma\subseteq \Qq\times \Qq$ such that the mismatch of $\sigma$ with respect to $\Rr$ is at most $3m$.

	Moreover, given $G$, $\Qq$, $\Rr$, and $\rho$, such a relation $\sigma$ can be computed in time $\Oh(n^2)$.
\end{lemma}
\begin{proof}
	Call a pair of parts $Q,Q'\in \Qq$ \emph{compatible} if
	\[(Q,Q')\in \rho\quad\textrm{if and only if}\quad (Q',Q)\in \rho,\]
	and \emph{incompatible} otherwise. Note that if $Q,Q'$ are incompatible, then $Q\neq Q'$.

	For every part $R\in \Rr$ we choose an arbitrary vertex $x_R\in R$ and call it the \emph{representative} of $R$.
	For a part $Q\in \Qq$, by $\Rr[Q]$ we denote the set of all parts $R\in \Rr$ such that $R\subseteq Q$; thus $\{\Rr[Q]\colon Q\in \Qq\}$ is a partition of $\Rr$. For parts $Q,Q'\in \Qq$, $R\in \Rr[Q]$, and $R'\in \Rr[Q']$, say that $(R,R')$ \emph{agrees} with $(Q,Q')$ if
	\[x_R\textrm{ and }x_{R'}\textrm{ are adjacent in }G\qquad\textrm{if and only if}\qquad (Q,Q')\in \rho,\]
	and $(R,R')$ \emph{disagrees} with $(Q,Q')$ otherwise.

	We note the following two observations about this definition.

	\begin{claim}\label{cl:noncmp-symdiff}
		Suppose $Q,Q'\in \Qq$, $R\in \Rr[Q]$, and $R'\in \Rr[Q']$ are such that $R\neq R'$ and $(R,R')$ disagrees with $(Q,Q')$. Then
		\[R'\cap (N(x_R)\symdiff N_\rho(Q))\neq \emptyset.\]
	\end{claim}
	\begin{claimproof}
		Since $(R,R')$ disagrees with $(Q,Q')$, exactly one of the assertions ``$x_R$ and $x_{R'}$ are adjacent in~$G$'' and ``$(Q,Q')\in \rho$'' holds true. The first assertion is equivalent to $x_{R'}\in N(x_R)$ and the second assertion is equivalent to $Q'\subseteq N_\rho(Q)$. As $x_{R'}\in R'\subseteq Q'$, we thus have \hbox{$x_{R'}\in R'\cap (N(x_R)\symdiff N_\rho(Q))$.}
	\end{claimproof}

	\begin{claim}\label{cl:incompatible-xor}
		Suppose $Q,Q'\in \Qq$ are incompatible. Then for every pair of parts $R\in \Rr[Q]$ and $R'\in \Rr[Q']$, exactly one of the following assertions holds:
		\begin{itemize}
			\item $(R,R')$ agrees with $(Q,Q')$;
			\item $(R',R)$ agrees with $(Q',Q)$.
		\end{itemize}
	\end{claim}
	\begin{claimproof}
		As $Q,Q'$ are incompatible, exactly one of the assertions $(Q,Q')\in \rho$ and $(Q',Q)\in \rho$ holds. So exactly one of them has the same truth value as the assertion ``$x_R$ and $x_{R'}$ are adjacent in~$G$''.
	\end{claimproof}

	We now construct the desired symmetric relation $\sigma\subseteq \Qq\times \Qq$. Consider any pair of parts $Q,Q'\in \Qq$.
	\begin{itemize}
		\item If $Q,Q'$ are compatible, then add $(Q,Q')$ and $(Q',Q)$ to $\sigma$ if and only if $(Q,Q')\in \rho$ (or equivalently, $(Q',Q)\in \rho$).
		\item If $Q,Q'$ are incompatible, say $(Q,Q')\in \rho$ and $(Q',Q)\notin \rho$, then consider the two \hbox{numbers:}
		\begin{itemize}
			\item $w_{Q,Q'}$: the number of pairs $(R,R')\in \Rr[Q]\times\Rr[Q']$ such that $(R,R')$ agrees with~$(Q,Q')$;
			\item $w_{Q',Q}$: the number of pairs $(R,R')\in \Rr[Q]\times \Rr[Q']$ such that $(R',R)$ agrees with~$(Q',Q)$.
		\end{itemize}
		Note that by \cref{cl:incompatible-xor}, we have $w_{Q,Q'}+w_{Q',Q}=|\Rr[Q]|\cdot |\Rr[Q']|$.
		Then add $(Q,Q')$ and $(Q',Q)$ to $\sigma$ if $w_{Q,Q'}>w_{Q',Q}$, and do not add them if $w_{Q,Q'}<w_{Q',Q}$. If $w_{Q,Q'}=w_{Q',Q}$, break the tie arbitrarily.\\[0.1cm]
		If $(Q,Q')\notin \rho$ and $(Q',Q)\in \rho$, proceed symmetrically with the roles of $Q$ and~$Q'$~swapped.
	\end{itemize}
	Note that for any given $Q,Q'\in \Qq$, the numbers $w_{Q,Q'},w_{Q',Q}$ can be computed in time $\Oh(|\Rr[Q]|\cdot |\Rr[Q']|)$ by inspecting all the pairs $(R,R')\in \Rr[Q]\times \Rr[Q']$ one by one. Hence, the relation $\sigma$ can be computed in time $\Oh(\sum_{(Q,Q')\in \Qq^2} |\Rr[Q]|\cdot |\Rr[Q']|)=\Oh(|\Rr|^2)\leq \Oh(n^2)$, as promised.
	
	From the construction we immediately derive the following claim, which roughly says the following: pairs of parts of $\Qq$ on which $\rho$ and $\sigma$ differ are heavily loaded with disagreeing pairs of parts of $\Rr$.

	\begin{claim}\label{cl:char}
		Suppose $(Q,Q')\in \Qq\times \Qq$ is such that $(Q,Q')\in \rho\symdiff \sigma$. Then $Q\neq Q'$ and there are at least $|\Rr[Q]|\cdot |\Rr[Q']|/2$ pairs $(R,R')\in \Rr[Q]\times \Rr[Q']$ such that $(R,R')$ disagrees with~$(Q,Q')$.
	\end{claim}
	\begin{claimproof}
		Since $(Q,Q')\in \rho\symdiff \sigma$, we must have that $Q,Q'$ are incompatible, hence in particular $Q\neq Q'$. Moreover, we must have $w_{Q',Q}\geq w_{Q,Q'}$, which due to $w_{Q,Q'}+w_{Q',Q}=|\Rr[Q]|\cdot |\Rr[Q']|$ entails $w_{Q,Q'}\leq |\Rr[Q]|\cdot |\Rr[Q']|/2$. This means that at most half of the pairs $(R,R')\in \Rr[Q]\times \Rr[Q']$ agree with $(Q,Q')$, which leaves at least half of them disagreeing.
	\end{claimproof}

	By combining \cref{cl:noncmp-symdiff,cl:char} we may now establish an upper bound on the total number of parts of $\Rr$ contained in $N_{\rho\symdiff \sigma}(Q)$, for any part $Q$ of $\Qq$.

	\begin{claim}\label{cl:error-parts}
		For every $Q\in \Qq$ we have
		\[|\{R'\in \Rr\mid R'\subseteq N_{\rho\symdiff\sigma}(Q)\}|\leq 2m.\]
	\end{claim}
	\begin{claimproof}
		Fix any $Q\in\Qq$.
		Denote \[\Nn\coloneqq \{Q'\in \Qq\mid Q'\subseteq N_{\rho\symdiff\sigma}(Q)\}\qquad \textrm{and} \qquad \Ss\coloneqq \{R'\in \Rr\mid R'\subseteq N_{\rho\symdiff\sigma}(Q)\}=\bigcup_{Q'\in \Nn} \Rr[Q'].\]
		For every $R\in \Rr[Q]$, define
		\[\Ww_{R}\coloneqq \{R'\in \Rr\setminus \{R\}\mid (R,R')\textrm{ disagrees with }(Q,Q')\textrm{, where }Q'\in \Qq\textrm{ is such that }R'\in \Rr[Q']\}.\]
		By \cref{cl:noncmp-symdiff}, every $R'\in \Ww_R$ is intersected by $N(x_R)\symdiff N_\rho(Q)$. Since the mismatch of $\rho$ with respect to $\Rr$ is at most $m$, we conclude that
		\begin{equation}\label{eq:wilkolak}
			|\Ww_R|\leq m\qquad\textrm{for every }R\in \Rr[Q].
		\end{equation}

		Writing $(R,R')\disagrees (Q,Q')$ as a shorthand for $(R,R')$ \emph{disagrees with} $(Q,Q')$, we have:
		\begin{align*}
			|\Ss| & = \sum_{Q'\in \Nn} |\Rr[Q']| &\\
			& \leq \sum_{Q'\in \Nn} \frac{2}{|\Rr[Q]|}\cdot  |\{(R,R')\in \Rr[Q]\times \Rr[Q']\mid (R,R')\disagrees (Q,Q')\}|  &\textrm{(by \cref{cl:char})}\\
			& = \sum_{Q'\in \Qq\setminus \{Q\}} \frac{2}{|\Rr[Q]|}\cdot  |\{(R,R')\in \Rr[Q]\times \Rr[Q']\mid (R,R')\disagrees (Q,Q')\}|  &\textrm{(as $Q\notin \Nn$)}
		\end{align*}
		which by double counting (see \Cref{fig:double-counting}) is equal to:
		\begin{align*}
			\phantom{|\Ss|}& = \frac{2}{|\Rr[Q]|}\cdot\sum_{R\in \Rr[Q]} |\{R'\in \Rr\mid R'\in \Rr[Q']\textrm{ for some }Q'\neq Q\textrm{ and }(R,R')\disagrees (Q,Q')\}| &\\
			& \leq \frac{2}{|\Rr[Q]|}\cdot\sum_{R\in \Rr[Q]} |\{R'\in \Rr\setminus \{R\}\mid (R,R')\disagrees (Q,Q')\textrm{ where }Q'\in \Qq\textrm{ is s.t. }R'\in \Rr[Q']\}| &\\
			& = \frac{2}{|\Rr[Q]|}\cdot \sum_{R\in \Rr[Q]} |\Ww_R| \leq \frac{2}{|\Rr[Q]|}\cdot |\Rr[Q]|\cdot m=2m. &\rlap{\hspace{-2.7cm}\textrm{(by \eqref{eq:wilkolak})}}
		\end{align*}
	\end{claimproof}

	\begin{figure}[htb]
		\centering
		\includegraphics[scale=1]{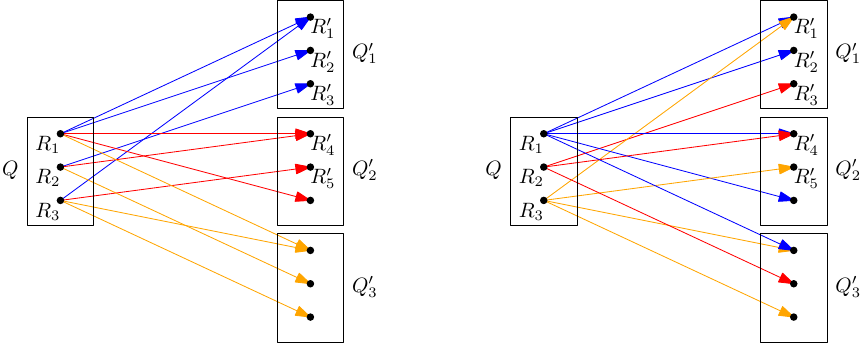}
		\caption{An illustration of the double counting argument in \Cref{cl:error-parts}. Depicted is a directed graph whose vertices are the parts of $\Rr$, where there is a directed edge between vertices $R$ and $R'$ if $(R,R')$ disagrees with the containing parts $(Q,Q')$ of $\Qq$.
		In \Cref{cl:error-parts}, the number of outgoing edges of $Q$ is counted in two ways:\\
		1) Summing over the incoming edges of each part $Q'$ (depicted left; colors group edges by parts $Q'$).\\
		2) Summing over the outgoing edges of each $R \in \Rr[Q]$ (depicted right; colors group edges by vertices $R$).}
		\label{fig:double-counting}
	\end{figure}

	The last claim links the mismatch of $\sigma$ with respect to $\Rr$ and the symmetric difference of $\rho$ and $\sigma$.

	\begin{claim}\label{cl:sigma-rho}
		Let $v,w$ be distinct vertices of $G$ and let $Q,Q'\in \Qq$, $R\in \Rr[Q]$, and $R'\in \Rr[Q']$ be such that $v\in R\subseteq Q$ and $w\in R'\subseteq Q'$. Suppose further that $w\in N(v)\symdiff N_\sigma(Q)$. Then at least one of the following conditions holds:
		\begin{itemize}
			\item $w\in N(v)\symdiff N_\rho(Q)$; or
			\item $(Q,Q')\in \rho\symdiff \sigma$.
		\end{itemize}
	\end{claim}
	\begin{claimproof}
		Suppose that the first condition fails:
		$w\notin N(v)\symdiff N_\rho(Q)$.
		As $w\in N(v)\symdiff N_\sigma(Q)$, it follows that $w\in N_\rho(Q)\symdiff N_\sigma(Q)$. Since $w\in Q'$, this is equivalent to $Q' \subseteq N_\rho(Q)\symdiff N_\sigma(Q)$, which is again equivalent to $(Q,Q')\in \rho\symdiff \sigma$.
	\end{claimproof}

	We may now bound the mismatch of $\sigma$ with respect to $\Rr$.
	Consider any vertex $v$ of $G$ and let $Q\in \Qq$ be such that $v\in Q$. By \cref{cl:sigma-rho}, every part of $\Rr$ that is intersected by $N(v)\symdiff N_\sigma(Q)$ is also intersected by $N(v)\symdiff N_\rho(Q)$ or is contained in $N_{\rho\symdiff \sigma}(Q)$. There are at most $m$ parts of $\Rr$ intersected by $N(v)\symdiff N_\rho(Q)$, due to the assumed bound on the mismatch of $\rho$, and at most $2m$ parts of $\Rr$ contained in $N_{\rho\symdiff \sigma}(Q)$, by \cref{cl:error-parts}. This gives us at most $3m$ parts of $\Rr$ that are intersected by $N(v)\symdiff N_\sigma(Q)$.
\end{proof}

By combining \cref{lem:mismatch-asymmetric,lem:symmetrization} we immediately get the following symmetric variant of \cref{lem:mismatch-asymmetric}.

\begin{corollary}\label{cor:mismatch-symmetric}
	Let $\Pp$ be a monotone and respectful partition selector, $G$ be a graph, and $\alpha,\beta,\gamma\in \N$ be such~that
	\begin{itemize}
		\item $\Pp$ has growth ratio at most $\gamma$ on $G$;
		\item $G$ has packing ratio at most $\beta$; and
		\item $G$ admits a vertex ordering $\preceq$ of alternation at most $\alpha$.
	\end{itemize}
	Then for every $q\in \N$ and non-empty set $A\subseteq V(G)$ there exists a set $S\subseteq V(G)$ with $|S|\leq |A|/q$ and a symmetric relation $\sigma\subseteq \Tt_1(G,S)\times \Tt_1(G,S)$ such that the mismatch of $\sigma$ with respect to $\Pp(G,A\cup S)$ is at most $264q\cdot (\alpha+\beta)\gamma$.

	Moreover, assuming that $\Pp$ is computable in time $t(n)$, there is an $\Oh((\beta+\gamma)|A|\cdot (\beta\gamma n^2+t(n)))$-time algorithm computing such a set $S$ and relation $\sigma$, given $q$, $G$, $A$, and $\preceq$.
\end{corollary}

\section{Proof of the Damping Theorem}\label{sec:damping}

Finally, we gather all the tools and complete the proof of the Damping Theorem (\cref{thm:static-s-flip}). The strategy is as expected: We take the set $S$ provided by \cref{lem:predamping} and the $S$-flip defined by the symmetric relation given by \cref{cor:mismatch-symmetric}, and verify that these two objects satisfy all the required properties. The only technical check missing is the translation of the mismatch condition asserted in \cref{cor:mismatch-symmetric} to the language of distance-$r$ reachability in the $S$-flip, used in the statement of \cref{thm:static-s-flip}. We execute this translation in the following lemma.

For a graph $G$, partition $\Qq$ of its vertex set, and  
symmetric binary relation $\sigma\subset \Qq\times \Qq$, let $G\oplus \sigma$ 
be the $\Qq$-flip of $G$
such that for all $X,Y\in \Qq$ and distinct vertices $x\in X,y\in Y$ we have that:
\begin{itemize}
	\item $xy\in E(G)\Leftrightarrow xy\in E(G\oplus\sigma)$\quad if $(X,Y)\notin\sigma$,
	\item $xy\in E(G)\Leftrightarrow xy\notin E(G\oplus\sigma)$\quad if $(X,Y)\in\sigma$.
\end{itemize}

\begin{lemma}\label{lem:mismatch-ball}
	Let $r\in \N$, $G$ be a graph, and $B$ be a subset of vertices of $G$. Denote $\Pp\coloneqq \Tt_{r+1}(G,B)$ and $\Qq\coloneqq \Tt_1(G,B)$.
	Further, let $\sigma\subseteq \Qq\times \Qq$ be a symmetric binary relation and $m$ be the mismatch of $\sigma$ with respect to $\Pp$. Then for every vertex $u$ of $G$, there are at most $(m+1)^r$ parts of $\Qq$ that contain a vertex that is at distance at most $r$ from $u$ in the graph $G\oplus \sigma$.
\end{lemma}
\begin{proof}
	Denote $H\coloneqq G\oplus \sigma$ and $\Pp_i\coloneqq \Tt_i(G,B)$,
	for $i\in [r+1]$. Thus 
	\[\Qq=\Pp_1\pgeq \Pp_2\pgeq \ldots \pgeq \Pp_r\pgeq \Pp_{r+1}=\Pp.\]
	The core observation is captured by the following statement.
	
	\begin{claim}\label{cl:telescope}
		For each $i\in [r]$ and part $P\in \Pp_{i+1}$, there are at most $m+1$ parts $R\in \Pp_i$ that intersect $N_H[P]$. 
	\end{claim}
	\begin{claimproof}
		Let $P'\in \Pp_i$ be the unique part of $\Pp_i$ that contains $P$. Consider any part $R\in \Pp_i$ that intersects $N_H[P]$ and is different from $P'$. Thus $P'\cap R=\emptyset$, for $R$ and $P'$ are different parts of $\Pp_i$. 
		
		Since $R$ is an $i$-type over $B$ and $P$ is an $(i+1)$-type over $B$, by the definition of $(i+1)$-types we have that one of the following three cases holds:
		\begin{itemize}
			\item all vertices of $P$ are complete to $R$;
			\item all vertices of $P$ are anticomplete to $R$; or
			\item all vertices of $P$ are non-homogeneous to $R$.
		\end{itemize}
		In the first two cases, $P$ and $R$ are homogeneous to each other in $G$, and as each of them is contained in a single part of $\Qq=\Pp_1$, they are also homogeneous to each other in $H=G\oplus \sigma$. Since $R\cap N_H[P]\neq \emptyset$, we must in fact have that $P$ and $R$ are complete to each other in $H$. In the third case, since each of $P$ and $R$ is contained in a single part of $\Qq=\Pp_1$, that a vertex $x\in P$ is non-homogeneous to $R$ in $G$ implies that $x$ is also non-homogeneous to $R$ in $H$.
		
		The reasoning above shows that in each of the three cases, for every vertex $x\in P$ there exists some vertex $x'\in R$ such that $x$ and $x'$ are adjacent in $H$. This implies that some part of $\Pp$ contained in~$R$ --- precisely, the part that contains $x'$ --- is intersected by $N(x)\symdiff N_\sigma(Q)$, where $Q$ is the unique part of $\Qq$ that contains $P$. The bound on the mismatch of $\sigma$ with respect to $\Pp$ implies that there can be at most $m$ such parts of $\Pp$ in total, for any $x\in P$. Hence, there are no more than $m$ parts $R\in \Pp_i$ satisfying $R\cap N_H[P]\neq \emptyset$ and $R\neq P'$. Taking into account also the single part $P'$ finishes the proof of the~claim. 
	\end{claimproof}
	
	With \cref{cl:telescope} established, a straightforward induction on $j$ proves the following statement:
	\begin{quote}
		For every $u\in V(G)$ and $j\in \{0,1,\ldots,r\}$, there are at most $(m+1)^j$ parts of $\Pp_{r+1-j}$ that intersect $\Ball_H^j[u]$.
	\end{quote}
	Applying this statement for $j=r$ completes the proof.
\end{proof}

We may now complete the proof of the Damping Theorem. We prove the following formulation that makes the dependence on $\alpha,\beta,\gamma$ explicit.

\begin{theorem}[Damping Theorem with explicit bounds]\label{thm:static-s-flip-explicit}
	Let $r\in \N$, $G$ be a graph, and $\alpha,\beta,\gamma\in \N$ be such~that
	\begin{itemize}
		\item $\Tt_{r+1}$ has growth ratio at most $\gamma$ on $G$;
		\item $G$ has packing ratio at most $\beta$; and
		\item $G$ admits a vertex ordering $\preceq$ of alternation at most $\alpha$.
	\end{itemize}
	Then for every $q\in \N$ and a non-empty vertex subset $A\subseteq V(G)$, there exist a set $S\subseteq V(G)$ and an $S$-flip $H$ of $G$ satisfying the following properties:
	\begin{itemize}
		\item $|S|\leq |A|/q$; and
		\item for every vertex $u\in V(G)$, $\Ball_H^r[u]$ intersects at most $(265q\cdot (\alpha+\beta)\gamma)^r$ parts of $\Tt_1(G,A\cup S)$.
	\end{itemize}
	Moreover, there is an algorithm computing such a set $S$ and 
	an $S$-flip $H$,
	 given $r$, $q$, $G$, $A$, and~$\preceq$,   in time $\Oh((\beta+\gamma)(\beta\gamma+r)\cdot |A|n^2)$.
\end{theorem}
\begin{proof}
	Apply \cref{cor:mismatch-symmetric} to $G$, $A$, $\preceq$, and the partition selector $\Tt_{r+1}$. Since $\Tt_{r+1}$ is computable in time $\Oh(rn^2)$ by \cref{lem:compute-Tr}, in this way we construct, within the promised time complexity, a set $S\subseteq V(G)$ and a symmetric relation $\sigma\subseteq \Tt_1(G,S)\times \Tt_1(G,S)$ such that $|S|\leq |A|/q$ and the mismatch of $\sigma$ with respect to $\Tt_{r+1}(G,A\cup S)$ is bounded by $m\coloneqq 264q\cdot (\alpha+\beta)\gamma$.
	
	Denote $B\coloneqq A\cup S$ and define a symmetric relation $\sigma'\subseteq \Tt_1(G,B)\times \Tt_1(G,B)$  as in the proof of \cref{obs:flips-monotone}: include a pair $(P,P')\in \Tt_1(G,B)\times \Tt_1(G,B)$ in $\sigma'$ if and only if $(Q,Q')\in \sigma$, where $Q$ and $Q'$ are the unique parts of $\Tt_1(G,S)$ that contain $P$ and $P'$, respectively. Note that for every vertex~$v$ we have $N_\sigma(Q)=N_{\sigma'}(P)$, where $Q$ and $P$ are the parts of $\Tt_1(G,S)$ and $\Tt_1(G,B)$ that contain $v$, respectively. So the mismatch of $\sigma'$ with respect to $\Tt_{r+1}(G,B)$ is exactly the same as that of $\sigma$; that is,~$m$.
	
	Let $H\coloneqq G\oplus \sigma=G\oplus \sigma'$. Thus $H$ is an $S$-flip of $G$. By \cref{lem:mismatch-ball} we conclude that for every vertex $u$ of $G$, there are at most $(m+1)^r\leq (265q\cdot (\alpha+\beta)\gamma)^r$ parts of $\Tt_1(G,A\cup S)$ that intersect $\Ball^r_H[u]$. This means that $S$ and $H$ satisfy all the required properties.
\end{proof}

\sct*
\begin{proof}
	By \cref{thm:mondep-t-complexity,cor:mondep-twinness,cor:mondep-welzl}, there are subpolynomial functions $\alpha(n),\beta(n),\gamma(n)\le \spoly_{\Cc,r}(n)$ such that
	\begin{itemize}
		\item $\Tt_{r+1}$ has growth ratio at most $\gamma(n)$ on $G$;
		\item $G$ has packing ratio at most $\beta(n)$; and
		\item $G$ admits a vertex ordering $\preceq$ of alternation at most $\alpha(n)$.
	\end{itemize}
	Moreover, a vertex ordering $\preceq$ as above can be computed deterministically in $\Oht_\Cc(n^3)$ time. With this, it suffices to apply \cref{thm:static-s-flip-explicit} and note that $(265q\cdot (\alpha(n)+\beta(n))\cdot\gamma(n))^r\leq q^r\cdot h(n)$, for some $h(n)\le \spoly_{\Cc,r}(n)$.
\end{proof}
\section{Conclusions}\label{sec:conclusions}

\cref{thm:main} provides an algorithm to compute a merge sequence of subpolynomial radius-$r$ width for any given graph $G$ drawn from a monadically dependent graph class $\Cc$. One may ask what can be guaranteed about the output of the algorithm provided we assume something more about $\Cc$. For instance, if $\Cc$ itself has bounded merge-width, what is the width of the merge sequence output by the algorithm of \cref{thm:main}? It will be convenient to ask this question in the greater generality of graph classes that are \emph{translinear}.

\begin{definition}
	A graph class $\Cc$ is \emph{translinear} if every graph class $\Dd$ transducible from $\Cc$ has \emph{linear neighborhood complexity}: There is a constant $\gamma_\Dd$ such that $\Tt_1$ has growth ratio at most $\gamma_\Dd$ on every $H\in \Dd$.
\end{definition}

Dreier and Toru\'nczyk~\cite{DreierT25} proved that whenever a graph class $\Dd$ is transducible from a graph class $\Cc$ of bounded merge-width, $\Dd$ again has bounded merge-width. Also, Bonamy and Geniet~\cite{BonamyG26} proved that classes of bounded merge-width have linear neighborhood complexity. Thus, the combination of these two results proves that classes of bounded merge-width are translinear. It would be consistent with our knowledge if the two concepts (bounded merge-width and translinearity) actually coincided. This is conjectured by Bonamy and Geniet \cite[Conjecture 1.7]{BonamyG26} and is implied by the conjecture of Dreier and Toruńczyk \cite[Conjecture 1.17, full version]{DreierT25}.

Consider any translinear class of graphs $\Cc$. In \Cref{sec:types-without-copying} we present an alternative proof of \Cref{thm:mondep-r-types} which allows us to deduce that 
assuming translinearity of $\Cc$, for every $r\in \N$ there is a constant $\gamma_r\le \Oh_{\Cc,r}(1)$ such that $\Tt_r$ has growth ratio at most $\gamma_r$ on any $G\in \Cc$
 (see \Cref{cor:translinear-growth}).
  Also, following the bounds of \cref{lem:haussler-application,cor:precise-welzl}, we conclude that any graph $G\in \Cc$ has packing ratio bounded by a constant $\beta\le \Oh_{\Cc}(1)$ and admits a vertex ordering $\preceq$ of alternation $\alpha\le \Oh_{\Cc}(\log n)$, and this bound cannot be improved in general\footnote{The class of cographs has bounded merge-width, so is translinear, but contains $n$-vertex graphs for which the smallest admissible alternation $\alpha$ of a vertex ordering is $\Omega(\log n)$  \cite{lax-214022}.}.
  Note that such an ordering $\preceq$ can be computed deterministically in $\Oh_\Cc(n^3\log^2 n)$ time by \cref{cor:precise-welzl}, and this can be done only once at the beginning of the whole argument. By plugging these bounds into the explicit formulation of the Damping Theorem (\cref{thm:static-s-flip-explicit}, with $q=2$ for simplicity), we obtain an $\Oh_{\Cc,r}(n^3)$-time algorithm to compute, for a given $n$-vertex graph $G\in \Cc$ (provided with a suitable $\preceq$) and a non-empty $A\subseteq V(G)$, a set $S\subseteq V(G)$ and an $S$-flip $H$ of $G$ such that
\begin{itemize}
	\item $|S|\leq |A|/2$; and
	\item for every $u\in V(G)$, $\Ball_H^r[u]$ intersects $\Oh_{\Cc,r}(\log^r n)$ parts of $\Tt_1(G,A\cup S)$.
\end{itemize}
With these bounds in place, the argument from the proof of \cref{thm:main-ordering} produces, in time $\Oh_{\Cc,r}(n^3\log n)$, a partition sequence of radius-$r$ width bounded by $k\le \Oh_{\Cc,r}(\log^{r+1} n)$. The additional $\log n$ factor is incurred by the length of the sequence $m\leq 1+\log_2 n$. As in the proof of \cref{thm:main}, if we apply the whole reasoning above for radius $r'\coloneqq 6r+1$ instead, then we can use \cref{thm:partition-mw} to turn the obtained partition sequence into a merge sequence of radius-$r$ width bounded by $k\cdot \pi_G(k)\le \gamma_1\cdot k^2\le \Oh_{\Cc,r}(\log^{12r+4} n)$.

Thus, the above discussion proves the following.

\begin{theorem}\label{thm:translinear-polylogarithmic}
	Let $r\in \N$ and $\Cc$ be a translinear class of graphs. Then there exists an $\Oh_{\Cc,r}(n^3\log^2 n)$-time algorithm that given an $n$-vertex graph $G\in \Cc$, constructs a merge sequence for $G$ of radius-$r$ width bounded by $\Oh_{\Cc,r}(\log^{12r+4} n)$.
\end{theorem}

Note that \cref{thm:translinear-polylogarithmic} can be used to compute merge sequences of polylogarithmic width in graph classes of bounded merge-width. It would be interesting to investigate whether this polylogarithmic approximation factor can be improved.

\section{AI methodology}
We obtained a first proof of the main result without AI assistance. Our original proof used an equivalent formulation of the Predamping Lemma, proved using an alternative algorithmic construction of online $\varepsilon$-nets, and a different proof of the Symmetrization Lemma. We subsequently used Claude Opus 5.5, Claude Fable 5.1, and GPT-6 to seek alternative proofs of these lemmas, providing a list of mathematical tools to consider. The suggestions included the present formulation of the Predamping Lemma (\cref{lem:predamping}), the replacement of $\eps$-nets by Welzl's vertex orderings of low alternation in its proof, and the present proof of the Symmetrization Lemma (\cref{lem:symmetrization}). AI also suggested the amortization argument presented in \cref{app:partition-mw}. We adapted these suggestions and wrote all technical content ourselves. We also used AI to improve the presentation and for proofreading.
The authors verified the correctness and originality of all content, including references.

\bibliographystyle{plain}
\bibliography{references}

\clearpage
\appendix
\section{Algorithmic aspects of \texorpdfstring{\cref{thm:partition-mw}}{Theorem \ref{thm:partition-mw}}}\label{app:partition-mw}

In this appendix we provide a brief discussion of the algorithmic claim of \cref{thm:partition-mw}, showing how to implement the method presented in \cite{DrabikDGNPT26} in $\Oh(n^3)$ time. We assume the reader's familiarity with the proof of \cite[Theorem~5.1]{DrabikDGNPT26}.

The key element of the proof is the refinement procedure described in \cite[Lemma 5.2]{DrabikDGNPT26}. A naive implementation of this refinement procedure works in $\Oh(n^2)$ time provided we have access to the following information about the input partition $\Pp$:
\begin{itemize}
	\item For every part $X\in \Pp$, the diameter of $G[X]$ and of its complement $\overline{G}[X]$.
	\item For every pair of distinct parts $X,X'\in \Pp$, the diameter of $G[X,X']$ and of its bipartite complement $\overline{G}[X,X']$.
\end{itemize}
For graphs $G[X,X']$ and $\overline{G}[X,X']$, we only need to maintain the maximum distance between a vertex $x\in X$ and a vertex $x'\in X'$. (We call these \emph{cross distances}.) This is because at the end, we are only interested in whether the diameter is at most $6$, which is equivalent to the maximum cross distance being at most $5$.

We now show that these values can be maintained efficiently along the construction of the partition sequence in the proof of \cite[Theorem~5.1]{DrabikDGNPT26} so that the total work associated with maintaining them is bounded by $\Oh(n^3)$. In fact, we maintain all the relevant distances:
\begin{itemize}
	\item For every part $X\in \Pp$, we maintain $\dist_{G[X]}(x,x')$ and $\dist_{\overline{G}[X]}(x,x')$ for all $x,x'\in X$.
	\item For every pair of distinct parts $X,X'\in \Pp$, we maintain $\dist_{G[X,X']}(x,x')$ and $\dist_{\overline{G}[X,X']}(x,x')$ for all $x\in X$ and $x'\in X'$.
\end{itemize}
The construction of the merge sequence in the proof of \cite[Theorem~5.1]{DrabikDGNPT26} can be expanded into a sequence of binary merges, where each binary merge replaces two distinct parts of the current partition with their union. Therefore, it suffices to show how to efficiently update the distances specified above during a single binary merge. We focus on distances in $G[X]$ and $G[X,X']$; the distances in $\overline{G}[X]$ and $\overline{G}[X,X']$ can be updated similarly by working in the complement $\overline{G}$.

Consider a single merge of two parts $A$ and $B$ into a new part $A\cup B$. Let $a\coloneqq |A|$ and $b\coloneqq |B|$, and by swapping $A$ and $B$ if necessary, assume that $a\leq b$. To compute the all-pairs distances in $G[A\cup B]$, we start with $G[B]$ (for which the all-pairs distances are already computed) and iteratively insert the vertices of $A$ into this graph. With each vertex $u$ inserted, we first compute the distances from $u$ to all the already present vertices (in time $\Oh((a+b)^2)\leq \Oh(b^2)$), and then update the distances between the already present vertices by checking if there is a shorter path going through $u$ (this takes $\Oh(1)$ time per pair of already present vertices, so again $\Oh((a+b)^2)\leq \Oh(b^2)$ time in total). By this method, the all-pairs distances in $G[A\cup B]$ can be computed in time $\Oh(ab^2)$. Next, for every other part $C\notin \{A,B\}$, we compute the cross distances in $G[A\cup B,C]$ by starting with the graph $G[B,C]$ (for which the cross distances are already computed) and iteratively add the vertices of $A$ one by one. As before, updating the cross distances upon insertion of a new vertex $u\in A$ takes time $\Oh(bc)$ where $c\coloneqq |C|$, so we can compute the cross distances in $G[A\cup B,C]$ in total time $\Oh(abc)$. Summing over all parts $C$, the total time complexity of updating the maintained distances during the merge of $A$ and $B$ is bounded by $\Oh(ab^2)+\Oh(abn)\leq \Oh(abn)$.

It now remains to observe that summing the values $ab$ over all the binary merges in a merge sequence yields exactly $\binom{n}{2}$, because each pair of distinct vertices falls into the same part exactly once during the sequence. Therefore, the total work associated with updating the distances can be bounded by $\Oh(n^3)$.

\section{Counting types of formulas}\label{sec:formula-types}
In this appendix, we prove \Cref{lem:mondep-formula-types}, which is used in the proof of \Cref{thm:mondep-r-types}. In \Cref{sec:types-without-copying} we also provide an alternative proof of \Cref{thm:mondep-r-types}, which also applies to translinear classes, discussed in the conclusions in \Cref{sec:conclusions}.

For a positive integer $k$, let $C_k(G)$ be the graph formed from $k$ disjoint copies of $G$ by joining the copies of each vertex into a clique. Thus its vertex set is $V(G)\times[k]$, and the additional \emph{matching edges} are $(u,i)(u,j)$ for $u\in V(G)$ and distinct $i,j\in[k]$. For every fixed $k$, if $\Cc$ is monadically dependent, then so is $\{C_k(G)\colon G\in\Cc\}$~\cite[Corollary~19.10, full version]{DreierMT24}.

\begin{proof}[Proof of \cref{lem:mondep-formula-types}]
	If $\Gamma=\emptyset$, then $\Tt_\Gamma(G^+,B)$ has at most one part, so we may take $\gamma_\Gamma(n)=1$. Hence assume that $\Gamma=\{\varphi_1,\ldots,\varphi_m\}$ with $m\geq 1$.
	Given an expansion $G^+$ of $G\in\Cc$ by the colors occurring in $\Gamma$, take $m+1$ copies $V_0,V_1,\ldots,V_m$ of $V(G)$, writing $u_i$ for the copy of $u$ in $V_i$. Construct a graph $H$ by putting, for $i\in[m]$ and $u,v\in V(G)$,
	\[u_0v_i\in E(H)\qquad\textrm{if and only if}\qquad \varphi_i(u,v)\text{ holds in $G^+$},\]
	and no other edges. On $C_{m+1}(G)$, distinguish the copies by colors and encode the colors of $G^+$ on $V_0$. The adjacency of $H$ is then defined by a fixed first-order formula $\psi_\Gamma$: restrict the quantifiers of each $\varphi_i$ to $V_0$ and identify the copy-zero representative of $v_i$ as its unique neighbor in $V_0$. Let $\Dd$ be the class of all graphs $H$ obtained this way. Then $\Dd$ is transducible from $\{C_{m+1}(G)\colon G\in\Cc\}$, so it is monadically dependent and depends only on $\Cc$ and $\Gamma$.
	
	Let $B_i\coloneqq\{b_i\colon b\in B\}$ and $R\coloneqq\bigcup_{i=1}^m B_i$, so $|R|=m|B|$. For $u,v\in V(G)$, the neighborhoods of $u_0$ and $v_0$ in $R$ are equal if and only if $u$ and $v$ have the same $\Gamma$-type over $B$ in $G^+$. Since $V_0\cap R=\emptyset$, it follows from \cref{thm:mondep-1-types} that
	\[|\Tt_\Gamma(G^+,B)|\leq |\Tt_1(H,R)|\leq m|B|\cdot\gamma_\Dd(m|B|),\]
	where $\gamma_\Dd$ is a subpolynomial function depending only on $\Dd$. We may thus set $\gamma_\Gamma(n)\coloneqq m\gamma_\Dd(mn)$, which is subpolynomial, as required.
\end{proof}

\subsection{A reduction without copying}\label{sec:types-without-copying}

For the application to translinear classes in \cref{sec:conclusions}, we provide an alternative proof of \Cref{thm:mondep-r-types}, based directly on transducibility (without copying). 
This reduction allows us to deduce bounds on the growth ratios of $\Tt_r$ in the case of translinear classes.

For $s\geq 1$, let $\gamma_s^*(G)$ denote the growth ratio of $\Tt_s$ on a non-empty graph $G$, and set $\gamma_0^*(G)\coloneqq 1$.

\begin{lemma}\label{lem:types-without-copying}
For every $r\in\N$, there is a first-order formula $\varphi_r^*(x,y)$ in the signature of $5$-colored graphs, with unary predicates $A$, $R_0$, $R^+$, $R^-$, and $R^\in$, with the following property. For every graph $G$ and non-empty set $A\subseteq V(G)$, there exist a $5$-coloring $G^+$ of $G$, interpreting the predicate $A$ as the given set, and a non-empty set $R\subseteq V(G)$ such that, for $H\coloneqq\varphi_r^*(G^+)$, the partition $\Tt_1(H,R)$ refines $\Tt_r(G,A)$ and
\[
    |\Tt_r(G,A)|\leq |\Tt_1(H,R)|
    \qquad\text{and}\qquad
    |R|\leq 3\gamma_{r-1}^*(G)|A|.
\]
For $r\geq 2$, one can moreover ensure $|R|\leq 3|\Tt_{r-1}(G,A)|$.
\end{lemma}
\begin{proof}
For $r=1$, take $\varphi_1^*(x,y)\coloneqq\adj(x,y)$ and $R\coloneqq A$, so that $H=G$ and the conclusion is immediate. Hence assume $r\geq 2$, fix $G$ and a non-empty set $A\subseteq V(G)$, and write $\Qq\coloneqq\Tt_{r-1}(G,A)$.

Let $\Qq_{\leq 2}$ and $\Qq_{\geq 3}$ be the sets of parts of $\Qq$ of size at most two and at least three, respectively, and set $R_0\coloneqq\bigcup\Qq_{\leq 2}$. For each $Q\in\Qq_{\geq 3}$, choose three distinct vertices $r_Q^+,r_Q^-,r_Q^\in\in Q$. For $t\in\{+,-,\in\}$, let $R^t\coloneqq\{r_Q^t\colon Q\in\Qq_{\geq 3}\}$, and define
\[
    R\coloneqq R_0\cup R^+\cup R^-\cup R^\in.
\]
These four sets are pairwise disjoint. Every part of $\Qq$ contributes at least one and at most three vertices to~$R$, so $0<|R|\leq 3|\Qq|$.

Construct a bipartite graph $H$ on $V(G)$ with sides $R$ and $V(G)\setminus R$ as follows. For $u\notin R$ and $v\in R$, put $uv\in E(H)$ according to the following rules:
\begin{itemize}
    \item If $v\in R_0$, then $uv\in E(H)$ if and only if $uv\in E(G)$.
    \item If $v=r_Q^+\in R^+$, then $uv\in E(H)$ if and only if $u$ is complete to $Q\setminus\{u\}$ in $G$.
    \item If $v=r_Q^-\in R^-$, then $uv\in E(H)$ if and only if $u$ is anticomplete to $Q\setminus\{u\}$ in $G$.
    \item If $v=r_Q^\in\in R^\in$, then $uv\in E(H)$ if and only if $u\in Q$.
\end{itemize}
There are no other edges in $H$.

To express this construction by a first-order formula, expand $G$ by five colors distinguishing $A$, $R_0$, $R^+$, $R^-$, and $R^\in$, and let $\tau_{r-1}$ be the formula from \cref{cl:types-definable}. Membership in $R$ abbreviates membership in the union of the four corresponding colors. Define
\begin{align*}
    \varphi_r^*(x,y)\coloneqq\, & (x\notin R)\wedge(y\in R)\wedge \\
    & (y\in R_0\rightarrow\adj(x,y))\wedge \\
    & \bigl(y\in R^+\rightarrow\forall z\,((z\neq x\wedge\tau_{r-1}(y,z))\rightarrow\adj(x,z))\bigr)\wedge \\
    & \bigl(y\in R^-\rightarrow\forall z\,((z\neq x\wedge\tau_{r-1}(y,z))\rightarrow\neg\adj(x,z))\bigr)\wedge \\
    & (y\in R^\in\rightarrow\tau_{r-1}(x,y)).
\end{align*}
Writing $G^+$ for this expansion, we have $H=\varphi_r^*(G^+)$. Note that the formula $\varphi_r^*$ depends only on $r$.

We next verify that $\Tt_1(H,R)$ refines $\Tt_r(G,A)$. Every vertex in $R$ is a singleton part of $\Tt_1(H,R)$, so it suffices to consider $u,u'\notin R$ with $N_H(u)\cap R=N_H(u')\cap R$.
\begin{enumerate}
    \item Each of $u,u'$ belongs to a part of $\Qq_{\geq 3}$, since all vertices of smaller parts lie in $R_0$. For every $Q\in\Qq_{\geq 3}$, adjacency to $r_Q^\in$ records membership in $Q$. Thus $u,u'$ belong to the same part of $\Qq$, that is, have the same $(r-1)$-type over $A$.
    \item If $Q\in\Qq_{\leq 2}$, then $Q\subseteq R_0$, and the construction gives
    \[
        N_G(u)\cap Q=N_H(u)\cap Q=N_H(u')\cap Q=N_G(u')\cap Q.
    \]
    Since $u,u'\notin Q$, this implies that $u$ is complete to $Q\setminus\{u\}$ if and only if $u'$ is complete to $Q\setminus\{u'\}$, and likewise for anticompleteness.
    \item If $Q\in\Qq_{\geq 3}$, equality of adjacency to $r_Q^+$ shows that $u$ is complete to $Q\setminus\{u\}$ if and only if $u'$ is complete to $Q\setminus\{u'\}$. Equality of adjacency to $r_Q^-$ gives the analogous equivalence for anticompleteness.
\end{enumerate}
These are exactly the three conditions defining equality of $r$-types, proving the refinement and hence the claimed inequality between the numbers of parts. Finally,
\[
    |R|\leq 3|\Tt_{r-1}(G,A)|\leq 3\gamma_{r-1}^*(G)|A|,
\]
which completes the proof.
\end{proof}

We now make the resulting bounds explicit. Let $\CC$ be a graph class and for each $r\geq 2$, let $\Dd_r$ be the class of all graphs defined by $\varphi_r^*$ from arbitrary $5$-colored expansions of graphs in $\Cc$. This class is transducible from $\Cc$ and contains every graph $H$ supplied by \cref{lem:types-without-copying}. If $\Cc$ is monadically dependent, then so is $\Dd_r$. By \cref{thm:mondep-1-types}, there is a non-decreasing subpolynomial function $\xi_r$, depending only on $\Cc$ and $r$, such that
\[
    |\Tt_1(H,B)|\leq |B|\cdot\xi_r(|B|)
    \qquad\text{for every }H\in\Dd_r\text{ and non-empty }B\subseteq V(H).
\]
Let $\gamma_1$ be a non-decreasing function supplied by \cref{thm:mondep-1-types} for $\Cc$, and define inductively, for $r\geq 2$,
\[
    \gamma_r(n)\coloneqq 3\gamma_{r-1}(n)\cdot\xi_r\bigl(3\gamma_{r-1}(n)n\bigr).
\]
Writing $a\coloneqq|A|$, \cref{lem:types-without-copying} and the induction hypothesis give
\begin{align*}
    |\Tt_r(G,A)|
    &\leq |\Tt_1(H,R)|
    \leq |R|\cdot\xi_r(|R|)\\
    &\leq 3\gamma_{r-1}(a)a\cdot\xi_r\bigl(3\gamma_{r-1}(a)a\bigr)
    =a\cdot\gamma_r(a).
\end{align*}
For each fixed $r$, this recurrence preserves subpolynomiality and hence also proves the bound in \cref{thm:mondep-r-types}.

If $\Cc$ is translinear, then $\gamma_1$ can be chosen constant, and linear neighborhood complexity of $\Dd_r$ allows us to choose $\xi_r$ to be a constant $c_r$. The recurrence then reduces to
\[
    \gamma_r=3c_r\gamma_{r-1}.
\]
Consequently, for every fixed $r$, the growth ratio of $\Tt_r$ on $\Cc$ is bounded by a constant depending only on $\Cc$ and $r$, as used in \cref{sec:conclusions}.

\begin{corollary}\label{cor:translinear-growth}
	Let $\CC$ be a translinear graph class and $r\in\N$. Then the growth ratio of $\Tt_r$ on $\CC$ is bounded by a constant $\gamma_r\le \Oh_{\Cc,r}(1).$
\end{corollary}
\section{Algorithmic claim of \texorpdfstring{\cref{thm:alternation-general}}{Theorem \ref{thm:alternation-general}}}\label{app:welzl}

In this section we argue that the proof of \cref{thm:alternation-general} presented in \cite{BonnetD0Z25} gives rise to an $\Oh(\alpha^2 n^3)$-time algorithm that given a graph $G$, computes a vertex ordering $\preceq$ with alternation bounded by $\alpha$, where $\alpha$ is the quantity specified in the statement of \cref{thm:alternation-general}. The part relevant to us is the method presented in the proof of \cite[Theorem~26]{BonnetD0Z25}. This proof works in the generality of set systems; here, we specialize it to the set system of closed neighborhoods in a~graph.

The method presented in \cite{BonnetD0Z25} follows the classic approach of Welzl~\cite{Welzl89}; the contribution of \cite{BonnetD0Z25} lies in analyzing the construction to derive the explicit bounds stated in \cref{thm:alternation-general}. The method first constructs a tree $T$ on the vertex set of $G$ whose \emph{crossing number} is bounded by $\alpha'\coloneqq 
\log_2 n+5d\cdot \sum_{i=1}^n \frac{1}{f^{-1}(i/2)}$ in the following sense: every vertex $x$ of $G$ crosses at most $\alpha'$ edges $uv\in E(T)$, where $x$ \emph{crosses} $uv$ if exactly one of $u$ and $v$ belongs to $N[x]$. The vertex ordering $\preceq$ is then obtained by running depth-first search on $T$ and taking $\preceq$ to be the preorder of $T$ yielded by this search. As argued in \cite{BonnetD0Z25}, the crossing number of $\preceq$ (interpreted as a path consisting of edges joining vertices that are consecutive in $\preceq$) is at most doubled compared to $T$. This implies a bound of $2\alpha'$ on the crossing number of $\preceq$, which translates to a bound of $\alpha=2+2\alpha'$ on the alternation of $\preceq$ in our terminology (the additive $+2$ comes from accounting for the two neighbors of a vertex $x$ in $\preceq$). As running a depth-first search on $T$ can be done in $\Oh(n)$ time, it remains to recall the procedure constructing $T$ and argue that it can be implemented in $\Oh(\alpha^2 n^3)$ time.

The procedure performs $n-1$ steps, where the $i$th step adds one new edge $u_iv_i$ to $T$. During the procedure we maintain a weight function $\wei\colon V(G)\to \N$ that satisfies the following invariant after every $i$th step:
\begin{equation}\label{eq:kolczatka}
	\wei(x)=2^{k},\qquad\textrm{where }k\textrm{ is the number of edges in }\{u_1v_1,\ldots,u_iv_i\}\textrm{ crossed by }x.
\end{equation}
Note that since at the end, every vertex $x$ crosses at most $\alpha'$ edges of $T$, the weights are bounded by $2^{\alpha'}$ at all times.

The $i$th step proceeds as follows:
\begin{itemize}
	\item Select a pair of distinct vertices $u,v\in V(G)\setminus \{u_1,\ldots,u_{i-1}\}$ that, among such pairs, minimizes~the~value
	\[\xi_{uv}\coloneqq \sum_{x\in V(G)\colon x\textrm{ crosses }uv} \wei(x).\]
	\item Set $u_iv_i\coloneqq uv$ and update the weight function $\wei$ by doubling the weight of every vertex $x$ that crosses $uv$, so that invariant \eqref{eq:kolczatka} is maintained.
\end{itemize}
Observe that the condition $u,v\notin \{u_1,\ldots,u_{i-1}\}$ guarantees that orienting the edges of $T$ towards their $u$-endpoints yields an acyclic orientation in which every vertex has indegree at most $1$; so at the end, after adding $n-1$ edges, $T$ becomes a tree spanning the whole vertex set of $G$. As argued in \cite{BonnetD0Z25}, the crossing number of the final tree $T$ is bounded by $\alpha'$. Note that the algorithm works with access to $G$ only and is oblivious to the function $f$.

To implement the algorithm efficiently, we explicitly maintain the weights $\wei(x)$ for $x\in V(G)$ and the values $\xi_{uv}$ for $uv\in \binom{V(G)}{2}$. Note that every value $\xi_{uv}$ is bounded by $2^{\alpha'}\cdot n$ at all times, and assuming the RAM model with words of length $\Oh(\log n)$, we can perform additions and comparisons on numbers of such magnitude in time $\Oh(\alpha')$ per operation. Initially all the weights $\wei(x)$ are set to $1$ (this takes $\Oh(n)$ time) and initializing the values $\xi_{uv}$ can be done in $\Oh(n)$ time per value, so in $\Oh(n^3)$ time in total.

In the $i$th step of the procedure, we first choose a suitable pair $uv$ using $\Oh(n^2)$ arithmetic operations by iterating through all the relevant pairs and choosing one that minimizes $\xi_{uv}$. Next, we construct a list $L_i$ consisting of all the vertices $x$ that cross $uv$; this takes time $\Oh(n)$. For every vertex $x\in L_i$, we double the weight of $x$ and update the values $\xi_{u'v'}$ for all pairs $u'v'\in \binom{V(G)}{2}$ that are crossed by $x$; the latter is done by adding the pre-update weight of $x$ to $\xi_{u'v'}$. This requires $\Oh(n^2)$ arithmetic operations per vertex $x$ on $L_i$, hence $\Oh(|L_i|\cdot n^2)$ arithmetic operations in total.

Here is the key observation of the analysis. The sum of the lengths of the lists $L_i$ counts the total number of times the crossing number of some vertex increases; but at the end, every vertex $x$ crosses at most $\alpha'$ edges of $T$. Therefore we have $\sum_{i=1}^{n-1} |L_i|\leq \alpha' n$. It follows that the whole iteration uses $\Oh(n^2\cdot \sum_{i=1}^{n-1} |L_i|)\leq \Oh(\alpha'\cdot n^3)$ time and arithmetic operations in total. As every arithmetic operation takes $\Oh(\alpha')$ time and $\alpha'\leq \alpha$, we arrive at the promised time complexity of $\Oh(\alpha^2n^3)$.

\end{document}